%% file: paper.tex
\documentclass[lettersize,journal]{IEEEtran}
\usepackage{cite}
\usepackage{amssymb}
\usepackage{standalone}
\usepackage{algorithm}
\usepackage{algpseudocode}

\usepackage{subcaption,tikz,lipsum}
\usepackage{lscape}  
\usepackage{booktabs}
\usepackage[utf8]{inputenc}
\usepackage{longtable}
\usepackage[textsize=scriptsize,textwidth=1.4cm]{todonotes}
\usepackage{makecell}
\usepackage{wrapfig}

\usepackage{amsmath,amsfonts}
\usepackage{array}
\usepackage{textcomp}
\usepackage{stfloats}
\usepackage{url}
\usepackage{verbatim}
\usepackage{graphicx}
\def\BibTeX{{\rm B\kern-.05em{\sc i\kern-.025em b}\kern-.08em
    T\kern-.1667em\lower.7ex\hbox{E}\kern-.125emX}}
\usepackage{balance}

\def\BibTeX{{\rm B\kern-.05em{\sc i\kern-.025em b}\kern-.08em
    T\kern-.1667em\lower.7ex\hbox{E}\kern-.125emX}}
\usepackage{scalerel,stackengine}
\stackMath
\newcommand\reallywidehat[1]{%
\savestack{\tmpbox}{\stretchto{%
  \scaleto{%
    \scalerel*[\widthof{\ensuremath{#1}}]{\kern.1pt\mathchar"0362\kern.1pt}%
    {\rule{0ex}{\textheight}}
  }{\textheight}%
}{2.4ex}}%
\stackon[-6.9pt]{#1}{\tmpbox}%
}

\newcommand{\defsign}{\triangleq}
\newcommand{\Radon}{\mathcal{R}}
\newcommand{\FBP}{\reallywidehat{\Radon^{-1}}}

\begin{document}

\title{Scatter Correction in X-Ray CT \\by Physics-Inspired Deep Learning}
\author{
Berk Iskender and Yoram Bresler
\thanks{This work was supported in part by the NSF under Grant IIS 14-47879, and by US Army MURI Award W911NF-15-1-0479.
 } 
\thanks{Berk Iskender and Yoram Bresler are with the Department of Electrical and Computer Engineering, and the Coordinated Science Laboratory, University of Illinois at Urbana-Champaign, Urbana, IL 61801 USA. 
(e-mail: berki2@illinois.edu; ybresler@illinois.edu)}
 \vspace*{-1cm}}

\maketitle

\input{abstract}


\input{intro}

\input{background}

\input{proposed_algorithms}

\input{methodology}

\input{experiments}

\input{conclusions}

\bibliographystyle{IEEEtran}
\small{\bibliography{refs}}

\input{appendix}

\end{document}

%% file: abstract.tex
\begin{abstract} Scatter due to interaction of photons with the imaged object is a fundamental problem in  X-ray Computed Tomography (CT). It manifests as various artifacts in the reconstruction, making its abatement or correction   critical  for image quality.
Despite success
in specific settings, hardware-based methods require modification in the hardware, or increase in the scan time or dose. This accounts for the great interest in software-based methods, including Monte-Carlo based scatter estimation, analytical-numerical, and kernel-based methods, with  data-driven learning-based approaches demonstrated recently. 
In this work, two novel physics-inspired deep-learning-based methods, PhILSCAT and OV-PhILSCAT, are proposed. The methods estimate and correct for the scatter in the acquired projection measurements. Different from previous works, they incorporate both an initial reconstruction of the object of interest and the scatter-corrupted measurements related to it, and use a deep neural network architecture and cost function, both specifically tailored to the problem.     
Numerical experiments with data generated by Monte-Carlo simulations of the imaging of phantoms reveal
consistent
improvement over a recent purely projection-domain deep neural network scatter correction method.
\end{abstract}

%% file: intro.tex
\section{Introduction}
\label{sec:intro}
\IEEEPARstart{C}{T} is widely used for imaging internal structures of the body, for pre-clinical imaging, and for non-destructive evaluation \cite{floyd1984energy}. However, scatter occurring due to the interaction of radiation with the imaged object degrades the reconstruction  causing streaks, cupping, shading artifacts and decrease in contrast. These severe artifacts due to scatter make its prevention or correction a critical component in any CT system.

De-scattering methods fall into two main categories: hardware-based; and software-based. Hardware-based methods include collimation, introducing a bow-tie filter in front of the X-ray source, increasing the distance between the detector and scattering object, the use of an anti-scatter grid, etc.  \cite{ruhrnschopf2011general}. These methods are successful in particular settings. However, they are either costly to implement or they subject the patient to a greater X-ray dose, which may pose a health risk. 
This motivates the great interest in software-based scatter estimation and correction methods, 
which in turn, fall into two classes: (i) methods that estimate the scatter for an initial, scatter corrupted reconstruction of the  object using a forward solver, and subtract out the computed scatter contribution ; and (ii) methods that work directly on the total projection data. 

We review Class (i) methods and their forward solvers first. 
One of the main approaches to model scatter in X-ray CT are Monte-Carlo (MC) solvers,
which stochastically sample photon propagation to estimate the scatter for a given object.
However, despite their potential to produce gold-standard estimates of scatter when a large number photons is used \cite{poludniowski2009efficient}, for clinical purposes the resulting computational costs and runtimes of MC-based methods are prohibitive.

Unlike the MC stochastic modeling of  scatter, the linear Boltzmann transport equation (LBTE) specifies a solution for the expected value of the scatter. 
To mitigate the computational cost of solving the LBTE, which is an integro-differential equation in a seven dimensional space, analytical-numerical forward solvers use  simplifying assumptions to find an approximate solution \cite{maslowski2018acuros,shiromascatter}. 
Although faster than MC, this approach involves a trade-off between 
discretization and the maximum order of scatter that is modeled, and the accuracy of the scatter estimate, therefore it too can be computationally expensive.

The third method for the forward problem  is the slice-by-slice method \cite{bai2000slice}, which models the scatter as a distance-dependent incremental blurring effect represented by analytically computed kernels.
This method has a reduced computational cost compared to the first two. However, it is unable to keep track of the angular distributions of the incoming photons to a slice and does not account for any multiple order scattering events in the medium, which limit its effectiveness.

In contrast to the forward-solver based methods, the Class (ii) methods work directly on the total, scatter corrupted projection data. They can be further classified, in turn, into two categories: kernel-based scatter estimation; and data-driven scatter estimation in the projection domain.

Given scatter-corrupted total measurements $\tau(t,\theta)$, kernel-based scatter estimation methods attempt to determine the scatter in 
$\tau(t,\theta)$ by convolving its weighted version $\tilde{\tau}(t,\theta)$
with a specific function of $t$ - a ``kernel" \cite{ohnesorge1999efficient,zhao2015patient,star2009efficient} for each view $\theta$.
These estimation methods are computationally efficient but prior assumptions such as neglecting the contribution of scatter to 
$\tau$ and pre-defined kernels with few degrees of freedom restrict their effectiveness. A hybrid method combining a MC and a kernel-based approach is studied in \cite{Baer2012HybridSC}.

Instead, data-driven approaches utilize neural networks to estimate scatter. Image domain methods that correct scatter using scatter-corrupted reconstructions include
 a method with two-step registration of CT-CBCT pairs \cite{xie2018scatter}, and a method that estimates and subtracts out the scatter corruption in the image domain using a deep residual CNN \cite{scatter_corr_drcnn}. 
Projection domain methods that estimate the scatter component from total scatter-corrupted measurements include
a  CBCT method using a scatter library of breast CT to estimate scatter \cite{shi2016library}, a two-network approach that learns scatter in the projection domain by separating it into high and low frequency components \cite{lee2019deep}, and
a method called Deep Scatter Estimation (DSE in short) \cite{maier2018deep,maier2019real} that operates on the projection domain and uses a modified U-net \cite{ronneberger2015u} architecture with an additional average pooling path for better extraction of features. {A similar approach can also be found in \cite{nomura2019projection}.} 
The image-domain methods do not have direct access to the scatter-corrupted measurements, nor do they estimate the scattered X-rays directly, and therefore are not interpretable, and are difficult to relate to the physics of the problem. The projection domain methods are less subject to these limitations, but because they use little or no information of the 3D object structure, which ultimately determines the scatter, their effectiveness is limited. 

\subsubsection*{Contributions} 
\label{sec:intro_contributions}
In this work (see also  \cite{iskender_bresler2020ISBI,iskenderx}),
unlike previous Deep Neural Network (DNN)-based approaches, we present scatter correction algorithms for X-ray CT based on a deep CNN that use both the raw projection data and an initial reconstructed image simultaneously. The data processing pipelines, network architectures, and loss function design for training of the proposed methods, are all inspired by the physics of  X-ray scatter. The tailored loss function  expresses the norm of a reconstruction domain error in the projection domain, avoiding the need to compute gradients (backpropagate) across
the filtered backprojection algorithm for every training sample, resulting in efficient network training. This loss function may therefore be of independent interest in other work on deep-learning methods in tomography.
As a benchmark for comparison, we use the projection domain DSE method, which is also physics inspired, in that it can be interpreted as a learned kernel-based scatter estimation method \cite{maier2018deep,maier2019real}.

We study not only the widely used polychromatic X-ray CT, but also the monochromatic case, for two reasons. First, it provides substantial insight into the problem and its solution, because in the monochromatic case, the only deviation from the ideal (post-log) linear measurement model is due to scatter. 
Hence, the effect of scatter and its mitigation can be clearly evaluated. In contrast, in the polychromatic case, beam hardening (another nonlinearity) confounds the problem and the interpretation of the results. 
In fact, the distinct subject of mitigating beam hardening has received much research attention.
Second,
the monochromatic case is of great independent practical interest, thanks to the unique applications it enables, and the recent  availability of compact low-cost monochromatic sources \cite{hornberger2019compact,bazzani2020bocxs}. Likewise, while the cone-beam  geometry is common in 3D CT, the parallel-beam geometry arises in monochromatic synchrotron CT imaging with its unique and important applications \cite{hwu2017q,Newham2020SynchrotronRX,baba2013improving,voltolini2017emerging}, and is therefore of considerable practical interest. We present a specialized version of our scatter reduction algorithm for the parallel-beam geometry, which takes advantage of additional structure in the scatter physics, to further mitigate the scatter.  

The paper is organized as follows.
We set up the scatter correction problem in Sec.~\ref{sec:background_and_related_work}, 
describe the proposed algorithms in Sec.~\ref{sec:physics_motivated_algorithms}, and provide the methodology and framework for the numerical experiments in Sec.~\ref{sec:methodology}, with the results of the various numerical experiments  in Sec.~\ref{sec:experiments}. We conclude and indicate possible future directions for research in Sec.~\ref{sec:conclusion}.

%% file: background.tex
\section{Problem Setup}
\label{sec:background_and_related_work}
\subsubsection{X-Ray CT}
\label{sec:xray_ct}
\begin{figure}[t]
    \centering
    \begin{subfigure}[t]{0.24\textwidth}
        \centering
        \includegraphics[width=0.99\textwidth]{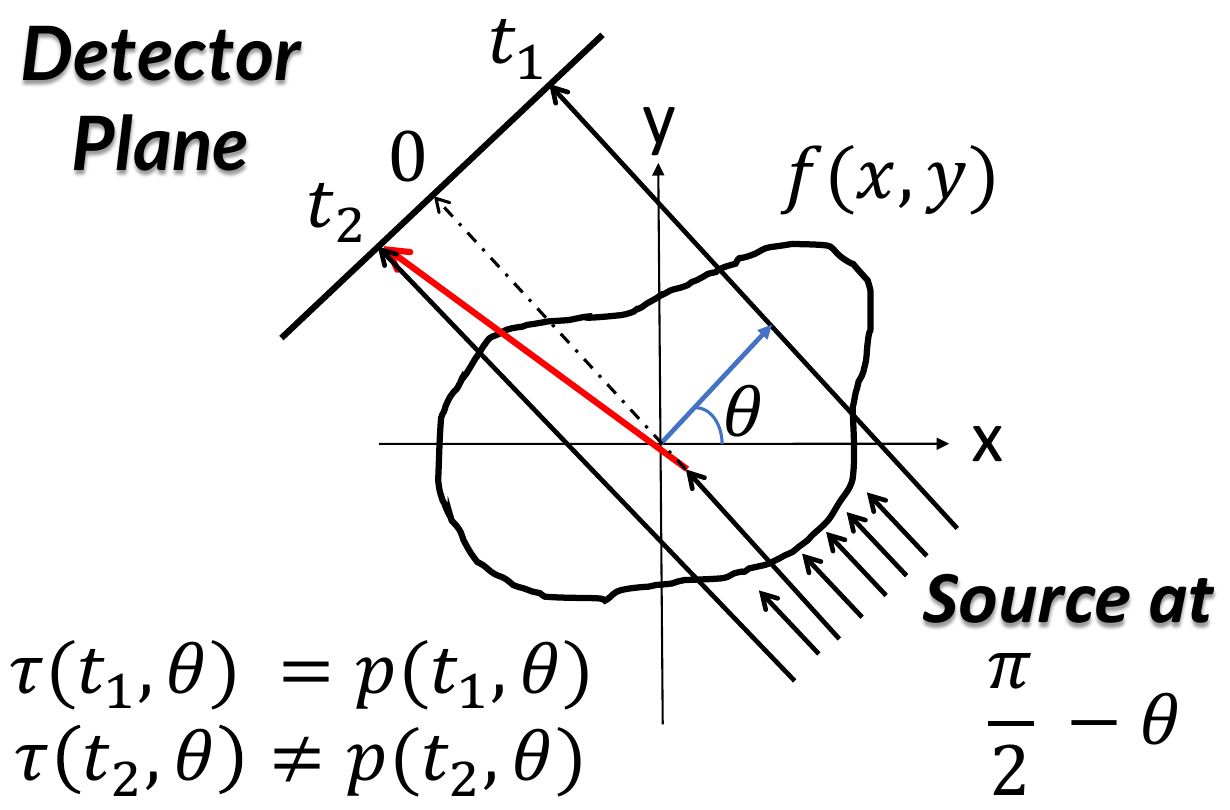}
        \caption{}
    \end{subfigure}%
    ~ 
    \begin{subfigure}[t]{0.24\textwidth}
        \centering
        \includegraphics[width=0.99\textwidth]{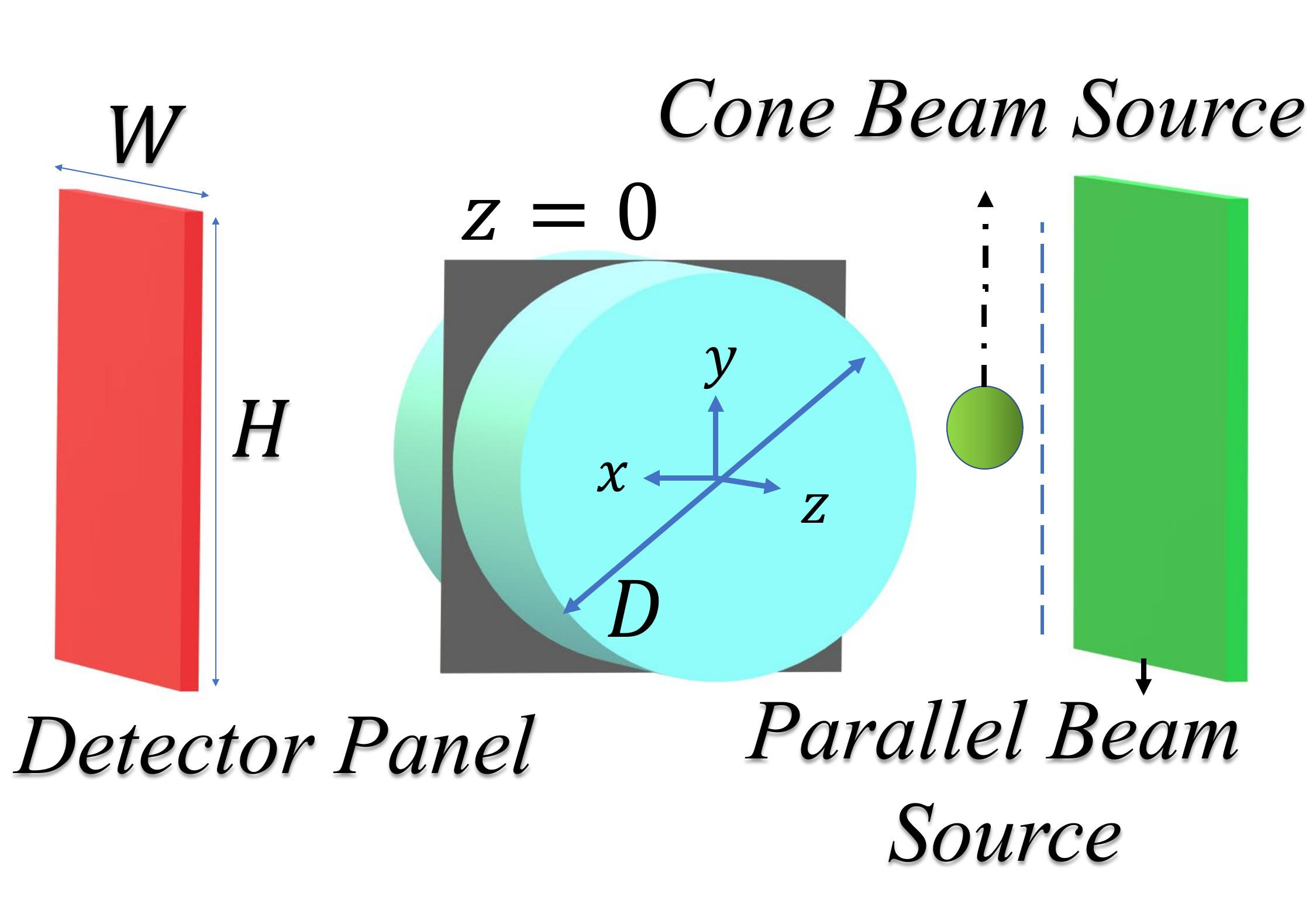}
        \caption{}
    \end{subfigure}
    \caption{\small  CT geometries and  X-ray  scatter. (a) 2D parallel-beam geometry.
    A scattered x-ray reaches the detector at $(t_2,\theta)$ instead of $(0, \theta)$, making the total measurement $\tau(t_2, \theta)$ differ from the primary $p(t_2, \theta)$.
    (b) 3D imaging geometries for cone and parallel beam CT reconstruction experiments shown together. The central axial 2D slice on the x-y plane indicated by the gray rectangle at $z=0$.}
    \vspace{-0.4cm}
    \label{fig:scatter_analytical_level_illustration}
\end{figure}

In a 2D setting with a parallel beam source, shown in Fig.~\ref{fig:scatter_analytical_level_illustration}-(a), let $f$ denote the object (i.e., the desired image) with $f(\textbf{x})$ the linear attenuation coefficient at position $\textbf{x} \in \mathbb{R}^2$.
The line integral of $f$ along the ray parametrized by offset (detector position) $t$ and angle $\theta$ is denoted by $g(t,\theta)$. For fixed $\theta$, the function $g(\cdot,\theta)$ is a projection of $f$ at view angle $\theta$, and the mapping from $f$ to the complete set of line integral projections $g \triangleq \{ g(t,\theta), \theta \in [0,\pi), t \in \mathbb{R}\}$
is the 2D Radon transform, a linear operator denoted by $\Radon$, 
\begin{equation}
\label{eq:radon_transform}
    g(t,\theta) = (\Radon f)(t,\theta) \, .
\end{equation}

Using the standard setup, we assume projections are measured at a finite uniformly spaced set of view angles, 
$\theta \in \Theta = \{2k\pi/K, k= 0,1,\ldots,K-1\}$), and at a finite set of ray offsets $T\defsign \{t =i \Delta_t, i= -(d-1)/2, \ldots 0, \ldots (d-1)/2\}$
 per view,
resulting in a discrete set of projections,
$g \defsign \{ g(t,\theta), \theta \in \Theta, t\in T \}$.
We use $\mathbf{g}_{\theta} \in \mathbb{R}^d$  to denote the discrete projection at angle $\theta$ -- the vector of $d$ uniformly-spaced samples of $g(t,\theta)$ along the detector position coordinate, $t$.

The 
reconstruction problem is to compute the inverse Radon transform $ f(\textbf{x})=(\Radon ^{-1}g)(\textbf{x})$.
In the discrete data case, with the usual assumption that $f$ is essentially bandlimited and supported on a bounded set, and the sampling in $\theta$ and $t$ is dense enough, the discrete-index filtered backprojection (FBP) is a good numerical approximation to the inverse of the Radon transform \cite{doi:10.1137/9780898717792}. We 
denote the FBP by $\FBP $, to emphasize that we assume that the conditions for accurate reconstruction by FBP are satisfied, and focus on the error due to scatter. 
To account for the dependence  
on the energy spectrum the source, we consider the two types of sources: polychromatic (emitting photons with a broad range of energies), and monochromatic (emitting essentially monoenergetic rays). 

Consider a 2D object
with 
a parallel beam source. 
Denoting the energy-dependent linear attenuation coefficient \cite{huda2003review}  of the object 
at source energy $E$ by $f_E(\textbf{x})$,  its projection is
\begin{equation}
\label{eq:radon-poly}
    g_E(t,\theta) = (\Radon f_E)(t,\theta) .
\end{equation}
Using an energy-integrating detector, the \emph{primary measurement}
is, by Beer's law \cite{beer1852},
\begin{equation}
\label{eq:polychromatic}
    p(t,\theta) = I_0 \int c(E)e^{-g_{E}(t,\theta)} dE,
\end{equation}
where $I_0$ is the vacuum (or bright field) fluence measurement, and $c(E)$ is a  function of the source spectrum and energy-dependent detector response, with 
$\int c(E)dE = 1$. 

With a monochromatic source with photon energy $E_0$, \eqref{eq:radon-poly} 
reduces to \eqref{eq:radon_transform} and \eqref{eq:polychromatic} reduces to
\begin{equation}
    \label{eq:beers_law}
    p(t,\theta) = I_0 e^{-g(t,\theta)}, 
\end{equation}
where the dependence on $E_0$ is suppressed to simplify notation, 
and the line integral projection is readily extracted from the measurement by a logarithm, $g(t,\theta)=-\ln[p(t,\theta)/I_0]$.
On the other hand, in the polychromatic case,   the mapping \eqref{eq:polychromatic} from $g_E$ to the primary measurements $p$ involves another nonlinearity in addition to the exponential, which cannot be inverted by taking the logarithm. 
This nonlinearity, unless corrected, may manifest as beam hardening artifacts in the reconstruction. 
To focus on scatter correction only, we limit the discussion 
in the remainder of this section and in Secs.~\ref{sec:physics_motivated_algorithms}-\ref{sec:methodology} to the monochromatic case, where the only deviation from the ideal measurement model is due to scatter. However, the general approach 
can be extended to the polychromatic setting to handle scatter and beam hardening simultaneously. This is demonstrated in the numerical experiments in Sec.~\ref{sec:experiments}.

Given the primary measurements $p$, the projections $g$ determined by inverting \eqref{eq:beers_law} suffice to obtain an accurate reconstructions by FBP. However, as discussed next, due to X-ray scatter, the primary measurements are corrupted by an additive scatter component, which unless blocked in the first place by physical means, or corrected, results in artifacts in the reconstruction.

\subsubsection{X-Ray Scatter}

The only significant source of scatter at X-ray energies of 30 keV - 450 keV  used in pre-clinical and medical CT 
(\textless140 keV) and non-destructive tomography (NDT) 
(\textless450 keV), 
is Compton scatter, in which 
an incident photon  is scattered (Fig.~\ref{fig:scatter_analytical_level_illustration}-(a)) by an electron, with both energy and the propagation direction of the photon significantly modified.
With many such scattering events, the \emph{total measurement} (detector reading) at angle $\theta$ that is obtained is 
\begin{equation}
\label{eq:primary+scatter}
    \tau(t,\theta) = p(t,\theta) + s(t,\theta),
\end{equation}
where $s(t,\theta)$ is an additive \emph{scatter term}, which is a nonlinear function of the object.
It is the contribution of this additive term that leads to artifacts in conventional reconstruction where FBP is directly implemented using the total, scatter corrupted, measurement $\tau$ instead of the ideal primary measurement $p$ to obtain an estimate $f^* \in \mathbb{R}^{d^2}$ of the image. 

While any practical CT measurements also include random noise due to finite number of photons and electronic noise at the detector, in our discussion we assume sufficient photon counts and sufficiently small electronic noise in the measurements that the reconstruction error is dominated by the deterministic bias due to scatter.
\subsubsection{Problem Statement}
We assume that we are given a set of total measurements 
$ \boldsymbol{\tau} \defsign \{ \boldsymbol{\tau}_\theta, \theta \in \Theta \}$
which, in the absence of the scatter component $s$, would suffice for accurate reconstruction of the object $f$ by FBP. Our goal is to produce a reconstruction $f^\ast$ that approximates the FBP reconstruction $\hat{f} =\FBP g$ that would be obtained from $g(t,\theta)=-\ln[p(t,\theta)/I_0]$, where $p$ is the, scatter-free, set of primary measurements.

%% file: proposed_algorithms.tex
\section{Physics-Inspired Scatter Correction}
\label{sec:physics_motivated_algorithms}

To provide invariance to source intensity or exposure time,
the problem and method are formulated in terms of the normalized quantities, $\bar{\tau}_\theta \triangleq \tau_\theta/I_0$, $\bar{s}_\theta \triangleq s_\theta /{I_0}$, and  $\bar{p}_\theta \triangleq p_\theta/I_0$. We denote the set of normalized total measurements by $\bar{\tau}  \triangleq \{\bar{\tau}_\theta, \theta \in \Theta\}$, with $\bar{\tau}_\theta \in \mathbb{R}^d$.  Given $\bar{\tau}$, the problem is to produce an estimate $\bar{p}^*$ of the scatter-free primary, which is then used to reconstruct the estimate $f^*$ using FBP. 

\subsection{PhILSCAT: Physics-Inspired Learned Scatter Correction Algori{T}hm}
\label{sec:ISBI_alg}

\begin{figure}
    \centering
    \includegraphics[scale=0.38]{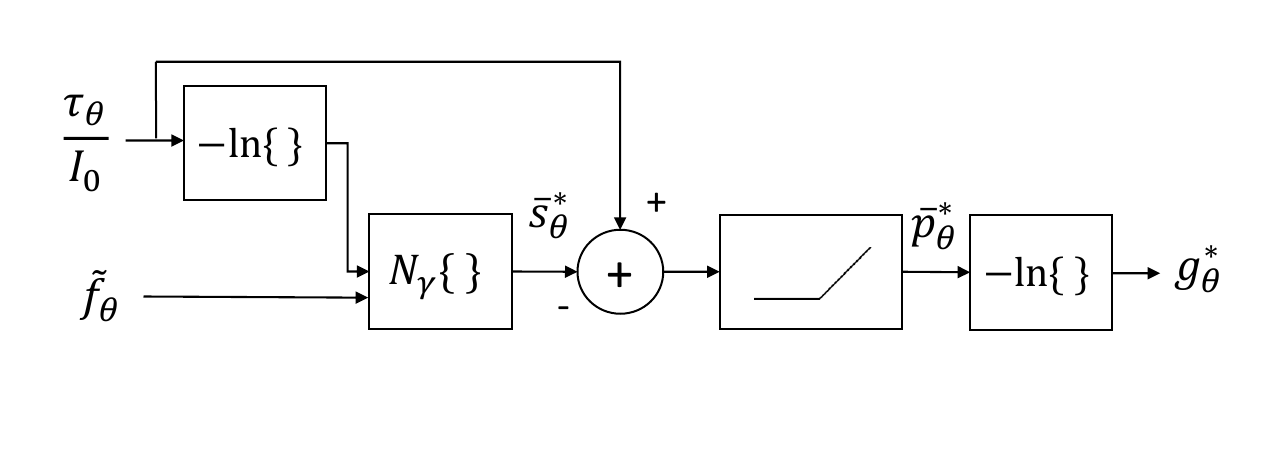}
    \vspace{-0.5cm}
    \caption{\small Block diagram of PhILSCAT.}
    \vspace{-0.5cm}
    \label{fig:block_diagram_of_the_ISBI_algorithm}
\end{figure}
The key idea in the proposed approach \cite{iskender_bresler2020ISBI}, illustrated in Fig.~\ref{fig:block_diagram_of_the_ISBI_algorithm}, is that because scatter in any one direction depends on the \emph{entire} object in a nonlinear fashion,
the measurement in one direction cannot be used to fully determine the  scatter in that direction. Instead, information about the entire object, which aggregates the information of all views is required. Thus, given a set of 
total measurements, $\bar{\tau}$, an initial reconstruction estimate $\tilde{f} \in \mathbb{R}^{d^2}$ of the image is computed by FBP
\begin{equation}
    \label{eq:initial_recon_estimate}
    \tilde{f}(\textbf{x}) = (\FBP \tilde{g})(\textbf{x}),
\end{equation}
using the initial estimate of the line integral projection 
\begin{equation}
    \label{eq:initial_g}
\tilde{g}(t,\theta) = -\ln \min \{\bar{\tau}(t,\theta),1\} ,
\end{equation}
where $\tau(t,\theta)$ is used as a surrogate for the primary measurement $p(t,\theta)$. By the inherent physics in \eqref{eq:polychromatic} or \eqref{eq:beers_law}, $p(t, \theta)$ must be smaller than the mean bright field fluence $I_0$. Hence, this physical constraint is used in \eqref{eq:initial_g} to improve the initial reconstruction, by clipping of $\bar{\tau}(t,\theta)$ to $1$.

The deep CNN (DCNN), $\mathcal{N}_\gamma$ with network parameters $\gamma$, operates view-by-view. It accepts two different inputs:

\begin{enumerate}
    \item[(i)] 
    A normalized post-log total measurement $-\ln \bar{\tau}_\theta$ at view angle $\theta$. Here, unlike \eqref{eq:initial_g}, 
    $\bar{\tau}_\theta$ at the input to the DCNN is not upper bounded by $1$, because values greater than $1$ are physically possible after the normalization by $I_0$ and they provide useful information for the estimation of the normalized scatter term $\bar{s}_\theta$.
    \item[(ii)] 
    A version $\tilde{f}_\theta \in \mathbb{R}^{d^2}$ of the initial reconstruction estimate that is rotated by the same angle $\theta$ of the projection being processed. As a consequence of the rotation, a projection of $\tilde{f}_\theta$ at zero angle yields the projection of $\tilde{f}$ at angle $\theta$, allowing the DCNN to be agnostic to $\theta$.
\end{enumerate}

 The DCNN returns an estimate $\bar{s}_\theta^* \in \mathbb{R}^d$ of the scatter
\begin{equation}
    \label{eq:scatter_est_DCNN_ISBI}
    \bar{s}_\theta^* = \mathcal{N}_\gamma\left( \tilde{f}_\theta, -\ln \bar{\tau}_\theta \right).
\end{equation}

The normalized primary measurement is 
estimated as 

\begin{equation}
\label{eq:primary_measurement_algorithm}
    \bar{p}_\theta^* = \max\{ \bar{\tau}_\theta - \bar{s}_\theta^* , \epsilon \}.
\end{equation}
where $\epsilon$ is a small constant to prevent photon starvation artifacts.
The clipping of $\bar{p}^*$ is again due to the physical constraint imposed by \eqref{eq:polychromatic} or \eqref{eq:beers_law}.  Finally, an estimate $g_\theta^*$ of the projection is obtained as $g^*(t,\theta) = -\ln(\bar{p}^*(t,\theta))$. Once every view has been processed, 
the reconstructed image is obtained from $g^*$ by FBP with Shepp-Logan filtering \cite{shepp1974fourier}, 

\begin{equation}
    \label{eq:reconstruction_estimate_FBP}
    f^*(\textbf{x}) = (\FBP g^*)(\textbf{x}).
\end{equation}

\subsection{OV-PhILSCAT: Opposite-View processing PhILSCAT}
\label{sec:CTmeeting_alg}
This algorithm \cite{iskenderx}
is a variation on PhILSCAT, utilizing one more physical aspect of the tomographic measurement.
We take advantage of the following property of the 2D Radon transform:  the 
projections in opposite directions ($\pi$-\emph{opposite projections}) coincide up to a sign reversal (flip) in $t$: $ {g}(t,\theta)= {g}(-t,\theta+\pi) \triangleq \hat{g}(t,\theta)$. 
Combining 
with \eqref{eq:polychromatic} and \eqref{eq:beers_law} we have

\begin{equation}
\begin{aligned}
    {g}_\theta = \hat{{g}}_{\theta+\pi},\quad
    \bar{p}_\theta = \hat{\bar{p}}_{\theta+\pi},\\
    \bar{\tau}_\theta - \hat{\bar{\tau}}_{\theta+\pi} = \bar{s}_\theta - \hat{\bar{s}}_{\theta+\pi} \triangleq \Delta \bar{s}_\theta.\\
\end{aligned}
\end{equation}

It follows that the difference $\Delta \bar{s}_\theta$ of $\pi$-opposite scatter components can be determined \emph{exactly} from the available (scatter-corrupted) total measurements. Thus, estimating the average $\bar{b}_\theta \triangleq (\bar{s}_\theta + \hat{\bar{s}}_{\theta + \pi})/2$ suffices to fully determine the individual scatter components. This approach is motivated by the following hypothesis: 
\begin{itemize}
    \item [H1] Because the average $\bar{b}_\theta$ is typically smoother than $\bar{s}_\theta$ and  $\hat{\bar{s}}_{\theta + \pi}$,  $\bar{b}_\theta$ should be easier to learn by a neural network.
    
\end{itemize} 
In contrast, the difference $\Delta\bar{s}_\theta$ typically has higher bandwidth, however, we do not need to estimate it, since it can be extracted exactly from the total measurements (the measured data). 

 \begin{wrapfigure}{r}{0.21\textwidth}
\vspace{-0.3cm}
\begin{center}
\includegraphics[width=0.20\textwidth]{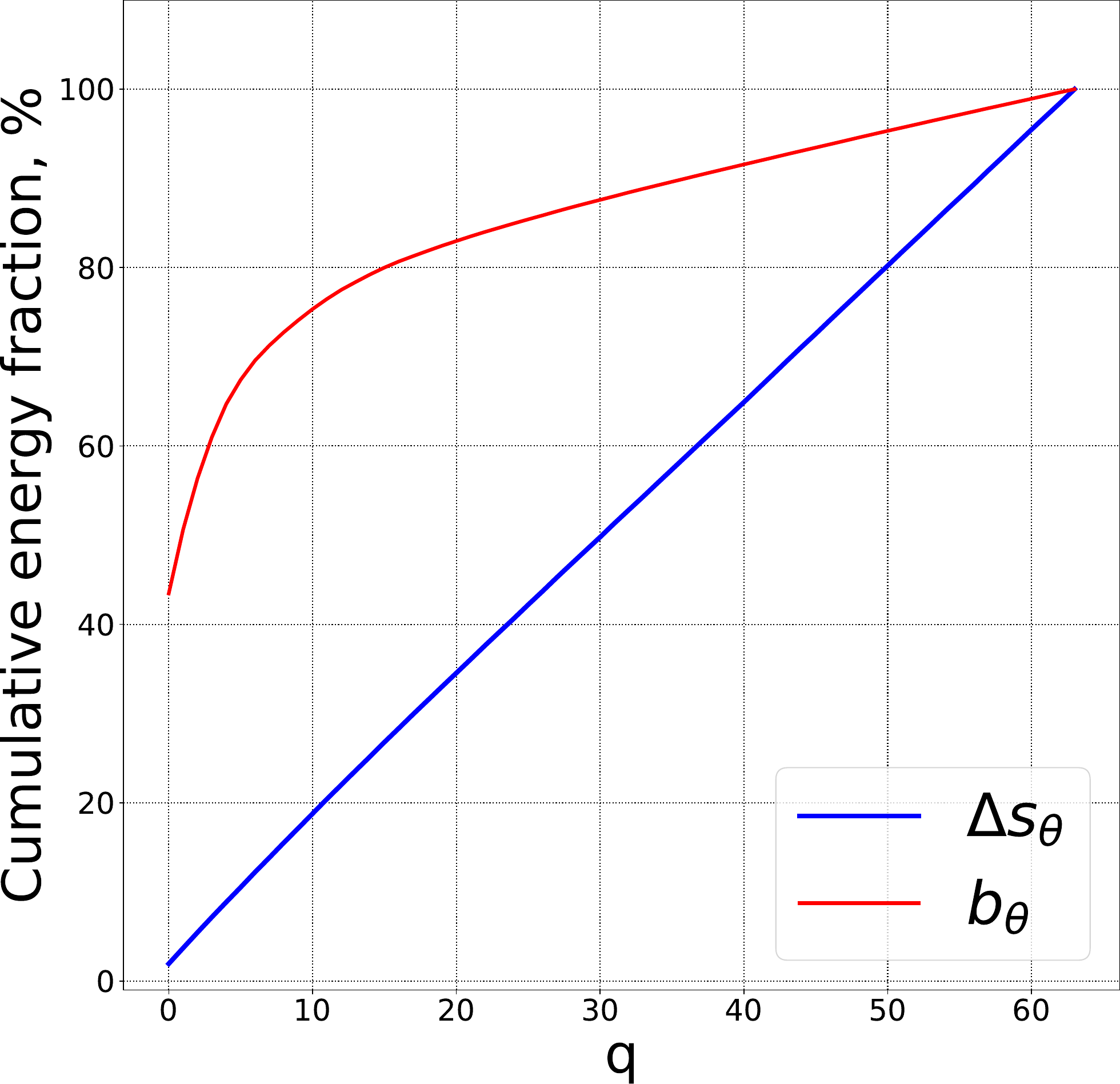}
\end{center}
\caption{\small 
Cumulative fraction of the total energy contained in the frequency components $[0,q]$. 
}
\label{fig:CTmeeting_b_delta_s_example}
\end{wrapfigure}
 The smoothness of $\bar{b}_\theta$ relative to $\Delta\bar{s}_\theta$ can be observed in Fig.~\ref{fig:CTmeeting_b_delta_s_example}, 
which depicts the normalized cummulative energy in the respective power spectral estimates computed over 27 different phantoms used in the
3D monochromatic parallel beam experiments (Sec.~\ref{sec:data_generation}).
As seen in Fig.~\ref{fig:CTmeeting_b_delta_s_example}, $\Delta s_{\theta}$ has a power spectrum close to that of white noise, with cumulative energy fraction increasing linearly with frequency. In contrast, $b_{\theta}$ has a large fraction of its energy concentrated in the first few frequency components: the DC component alone accounts for about 45\%, and the first 11 components capture about 75\% of the energy in $b_{\theta}$.

Given $b_\theta$, then, using the relation between the $\pi$-opposite primaries, it follows that 
 $   \bar{p}_\theta = 
    (\bar{\tau}_\theta + \hat{\bar{\tau}}_{\theta + \pi} )/{2} - \bar{b}_\theta$, so that the network only needs to estimate $\bar{b}_\theta$.

As in PhILSCAT, the DCNN $\mathcal{N}_\gamma$, operates view-by-view, but instead of $K$ times, ,
it is used here only $K/2$ times, for $\theta = k(2\pi/K)$, $k=0,1,\ldots,K/2 -1$.

The network takes two inputs:
\begin{enumerate}
    \item[(i)] the average $(\bar{\tau}_\theta + \hat{\bar{\tau}}_{\theta+\pi})/2$ of normalized pre-log $\pi$-opposite total measurements at view angle $\theta$,
    \item[(ii)] the initial reconstruction estimate rotated by $\theta$, $\tilde{f}_\theta \in \mathbb{R}^{d^2}$. Different than PhILSCAT, $\tilde{f}$ is obtained by using $(\bar{\tau}_\theta + \hat{\bar{\tau}}_{\theta + \pi})/2$ for $\theta = k(2\pi/K)$,  $k=0,1,\ldots,K/2 - 1$. 
\end{enumerate}

The DCNN produces an estimate of the average $\bar{b}_\theta^* \in \mathbb{R}^d$ of the normalized $\pi$-opposite scatter components
\begin{equation}
\label{eq:alg2_DCNN_output}
\bar{b}_{\theta}^{*}=\mathcal{N}_{\gamma}\left(\tilde{f}_{\theta}, -\ln[
(\bar{\tau}_{\theta}+\hat{\bar{\tau}}_{\theta+\pi})/2]
\right).
\end{equation}

The normalized primary projection $\bar{p}_\theta$ is then estimated as
\begin{equation}
\label{eq:Alg2-primary-est}
    \bar{p}_\theta^* = \max\left\{
    (\bar{\tau}_\theta + \hat{\bar{\tau}}_{\theta+\pi})/2
    - \bar{b}_\theta^*, 0 \right\},
\end{equation}
and, as in Section \ref{sec:ISBI_alg} to reflect the physical constraints, we constrain  $\bar{p}_\theta$ to be non-negative. Similarly, $\bar{\tau}_\theta$ is not clipped to $1$, since $\bar{\tau}_\theta > 1$ is physically possible. 

Finally, the reconstruction estimate is obtained using FBP with Shepp-Logan filtering \cite{shepp1974fourier}, as in \eqref{eq:reconstruction_estimate_FBP}, this time only using estimates of the projection $g^*_{\theta}$ from half of the angular range $\Theta = \{2k\pi/K,\,\, k = 0, 1, \ldots, K/2 - 1\}$.

Regardless of scatter correction, in scenarios with unavoidable subpixel misalignments, calibration and correction 
are required to obtain reasonable reconstructions. We propose that these corrections be applied as a first step, prior to the scatter correction. If 
successful, the initial 
(scatter-corrupted) and the scatter-corrected reconstructions can be obtained without 
calibration problems. 
However, the difference of conjugate projections $\Delta s_\theta$ could be more sensitive to residual subpixel misalignments than the sum signal $b_\theta$, or a regular reconstruction. Such misalignments could result in imperfect cancellation of the primary component in $\Delta s_\theta$ leading to “spike-like” artifacts at locations with sharp edges in the projection. We propose to add a step of median filtering of $\Delta s_\theta$ to mitigate these effects in such scenarios. Because without misalignment $\Delta s_\theta$ should contain only scatter signal, which would be relatively smooth, the median filtering will not affect it.

\input{LossFunctionNew}

\input{Two-NormNew}

\subsection{Extension to 3D} \label{sec:3D Extension}
In the 3D geometry (Fig.~\ref{fig:scatter_analytical_level_illustration}-b) the object axial coordinate $z$ is the rotation axis,  perpendicular to the cross-sectional $(x,y)$ plane.
In the case of a parallel beam source, we reconstruct a stack of 2D slices using 2D FBP, and for  a divergent (cone beam) source  we use the Feldkamp (FDK) reconstruction algorithm \cite{feldkamp1984practical}, both using the same Shepp-Logan ramp filtering. 

While the derivation of the image 2-norm in projection space in Sec.~\ref{sec:image_2norm} and the resulting formulation of the loss function in  
\eqref{eq:loss_fct} hold strictly only for the parallel beam case, they are applicable, to a good approximation, to CBCT with sufficiently small cone and angles. In fact, for cone angle small enough to avoid artifacts using the FDK reconstruction algorithm, the same approximation as in the FDK algorithm, which uses the 2D weighting and filtering, can be justified. Furthermore, for sufficiently small fan angles, the same filter response can be used, to a good approximation \cite{zhao2014fan}.  Accordingly,  denoting the  2D projections by $g_\theta(t,v)$, where $v$ is the detector coordinate parallel to $z$,
the loss function \eqref{eq:loss_fct} is extended to 3D by performing the convolution with $h$ along the transverse $t$ coordinate, and computing the 2-norms and 1-norms over the respective 2D projections.

One input of the  DNN is now the  3D initial reconstruction $\tilde{f}_\theta \in \mathbb{R}^{d \times d \times d}$ axially rotated by angle $\theta$. Coordinate $u$, which is perpendicular to the detector plane, selects coronal planes within $\tilde{f}_\theta$, each of which is indexed by detector coordinates $t,v$.
In the case of the parallel beam geometry, the $u$ coordinate coincides with the direction of photon propagation.

The second input to the DNN, indexed along $t,v$, is the 2D log total measurement   $-\ln \bar{\tau}_\theta\in \mathbb{R}^{d \times d}$ at one view angle for PhILSCAT, or the average of $\pi$-opposite views for OV-PhILSCAT, which is only used for the parallel beam geometry. 

\subsection{Network Structures}
\label{sec:network_structures}

\begin{figure*}[!hpt]
  \centering
  \vspace*{-0.75cm}
  \includegraphics[trim={0 0 0 0.65cm},clip,width=0.96\linewidth]{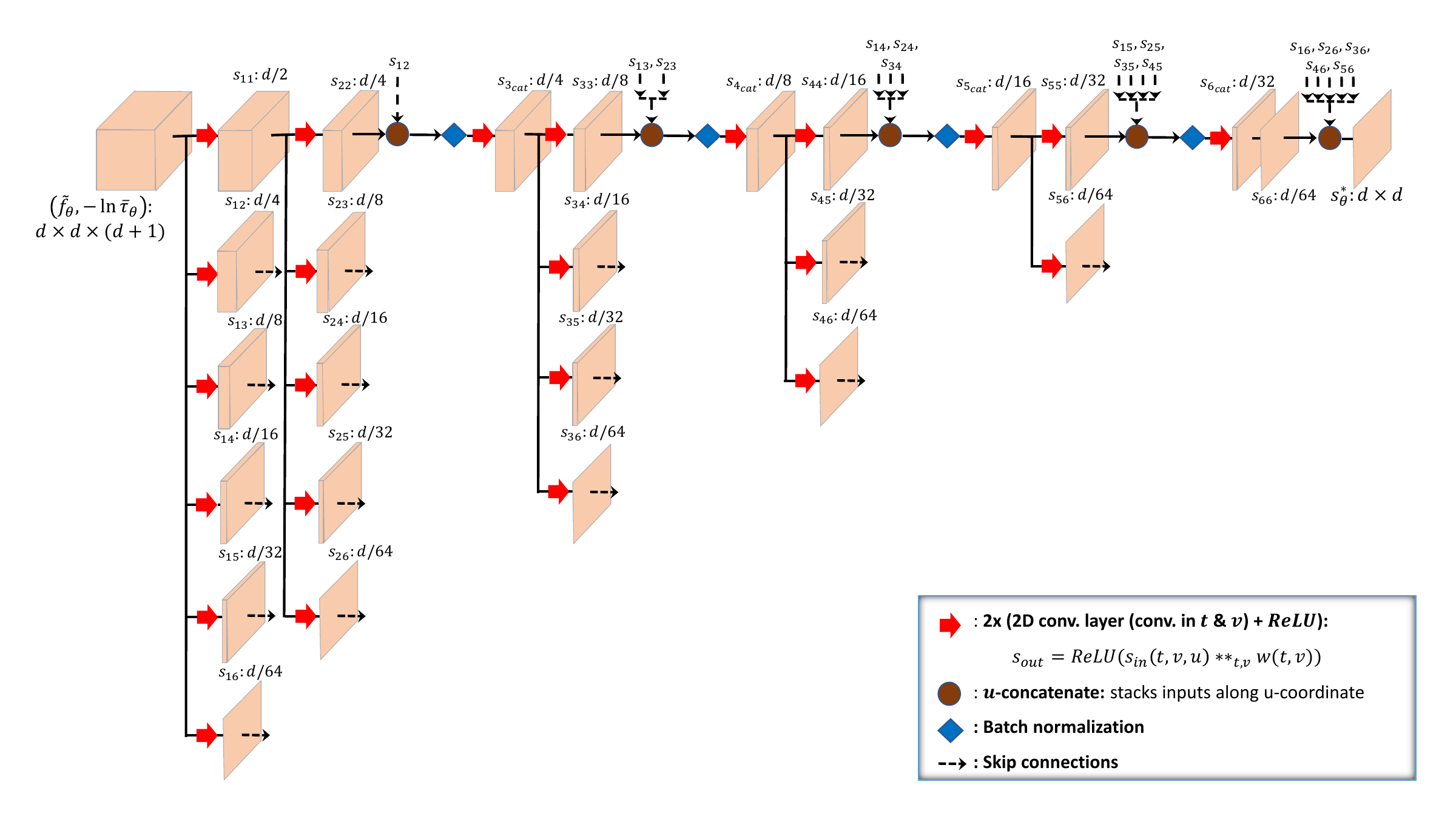}
  \vspace*{-0.5cm}
  \caption{\small Network architecture for PhILSCAT for 3D reconstruction (illustrated for $d=64$). Each red solid arrow represents two consecutive convolutional layers with ReLU nonlinearity modules. The number of channels in each intermediate output $s_\mathrm{ij} \in \mathbb{R}^{d \times d \times d/2^j}$ is provided next to its label in the diagram.}
  \vspace{-0.4cm}
  \label{fig:arch_alg_1}
\end{figure*}

The $d \times d \times (d+1)$ input to the DCNN (illustrated for $d=64$) in Fig. \ref{fig:arch_alg_1} is the axially rotated initial reconstruction $\tilde{f}_{\theta}$ concatenated along the channel coordinate $u$ with $-\ln{\bar{\tau}_\theta}$ (resp. 
$-\ln[ (\bar{\tau}_\theta+\bar{\tau}_{\theta+\pi})/2]$ for OV-PhILSCAT). The DCNN provides the $d \times d$ 2D output $s_\theta^*$ (resp. $b_\theta^*$ for OV-PhILSCAT).
The DCNN 
operates one view angle $\theta$ at a time, and has 2D convolutions in the
$(t,v)$ input planes, and a contracting structure in the channel coordinate $u$. 
The network architectures for PhILSCAT and OV-PhILSCAT are identical except for the input-output pairs as described earlier.

This specific structure is inspired by the slice-by-slice approach \cite{bai2000slice}, which divides the object into layers perpendicular of the primary X-ray propagation, and based on the Klein-Nishina formula \cite{klein1929streuung}, models scatter by a distance-dependent incremental blurring effect at each layer with pre-specified kernels to obtain the scatter estimate for the next layer. Accordingly, because after rotating the initial reconstruction $\tilde{f}$ by angle $\theta$ the slices of the input $\tilde{f}_\theta(\cdot,\cdot, u)$ correspond to object layers (coronal slices) perpendicular to propagation of primary X-rays, the network performs 2D convolutions (blurring) in the  plane of these slices, 
and contracts along the channel $u$. 

At the $m$-th step, for $m=1,\ldots,\log_2 d$, the network has a ``ladder" of convolutions compressing the input information by diadic factors to a set of outputs with number of channels equal to $ \{ 2^{-m}d, 2^{-(m+1)}d, \allowbreak \ldots, 1 \}$ using different convolutional layers, and conveys the output of those to the following steps via skip connections, concatenating all intermediate outputs from previous steps that have the same number of channels (e.g. $s_{14}$ is concatenated  with every $s_{m4}$, $m<4$ and $s_4$ at the end of the 4th step in the diagram, each having $2^{-4}d$ channels). After concatenation in the $m=2,3, \ldots $ step, another convolutional layer reduces the number of channels back to $2^{-m}d$. 

\subsection{Computational Cost in the Inference Phase}
\label{sec:computational_cost}

The computation in the DCNN is dominated by the cost of convolutions.
For 3D reconstruction, 2D convolutions with filters of size $L \times L$ are used. Assuming $d \times d$ detector panel readouts, convolutions in the network require about $L^2d^2$  operations per input and output channel in the convolution. The total cost of the DCNN can be shown to be bounded by $ 3L^2d^4 $. The DCNN is used $K$ times - once for each projection.
This results in total costs of approximately $3L^2Kd^4$ and $3L^2Kd^4/2$ for PhILSCAT and OV-PhILSCAT, respectively. The cost of FBP is $cd^4$ with $c \approx 1$ since it is performed for each slice of the $d \times d \times d$ dimensional phantom. Consequently, the total cost of PhILSCAT and OV-PhILSCAT in the 3D case can be bounded by $3L^2Kd^4$ and $3L^2Kd^4/2$, respectively. 

%% file: LossFunctionNew.tex
\subsection{Loss Function}
\label{sec:loss_fct}
The networks in both algorithms are trained by minimizing a common loss function with respect to the network parameters $\gamma$. In other work \cite{maier2018deep} on learning-based scatter estimation, the loss function is expressed in terms of estimation error  $p-p^*$ of the primary measurements. However, because the image produced by  FBP  depends logarithmically on $p$, it is the \emph{relative} error in $p^*$ that matters. Therefore, an error with the same magnitude in $p^*$ has greater impact on the reconstruction $f^*$ for smaller $p$ than for larger $p$. It follows that measuring the loss in terms of the error $p^*$ would not reflect the error in the reconstructed image, with the effect manifesting most strongly for rays passing through highly attenuating regions.

Instead, because our goal is to  approximate the scatter-free FBP reconstruction $\hat{f}$ by $f^*$, a reasonable choice for a loss function to represent the relevant deviation averaged over $M$ training samples would be 
\begin{align}
    L(\gamma) & = (1/M) \sum_{m=1}^M \hat{\Lambda}[\hat{f}_m,f^*_m(\gamma)] \nonumber \\
    \text{where } 
    \hat{\Lambda}[\hat{f},f^*(\gamma)]
    & \defsign \bigl \|Q[\hat{f}-f^*(\gamma)]\bigr \|_2^2 
     \nonumber \\ 
     \label{eq:Loss-f-to-g}
     &= \bigl \|\mathcal{Q} \FBP [g -g^*(\gamma)] \bigr \|_2^2 \, .
\end{align}
In \eqref{eq:Loss-f-to-g} $g_\theta^*(\gamma) = -\ln \bar{p}_\theta^*(\gamma)$  and $g_\theta = -\ln \bar{p}_\theta$ are the estimated and the true line integral projections of the object, with the dependence on $\gamma$ arising from
\eqref{eq:scatter_est_DCNN_ISBI}-\eqref{eq:primary_measurement_algorithm} for PhILSCAT, or
\eqref{eq:alg2_DCNN_output}-\eqref{eq:Alg2-primary-est} for OV-PhILSCAT. Operator $\mathcal{Q}$ represents possible
perceptually-oriented weighting of the reconstruction error to emphasize visually salient image features
in the image domain. In particular,  to account for the perceptual significance of  edges in CT images, we consider $\mathcal{Q}$ that corresponds to an image-domain filter with radially symmetric high-pass frequency response  $Q(\omega)$ (in polar coordinates).

A drawback of the loss function \eqref{eq:Loss-f-to-g}
is that to compute its gradients, it requires back-propagation across the FBP operator $\FBP$ for every training sample, which can be computationally expensive. Instead, we show in the next subsection that the loss \eqref{eq:Loss-f-to-g} can be expressed directly and exactly in terms of the projections, without the need for an FBP. This leads to the following modified loss function
\begin{equation}
\label{eq:loss_fct}
    \Lambda[g,g^*(\gamma)] = 
    \sum_{\theta\in \Theta} \bigl\|h * (g_{\theta} - g^*_{\theta})(\gamma)\bigr\|_2^2 + \lambda \bigl\|g_{\theta} - g^*_{\theta}(\gamma)\bigr\|_1.
\end{equation}
 The filtered $l2$-norm term is exactly equivalent to the loss in \eqref{eq:Loss-f-to-g}. 
 The filter $h$ is the simple two-tap filter 
\begin{equation}
    \label{eq:filter_h}
    h[n] = 0.5\delta[n+1] - 0.5\delta[n-1] ,
\end{equation}
resulting in a very efficient implementation of the loss function and its gradients.
Since $h$ is a high-pass filter, the $l_1$-norm term (with a small constant, $\lambda > 0$) is used to recover the zero frequency component of the projections. 
The specific choice of filter $h$ is explained in detail next. 

%% file: Two-NormNew.tex
\subsection{Image 2-Norm in Projection Space}
\label{sec:image_2norm}
 In this subsection we derive a simple expression for the reconstruction error in image space, in terms of the error in projection space. This enables to express the image-domain loss for training the neural network in terms of the estimated projections, and to obtain the gradients needed for training while avoiding the need to back-propagate across the FBP. The result in this subsection may therefore be of independent interest for other applications of deep learning in tomography.

Assuming  $f \in L_2(\mathbb{R}^2)$ is a finite energy image having line integral projection at angle $\theta$ denoted by $g(t,\theta)$, let $G(\omega,\theta)$ denote the one-dimensional Fourier transform of $g(t,\theta)$ with respect to the first variable. Then by  Parseval's relation for the Radon transform \cite{doi:10.1137/9780898717792}, the 2-norm  of the image $f$ is 
expressed  in terms its line integral projections $\{ g(t,\theta), -\infty < t < \infty, \theta \in [0,\pi] \}$ by

\begin{equation}
\label{eq:2norm_in_line_int}
\|f\|_{2}^{2}=\frac{1}{4 \pi^{2}} \int_{0}^{\pi} \int_{-\infty}^{\infty} \bigl \lvert G(\omega, \theta)\bigr \rvert^{2}|\omega| d \omega d \theta. \end{equation}

Note that although the inverse Radon transform to compute $f$ from $g$ involves applying the ramp filter $|\omega|$ to the Fourier transform $G(\omega, \theta)$ of the projections,  Parseval's relation in  \eqref{eq:2norm_in_line_int} involves applying the same ramp filter to the \emph{square} of $|G(\omega, \theta)|$ instead.

Now, we wish to express the 2-norm of an image $\hat{f}$, obtained from the  projections
by FBP using a weighted ramp filter $W(\omega)|\omega|$. Because the only difference to the inverse Radon transform is in the additional weight $W(\omega)$ included with the ramp filter to filter $G(\omega, \theta)$, it follows that $\|\hat{f}\|_2^2$ is given by replacing 
$G(\omega, \theta)$ in \eqref{eq:2norm_in_line_int} by its weighted version $W(\omega)G(\omega, \theta)$. 
 Finally, incorporating the perceptual weighting filter $Q$  with radially symmetric frequency response in the image domain, is equivalent, by the Radon convolution Theorem \cite{doi:10.1137/9780898717792} to filtering the projections $G(\omega, \theta)$ by the filter with frequency response $Q(\omega)$. It therefore follows that
\begin{equation}
\label{eq:2norm_filtered_f_equivalence}
\|\mathcal{Q}\hat{f}\|_{2}^{2} =\frac{1}{4 \pi^{2}} \int_{0}^{\pi} \int_{-\infty}^{\infty} \bigl \lvert W(\omega) Q(\omega)G(\omega, \theta) \bigr \rvert^{2}|\omega| d \omega d \theta
\end{equation}

In order to obtain an expression in terms of filtered projections, \eqref{eq:2norm_filtered_f_equivalence} can be rewritten as
\begin{equation}
\label{eq:2norm_in_line_int_v2}
\|\mathcal{Q}\hat{f}\|_{2}^{2}=\frac{1}{4 \pi^{2}} \int_{0}^{\pi} d \theta \int_{-\infty}^{\infty} \left| G(\omega, \theta) H(\omega)\right|^{2} d \omega , \end{equation}
\begin{equation}
\label{eq:filter_freq_resp_2norm}
\text{where} \quad    H(\omega) = Q(\omega)W(\omega)|\omega|^{\frac{1}{2}}e^{j\phi(\omega)}
\end{equation}
can be considered as the frequency response of a filter. The phase of $H$ has no effect on the result owing to the absolute value in \eqref{eq:2norm_in_line_int_v2}. Thus, $\phi(\omega)$ can be chosen arbitrarily - for example as identically zero. Using the standard Parseval's identity for the second integral in \eqref{eq:2norm_in_line_int_v2}, the filtering implied by \eqref{eq:2norm_in_line_int_v2} can be expressed and implemented as 
\begin{equation}
\label{eq:2norm_time_dom_imp}
\|\mathcal{Q}\hat{f}\|_2^2 = \frac{1}{2 \pi}  \int_{0}^{\pi} d \theta \int_{-\infty}^{\infty}|\hat{u}(t, \theta)|^{2} d t , \end{equation}
\begin{equation}
\label{eq:proj-filter}
  \text{with} \qquad  \hat{u}(t,\theta) = h(t) \ast g(t,\theta),
\end{equation}
where $h(t)$ is a filter with frequency response $H(\omega)$ and $*$ denotes convolution in the $t$ variable.

The implication of these results is that  a loss function defined in terms of a $\mathcal{Q}$-weighted $L_2$ error of an image reconstructed using FBP with ramp filter weighting $W(\omega)$, can be expressed in terms of a loss function defined on the projections, using \eqref{eq:2norm_time_dom_imp} - \eqref{eq:proj-filter}.

For practical implementation, we derive 
the discretized version of \eqref{eq:2norm_time_dom_imp} - \eqref{eq:proj-filter}. When $f$ has bounded 
support and $W$ is bandlimited such that the reconstruction $\hat{f}$ is bandlimited, and $P$ uniformly spaced view angles over  $[0,2\pi]$ are used,  \eqref{eq:2norm_filtered_f_equivalence}-\eqref{eq:proj-filter} can be replaced by their discrete analogues, yielding

\begin{align}
\label{eq:2norm_filtered_f_discrete_analogues}
\|\mathcal{Q}\hat{f}\|_{2}^{2} 
&=\frac{1}{P} \sum_{p=0}^{P-1} \sum_{n=-\infty}^{\infty}\bigl \lvert \hat{u}_{d}[n, p]\bigr \rvert^{2},
\end{align}
\begin{equation}
\label{eq:2norm_filtered_f_u_time_dom}
  \text{where} \quad  \hat{u}_d[n,p] = \hat{h}_d[n] * g_d\left(n,p(2\pi/P)\right),
\end{equation}
and where the convolution is over $n$ and $\hat{h}_d$ is a discrete-time filter with frequency response
\begin{equation*}
\label{eq:2norm_filtered_f_H_discrete}
    \hat{H}_d(\nu) = Q_d(\nu) W_d(\nu)|\nu|^{0.5},
\end{equation*}
and $g_d(n.\theta) \defsign g(n\Delta_t,\theta)$,  $W_d(\nu) \defsign W(\frac{\nu}{\Delta_t})$, $Q_d(\nu) \defsign Q(\frac{\nu}{\Delta_t})$ for $\nu \in [-\pi,\pi]$.


 Considering the special case of $Q(\omega) = |\omega|^{0.5}$, we obtain $\hat{H}_d(\nu) = W_d(\nu)|\nu|$. However, since we use the magnitude in \eqref{eq:2norm_filtered_f_discrete_analogues}, the phase of $\hat{H}_d$ does not matter, and we may use 

\begin{equation*}
    \hat{H}_d(\nu) = W_d(\nu)\nu.
\end{equation*}
The advantage of this particular filter selection is that it removes the jump discontinuity in the derivative of $\hat{H}_d$ at the origin, and can result in a short filter $\hat{h}_d[n]$. For instance, when $W_d(\nu) = \mathrm{sinc}(\nu)$ is selected, which corresponds to a Shepp-Logan ramp filter \cite{shepp1974fourier}, we obtain $\hat{H}_d(\nu) = \sin(\nu)$, for which the impulse response 
   $ \hat{h}_d[n] = 0.5\delta[n+1] - 0.5\delta[n-1]$ 
has only two non-zero values. This enables an easy time-domain implementation of the computation in \eqref{eq:2norm_filtered_f_discrete_analogues} and \eqref{eq:2norm_filtered_f_u_time_dom}.

%% file: methodology.tex
\section{Methodology}
\label{sec:methodology}
\subsection{Data Generation and Training}
\label{sec:data_generation}
In order to train the algorithms, total measurements $\tau$ of objects with known ground-truth primary measurements $p$ are needed. Obtaining such data requires a fully characterized CT device and many different phantoms with known material composition and geometry. In our case, it was impractical to obtain such data. Instead, $\tau$ for training, validation, and testing data were generated by MC-based simulations.

For each phantom,  
2D $d \times d$ detector panel readouts, i.e., projection total measurements $\tau_\theta$, were simulated at $K$ view angles $\theta$ uniformly spaced in $[0,2\pi)$, each having $P$ photons.
The simulated data was divided into training and test sets phantom-wise, grouping all measurements belonging to a phantom into the same set. Then, the total projections measurements and their corresponding initial reconstructions in the training set were used in a randomized order for training. The proposed algorithms were trained and tested using the loss function $L$ in \eqref{eq:loss_fct}.

\subsubsection{Parallel Beam CT Experiments}
\label{sec:data_generation_parallel_beam}
To obtain the total measurements $\tau_\theta$ for training and validation for  the parallel beam CT experiments, we used the GATE \cite{jan2004gate} software package. 
GATE encapsulates the GEANT4 MC \cite{agostinelli2003geant4} simulation libraries, which perform the simulation of particle propagation through matter. The modeling by GATE included scatter in the detector panel.
On the other hand, the primary projections $p_\theta$ of voxelized phantoms were generated by numerical forward reprojection to compute $g(t,\theta)$ and using Beer's law  \eqref{eq:beers_law}. 

As illustrated in 
Fig.~\ref{fig:scatter_analytical_level_illustration}(b), for these experiments we used a 200 keV monoenergetic parallel-beam source with rays perpendicular to the rotation axis $z$, and a flat detector panel, both of width $W=128$ cm and height $H=128$ cm,  with the detector panel  perpendicular to the rays of the source, and divided into $(d=128) \times (d=128)$ pixels. 
The source was not collimated, extending over the entire detector. Simulated objects (mathematical phantoms) were generated as a composition of distinct object components, all contained in a cylindrical air volume of diameter $D=128$ cm and height $L=128$ cm. The  parallel-beam projections of the  objects  were fully contained in the detector plane for each view, with a margin of approximately $24$ cm from the boundaries of the cylindrical geometry in
Fig.~\ref{fig:scatter_analytical_level_illustration}.

The object components in the phantoms were:  (i) rectangular prisms; (ii) cylinders with their long axis aligned with the z-axis; 
and (iii) spheres; with material assigned randomly as water, aluminium, or titanium. Positions and 
dimensions were randomized, ensuring the components are contained in the phantom volume.

We found in initial experiments that high noise in the simulated training measurements could lead to the networks learning to denoise the projection data rather than just estimating, as intended, the scatter component. This came at the price of reduced resolution of the reconstruction. To mitigate this effect without an expensive increase of the photon counts used in simulation, we introduced a simple pre-processing step, whose details are described in the Appendix, to decrease the level of noise in the simulated measurements. The idea is to identify the areas in the obtained 2D total measurements $\tau_\theta$ 
that contain only stochastic noise and little scatter, and smooth them to reduce the noise, without affecting the areas that contain object or significant scatter information. 

A parameter setting of $\lambda = 5 \cdot 10^{-2}$ was used in the loss function $L$ in \eqref{eq:loss_fct} for training of both parallel beam settings, after observing the best performance among all values used in a log scale sweep. All FBPs 
used the Shepp-Logan filter \cite{shepp1974fourier}. 

\smallskip
\subsubsection{  Cone Beam CT (CBCT) Experiments}
\label{sec:data_generation_cbct}
For the CBCT experiments, we used MC-GPU \cite{badal2009accelerating}, which is a GPU-accelerated X-ray photon transport MC simulation code for CBCT. The code performs the transport simulation in a voxelized geometry and uses CUDA programming supporting multiple GPUs. 
The code provides access to the total measurement $\tau_\theta$, scatter signal $s_\theta$, and primary measurement $p_\theta$ for each view $\theta$, eliminating the need to compute $p_\theta$ separately.

For training and testing each method in the polychromatic CBCT experiments, we used $\FBP g$ with $g_\theta = -\ln [p_\theta/I_0]$  as the ``gold reference" image, and similarly used  $\tilde{g}$, and $g^*$ for the corresponding reconstructions,  treating the measurements as in the monochromatic case. Thus, the primary measurement (scatter-free) reconstructions $\FBP g$ include some beam-hardening artifacts, and in our experiments the scatter correction methods do not perform beam-hardening correction, instead focusing on estimating and correcting the scatter. 

The geometry for the CBCT experiments is shown in 
Fig.~\ref{fig:scatter_analytical_level_illustration}(b)
with a divergent beam point source at source-to-detector distance, $d_{sd}=180$ cm, and source-to-origin distance, $d_{so}=130$ cm and with a
$64 \times 64 \,\, \text{cm}^2$ flat panel detector gridded to $128 \times 128$ pixels. 
The source had a $90$ keV monoenergetic spectrum for the monochromatic experiments, and the photons were sampled from a 
$120$ keV tungsten spectrum with 4.3mm Al filter for the polychromatic case.

Thanks to the small fan beam span angle $\xi = 0.1233  \approx \sin\xi = 0.1230$  for the given CBCT geometry, we expect the extension of the loss function to 3D described in Sec.~\ref{sec:3D Extension} to hold to a good approximation.

Again, $\lambda = 5(10^{-2})$ was used for the CBCT setting after conducting a similar sweep as in the parallel beam case. For each algorithm, the FDK method \cite{feldkamp1984practical} with a Shepp-Logan \cite{shepp1974fourier} ramp filter was used for reconstructions, with  a total of $K=360$ view angles for each phantom. 

The algorithms were compared on two types of phantoms. The first consisted of titanium rods with $2 \times 2$ $\text{cm}^2$  cross-sections placed in parallel with the z-axis with random center locations in the x-y plane sampled from a uniform distribution  $\mathcal{U}[-D/4 \,\text{cm},D/4 \,\text{cm}]$, where $D = 64\, \text{cm}$. Intersections between the objects were allowed with intersecting regions also consisting of titanium. 
The second type of phantoms were used for
experiments on medical data. Thirty phantoms extracted from the CT Lymph Nodes dataset \cite{roth2014new} from the cancer-imaging archive \cite{clark2013cancer} were tissue-mapped to five different materials and tissue types: (1) air, (2) lung, (3) adipose, (4) soft tissue and (5) bone. 

Since the GPU-accelerated MC code allowed us to simulate  photon numbers sufficient to obtain acceptable noise levels in the simulated measurements, a noise reduction scheme as in parallel beam case was not used for CBCT experiments.

\subsection{Setup}
\label{sec:framework}

The proposed algorithms were compared among themselves and with the recent data-driven projection-based correction method DSE \cite{maier2018deep,maier2019real} as described in Section~\ref{sec:intro}. Since DSE was shown to perform consistently better than a classic kernel-based scatter correction method, direct comparison with such a method was not performed.
The DSE method was implemented as described in \cite{maier2018deep} and trained using the same total measurements $\tau_\theta$ and primary measurements $p_\theta$ used in the training of the proposed methods. The network $N_\gamma$ in DSE seeks to minimize the MAE (mean absolute error) between the estimated and ground-truth scatter signals, and estimates the primary measurement as 

\begin{align*}
\label{eq:dse}
    s_\theta^* = \mathcal{N}_{\gamma^*} \left( \tau_\theta \right), \quad
    p_\theta^* = \tau_\theta - s_\theta^*,
\end{align*}
\begin{equation*}
 \text{where} \quad   \gamma^* = \arg\min_\gamma \sum_{\theta} \left| 
    [s_\theta - \mathcal{N}_{\gamma}\left( \tau_\theta \right)] / 
    {s_\theta} \right|.
\end{equation*}

Reconstruction quality, as compared to the reference images (FBP of numerically computed projections of test phantoms) is quantified using four metrics: PSNR (in dB) as the ratio of peak reconstruction value to root mean square error with higher values indicating better performance; MAE; PE (peak error), equal to the infinity norm of the error; and SSIM (structural similarity index) -- higher values for greater similarity. 

The networks for DSE and the proposed methods were implemented in Pytorch, and the Adam optimizer \cite{kingma2014adam} was used for all methods for training. No additional regularization was used for training, as we observed  close validation and training errors, indicating the absence of overfitting. Convergence for different algorithms was determined by flattening of training and validation loss curves as a function of training iterations.

Computations were performed on an NVIDIA GeForce GTX TITAN X GPU, and an Intel Core i7-4770K CPU with 32 GB RAM. For parallel beam reconstruction, both FBP and image rotations were implemented on the CPU, whereas for CBCT the FDK algorithm was implemented on GPU. 
The split of run-time averages are shown in Table~\ref{tab:rutimes-parallel}. 
Once both FBPs and image rotations  are migrated to the GPU, the DCNN runtimes, which in these experiments account for only a small fraction of the total runtime,  will dominate, resulting in total runtimes for the algorithms of 2 -- 4 seconds per reconstructed volume.
\begin{table}[hbt]
\centering
\begin{tabular}{|l|c|c|c|c|c|}
\hline
 \hspace*{0.5em} Algorithm & $K$  & Total &   FBP & Rotation & DCNN \\ \hline
  PhILSCAT Par-Beam & 360 & 28.2 s  & 7.7 s & 17.9 s & 2.6 s\\
  OV-PhILSCAT Par-Beam & 180 & 18.0 s  & 7.7 s & 9.0 s & 1.3 s\\
 PhILSCAT CBCT & 360 & 22.6 s & 2.1 s& 17.9 s & 2.6 s\\
  \hline
\end{tabular}
\caption{\small Average run times.}
\vspace{-0.4cm}
\label{tab:rutimes-parallel}
\end{table}

%% file: experiments.tex
\section{Experiments}
\label{sec:experiments}

\subsection{Monochromatic 3D Parallel Beam CT}
\label{sec:mono_parallel_CT}
To compare PhILSCAT, OV-PhILSCAT, and DSE \cite{maier2018deep, maier2019real}, each algorithm was trained for 100 epochs and tested on the same 27 and 3  phantoms, respectively, 
randomly generated as in Sec.~\ref{sec:data_generation}. 
Each phantom had 360 uniformly spaced views with 
$8 \times 10^6$ photons each.  

The synthesized 
measurements display an expected strong correlation between attenuation along a $(t, \theta)$-ray and the corresponding scatter-to-primary ratio
$s(t,\theta)/p(t,\theta)$: 
when averaged
over the areas in the test phantoms with $p(t,\theta) < \sqrt{I_0}$ we observed
$\operatorname{AVG}_{t,\theta}s(t,\theta)/p(t,\theta)\approx 59\%$ . The peak  value over all detector pixels 
is $\max_{t,\theta} s(t,\theta)/p(t,\theta) = 956\%$. 
This shows that the higher values of scatter-to-primary ratio are not restricted to few detector pixels.
The strong scatter scenario is also evident in the corruption of the $\tau$-reconstructions from uncorrected total measurements with peak reconstruction error 
of 1557 HU, or $46\%$ 
of the peak density of 3407 HU of the $p$-reconstructions from primary $p$. 

Average reconstruction accuracies 
are reported in Table \ref{table:recon_acc_mono_3D_recon} for the three test phantoms. DSE improves on  the  uncorrected case as expected. However, consistent with our observations, which will be discussed next, PhILSCAT and OV-PhILSCAT  perform significantly better than DSE in these experiments on all metrics. OV-PhILSCAT provides not only a slightly better PSNR than PhILSCAT, but is also twice as fast.

\begin{table}[htp!]
\setlength{\tabcolsep}{1pt}
\renewcommand{\arraystretch}{1}
\centering
\begin{tabular}{@{}lcccc@{}}
\toprule
\multicolumn{1}{c}{  }&\multicolumn{1}{c}{Uncor.}& \multicolumn{1}{c}{DSE}& \multicolumn{1}{c}{PhILSCAT}& \multicolumn{1}{c}{OV-PhILSCAT} \\
\cmidrule(r){1-1}\cmidrule(lr){2-2}\cmidrule(lr){3-3}\cmidrule(lr){4-4}\cmidrule(l){5-5}
PSNR (dB) & $38.6 \pm 0.3$ & $45.4 \pm 0.5$ & $51.1 \pm 0.2$ & $\bold{51.3 \pm 0.2}$ \\
\cmidrule(r){0-0}
SSIM & $0.964 {\small \pm 0.002}$ 
& $0.984$
& 
$\bold{0.998}$ 
& $\bold{0.998}$ \\
\cmidrule(r){1-1}
MAE (HU) & $16.4 \pm 1.0$ & $6.6 \pm 0.5$ & $\bold{3.6 \pm 0.1}$ & $\bold{3.6 \pm 0.1}$ \\
\cmidrule(r){1-1}
Peak$\,$Error$\,$(HU) & 1572 & 1228 & \textbf{514} & \textbf{510} \\
\bottomrule
\end{tabular}
\caption{\small Average reconstruction accuracies and standard deviations for three 3D test phantoms 
for the algorithms. Standard deviation up to $5 \cdot 10^{-4}$ is omitted for SSIM. The peak density value of $p$-reconstructions is 3407 HU.}
\vspace{-0.25cm}
\label{table:recon_acc_mono_3D_recon}
\end{table}

\begin{figure*}[!htp]
\centering
\setlength{\tabcolsep}{-0.05cm}
\renewcommand{\arraystretch}{0.1}
\vspace{-0.5cm}
\begin{tabular}{ccccc}
\hspace{-8mm} \textbf{Reference Recon} & \multicolumn{4}{c}{\textbf{Error Magnitudes}}\\
[-2mm]\includegraphics[width=.18\linewidth]{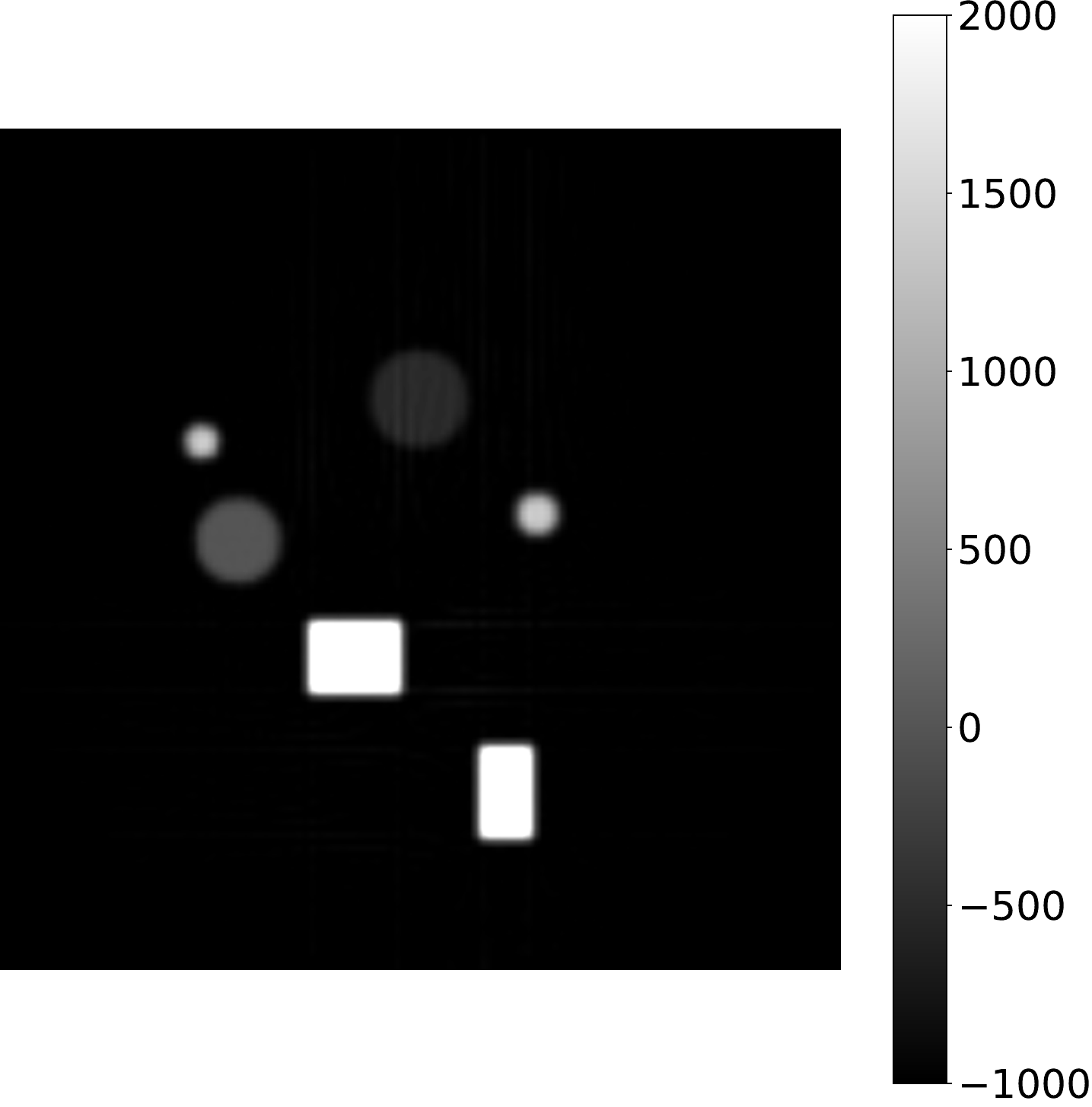} & \hspace{0.2cm}
\includegraphics[width=.18\linewidth]{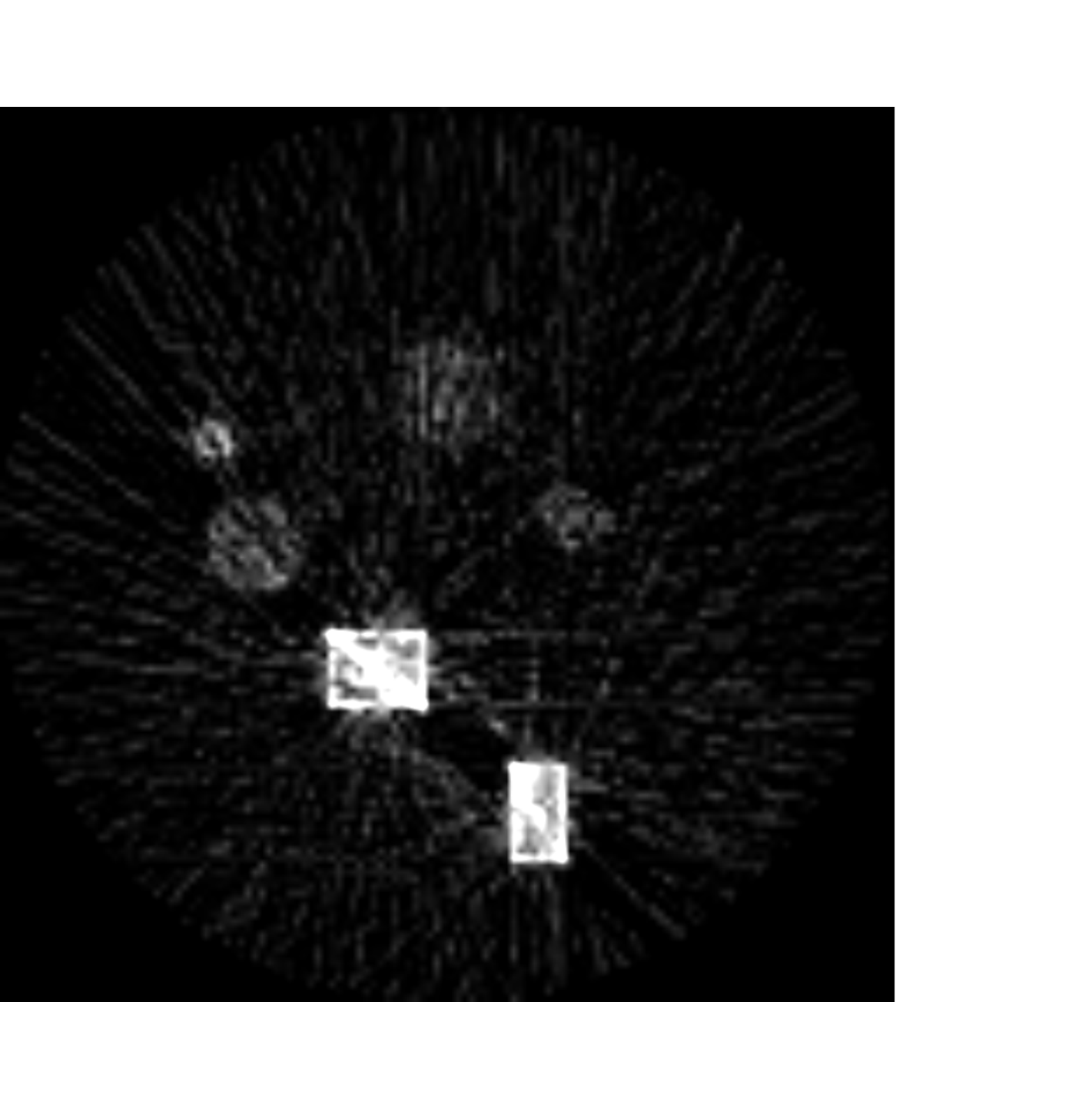} & \hspace{-0.3cm}
\includegraphics[width=.18\linewidth]{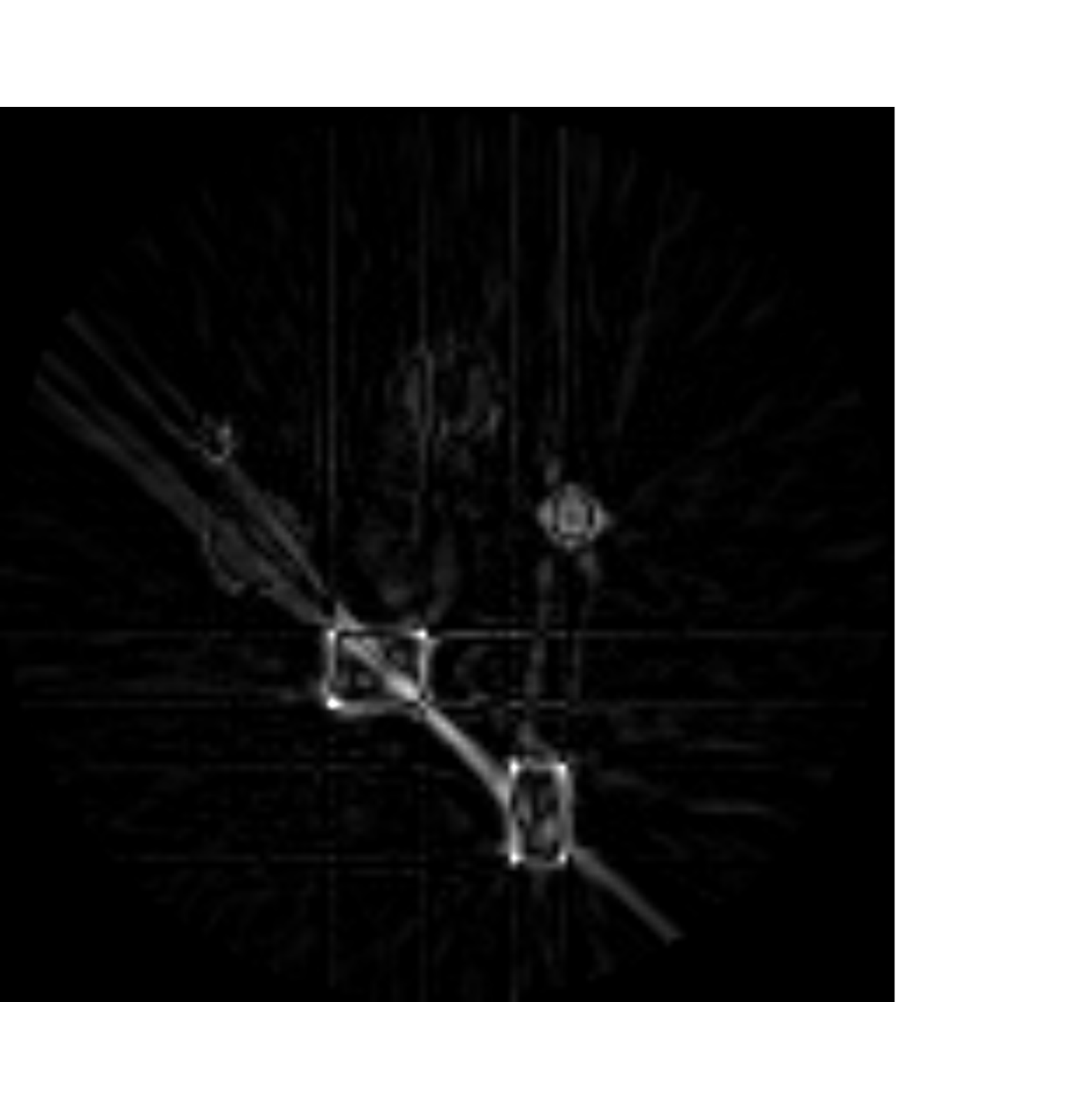} & \hspace{-0.3cm}
\includegraphics[width=.18\linewidth]{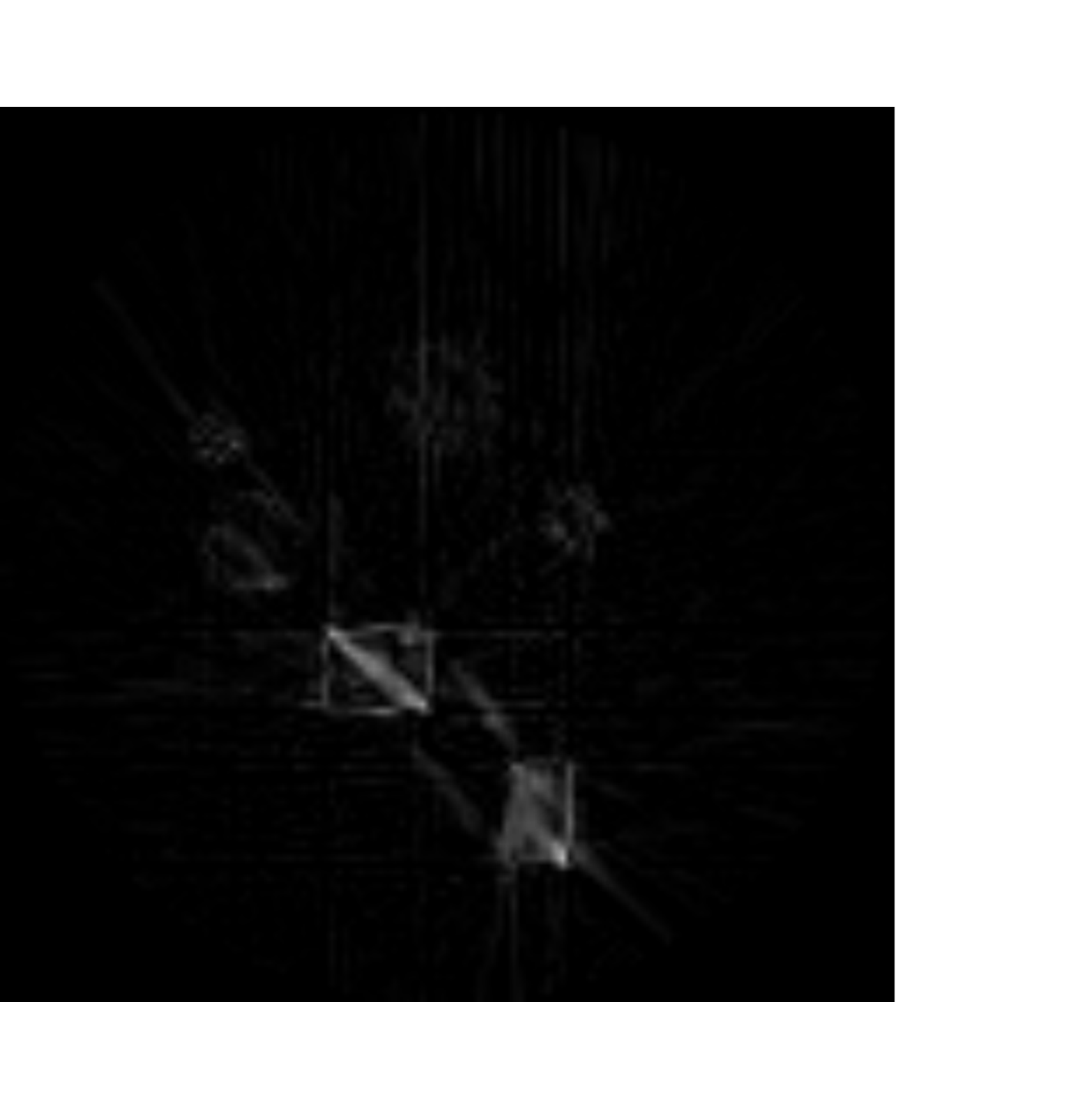} & \hspace{-0.3cm}
\includegraphics[width=.18\linewidth]{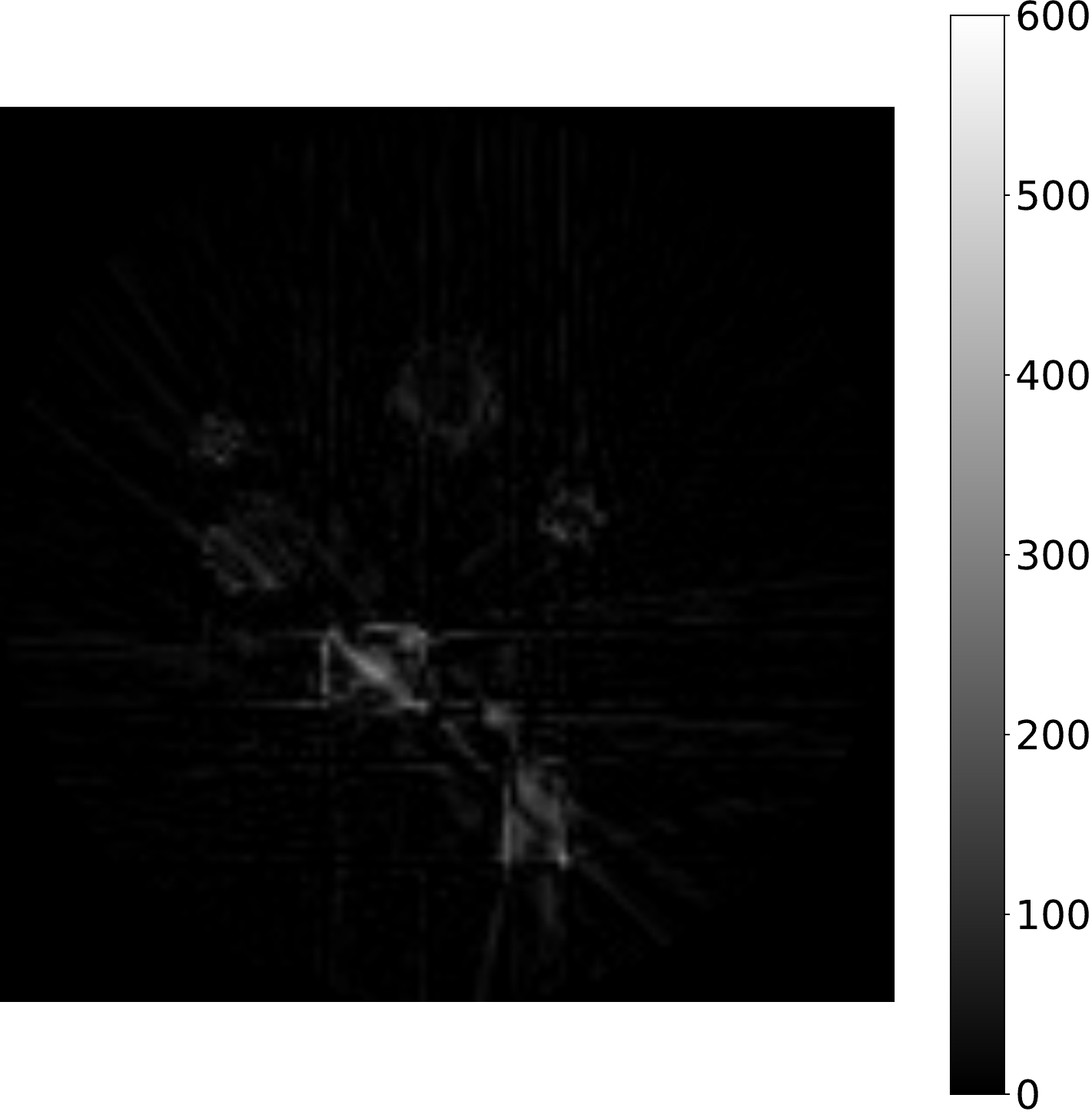}\\[-2mm] 
\hspace{-0.8cm} {\small (a) Axial} & \hspace{-0.5cm} {\small(c) Total} & \hspace{-1cm} {\small(e) DSE} & \hspace{-0.9cm} {\small(g) PhILSCAT} & \hspace{-0.9cm} {\small(i) OV-PhILSCAT} \\

\includegraphics[width=.18\linewidth]{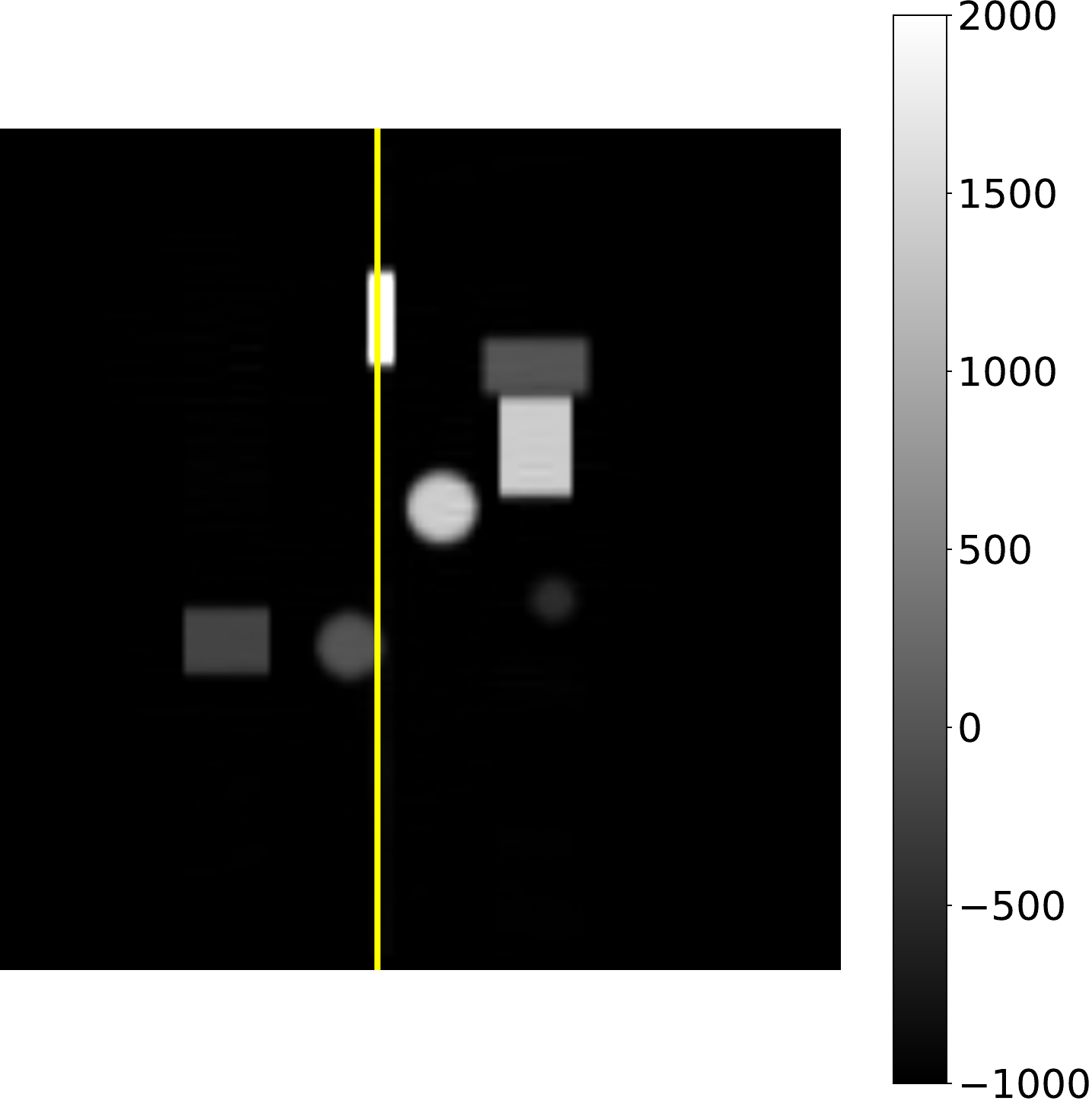} & \hspace{0.2cm}
\includegraphics[width=.18\linewidth]{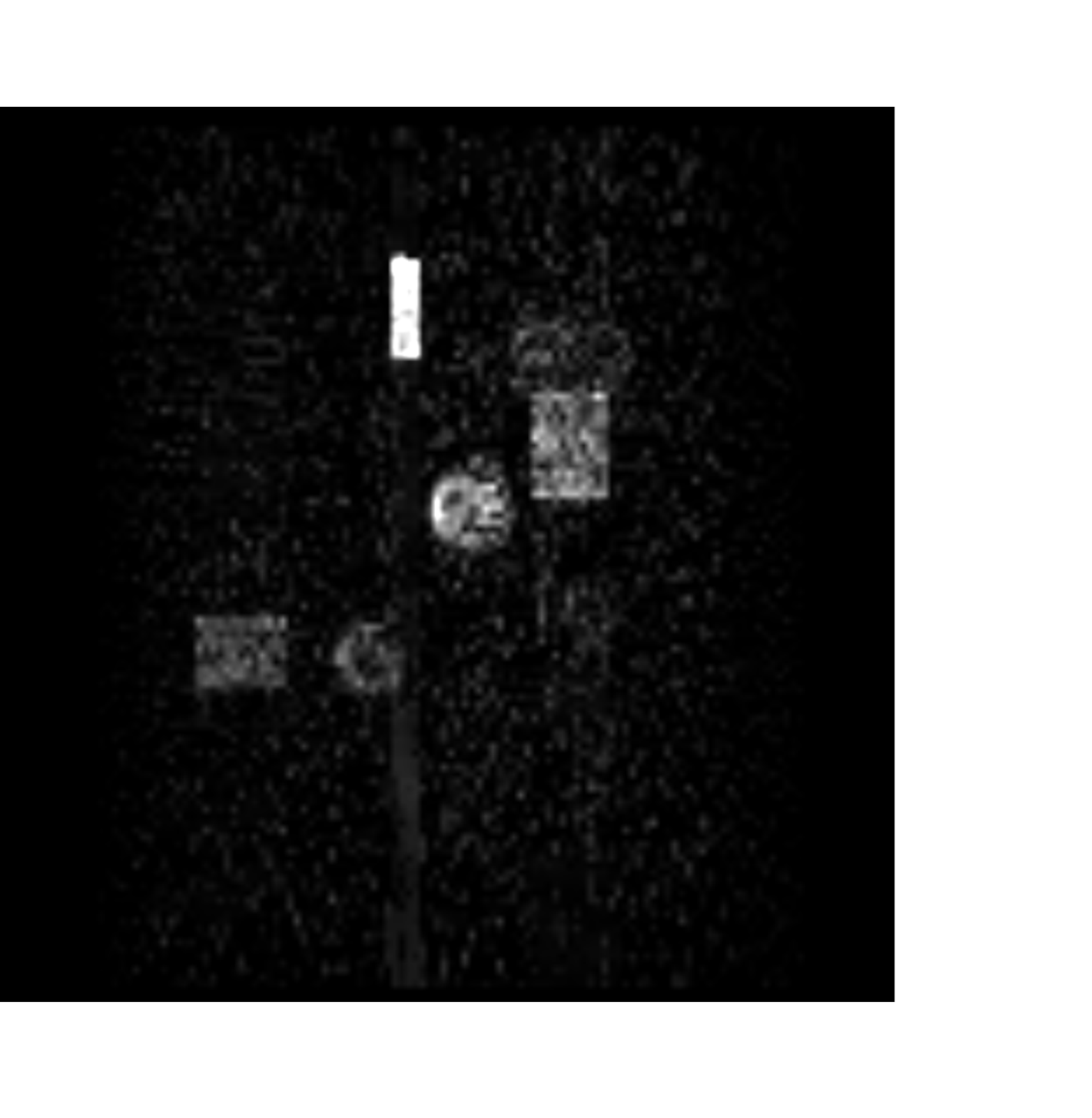} & \hspace{-0.3cm}
\includegraphics[width=.18\linewidth]{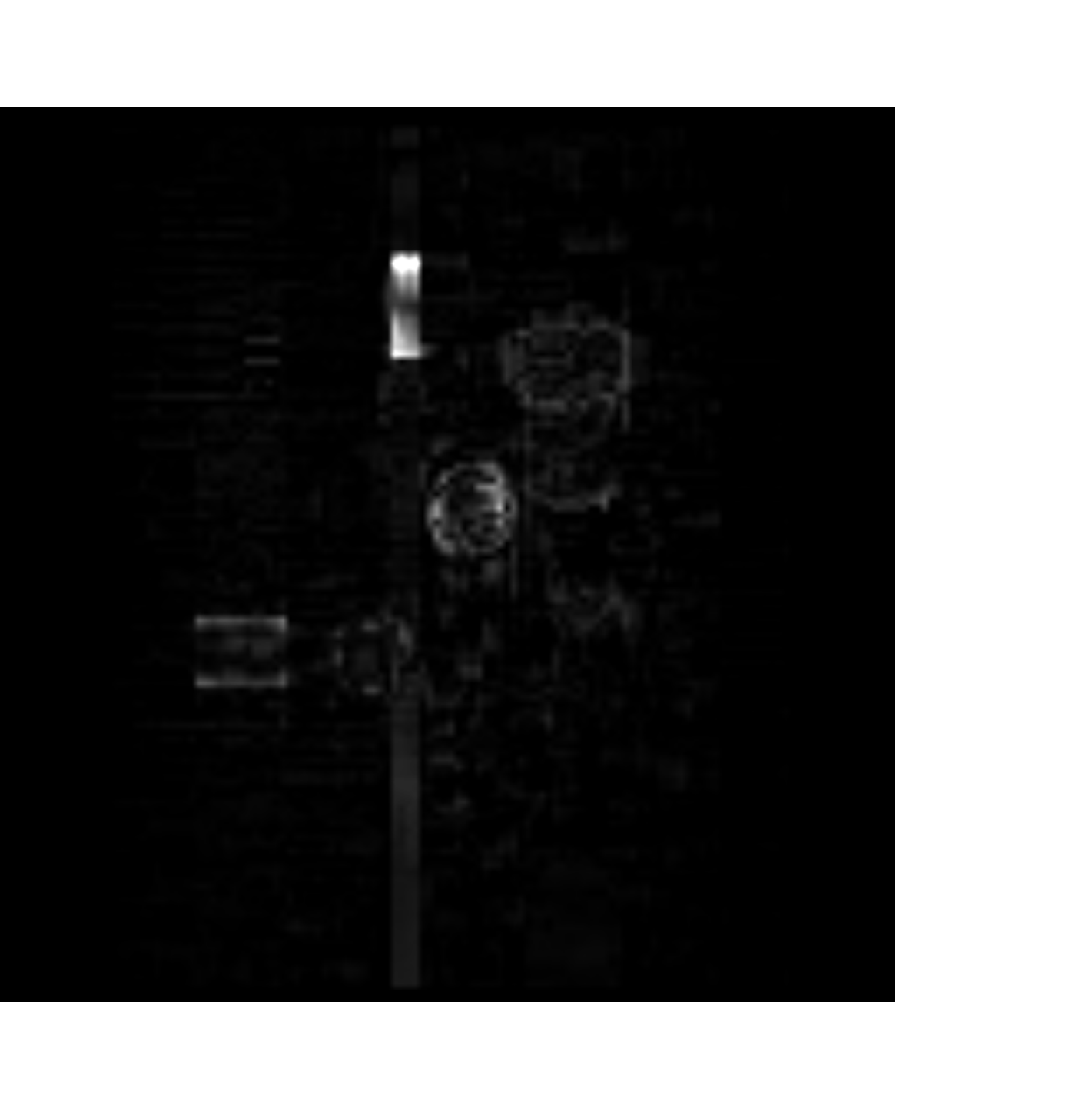} & \hspace{-0.3cm}
\includegraphics[width=.18\linewidth]{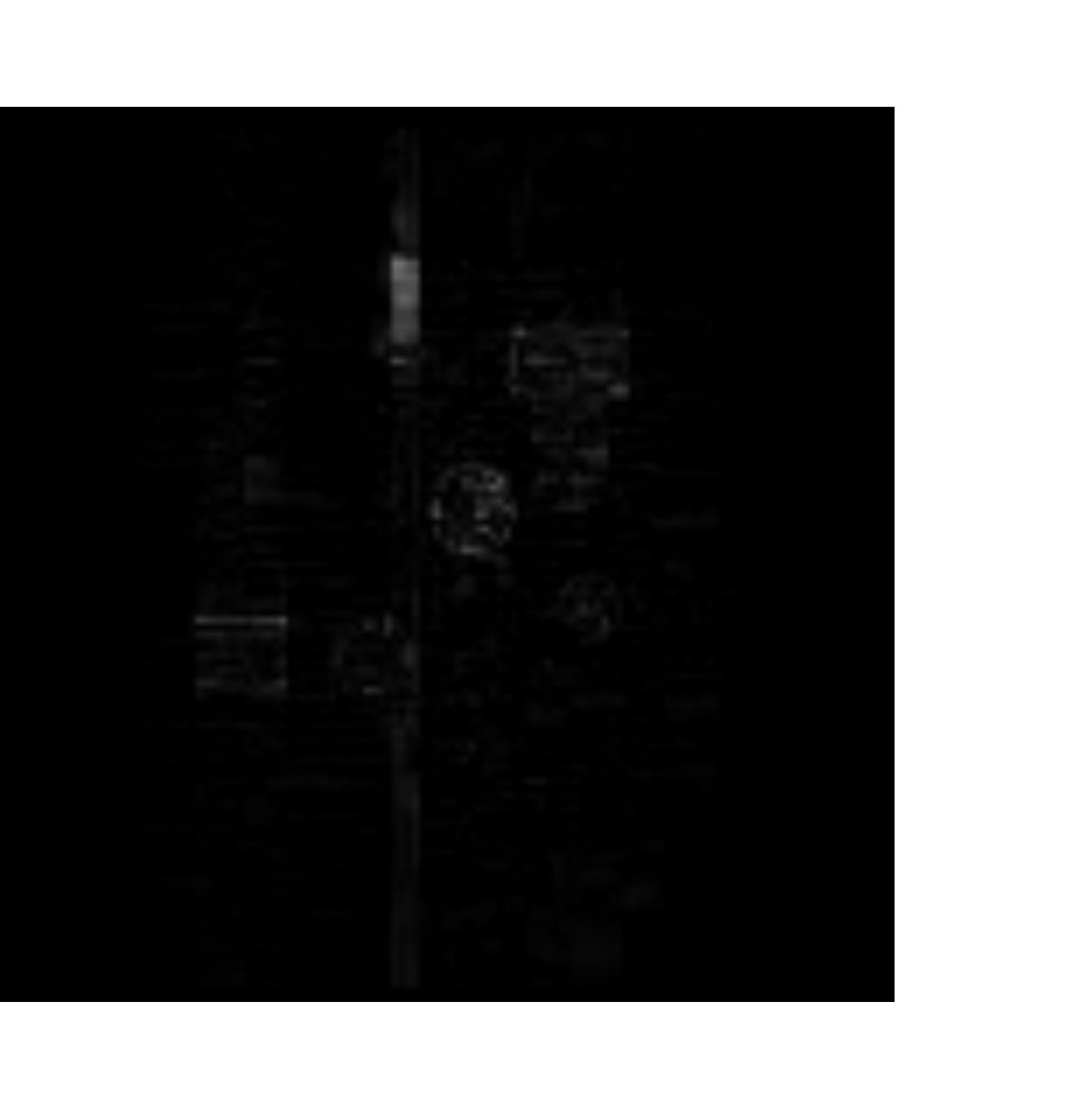} & \hspace{-0.3cm}
\includegraphics[width=.18\linewidth]{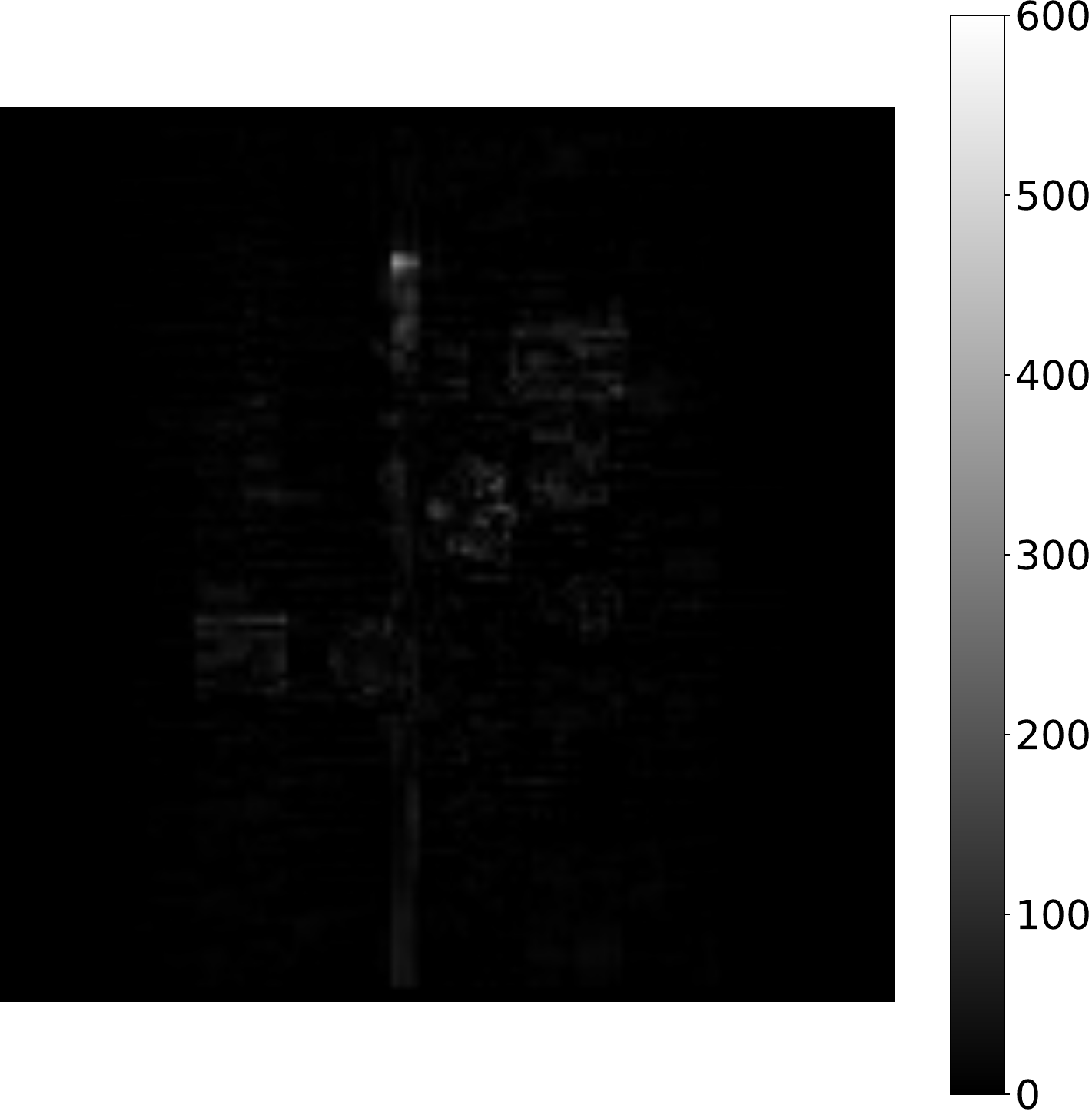}\\[-2mm] 
\hspace{-0.8cm} {\small (b) Sagittal} & \hspace{-0.5cm} {\small (d) Total} & \hspace{-1cm} {\small (f) DSE} & \hspace{-0.9cm} {\small (h) PhILSCAT} & \hspace{-0.9cm} {\small (j) OV-PhILSCAT} \\

\end{tabular}
\caption{\small Monochromatic parallel-beam reconstructions: (a), (b) axial and sagittal slices of $p_\theta$-reconstructions; (c), (d) error magnitude using total measurements $\tau_\theta$;  
vs. (e), (f) using primary measurements $p_\theta^*$ estimated by DSE; (g), (h) estimated by PhILSCAT; and (i), (j) estimated by OV-PhILSCAT.  Display windows in HU (different for the reconstructions and for the error maps) are indicated by the colorbars.}
\vspace{-0.25cm}
    \label{fig:recon_comp_3D_parallel_CT_sagittal} 
\end{figure*}

\begin{figure}
        \centering
        \begin{subfigure}[b]{0.22\textwidth}
            \centering
            \includegraphics[width=.8\textwidth]{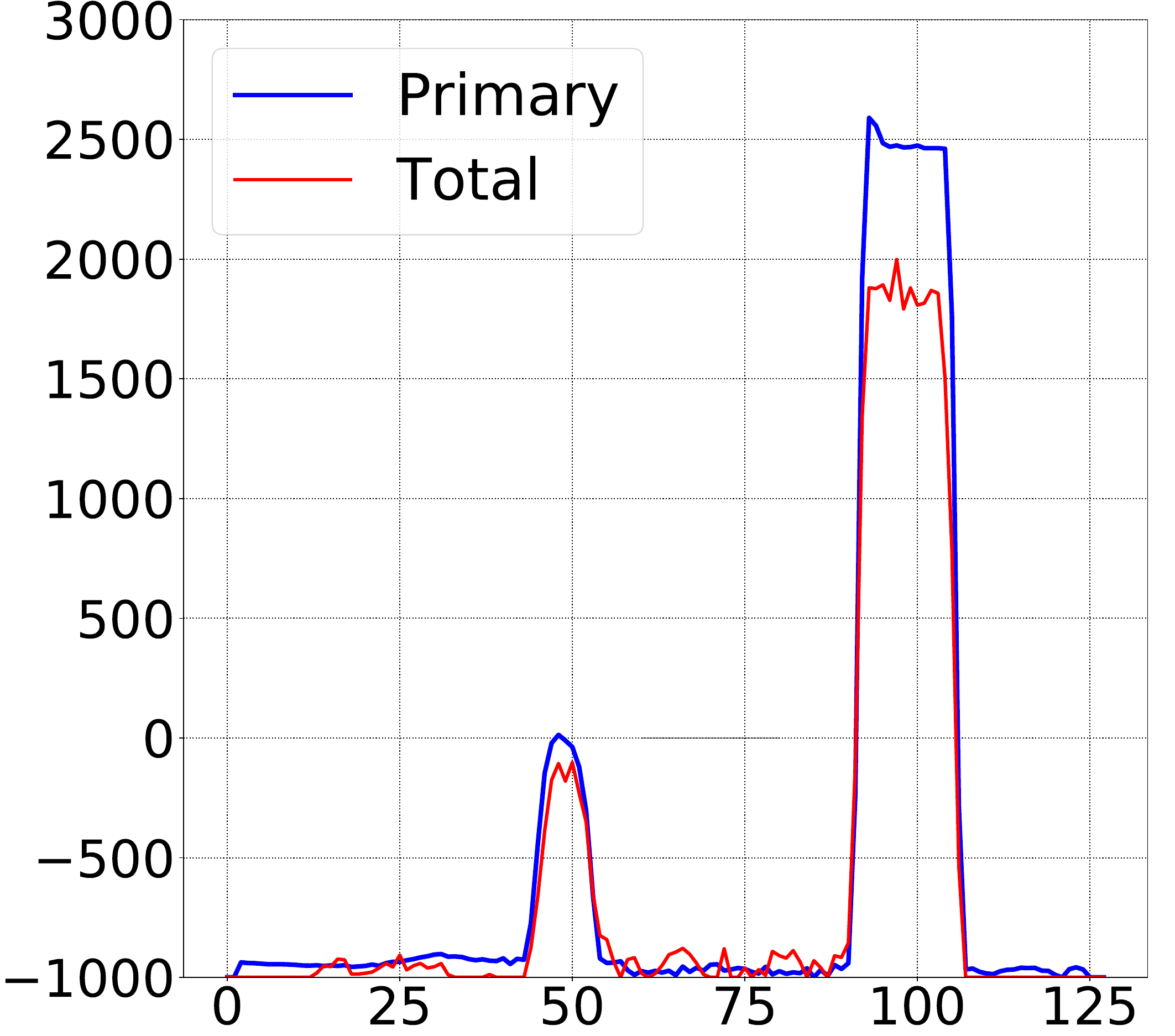}
            \caption[]%
            {{\small Total}}    
        \end{subfigure}
        \hfill
        \begin{subfigure}[b]{0.22\textwidth}  
            \centering 
            \includegraphics[width=.8\textwidth]{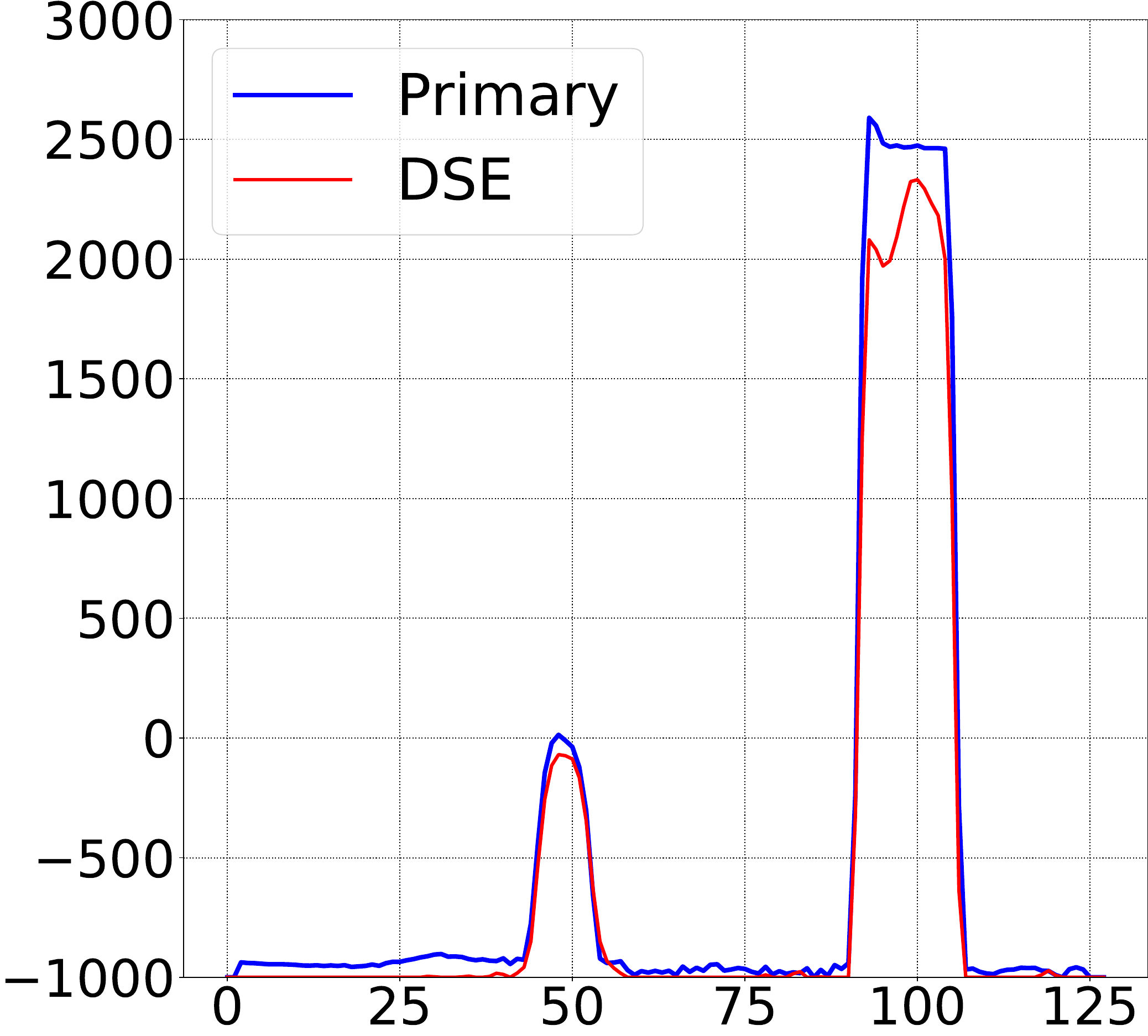}
            \caption[]%
            {{\small DSE}}    
        \end{subfigure}
        \\
        \begin{subfigure}[b]{0.22\textwidth} 
            \centering 
            \includegraphics[width=.8\textwidth]{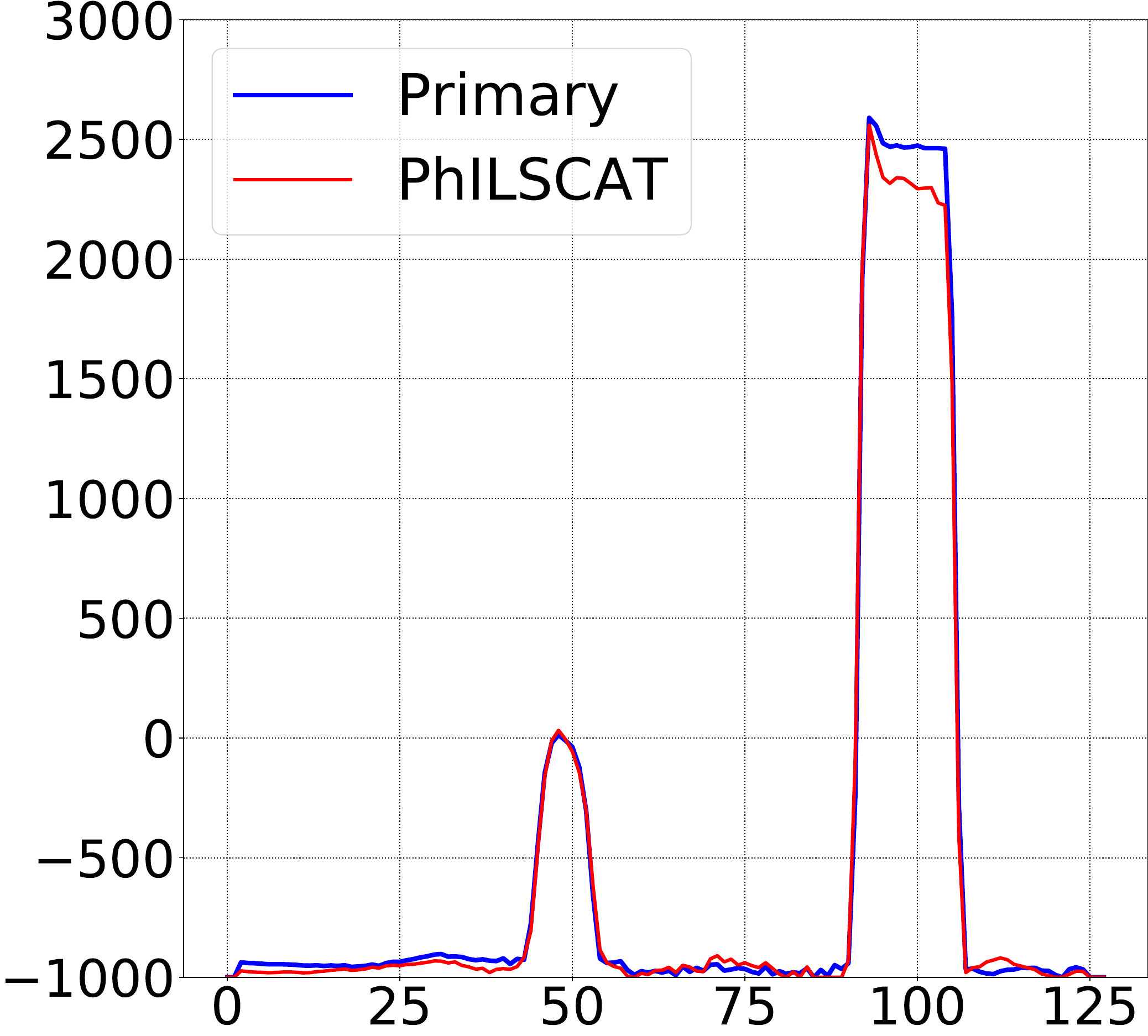}
            \caption[]%
            {{\small PhILSCAT}}    
        \end{subfigure}
        \hfill
        \begin{subfigure}[b]{0.22\textwidth}   
            \centering 
            \includegraphics[width=.8\textwidth]{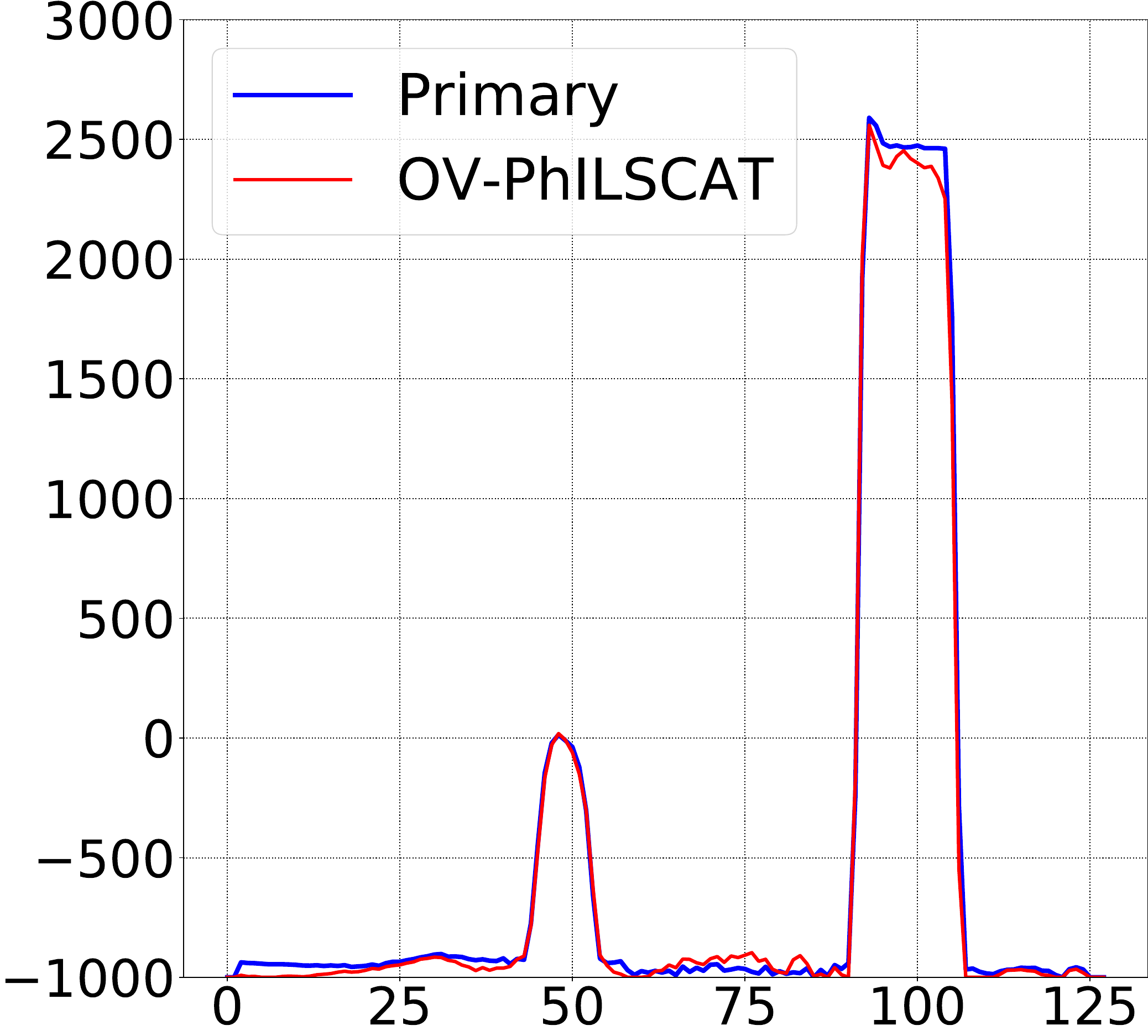}
            \caption[]%
            {{\small OV-PhILSCAT}}    
        \end{subfigure}
        \caption{\small Monochromatic parallel-beam reconstructions: vertical line profiles from Fig.~\ref{fig:recon_comp_3D_parallel_CT_sagittal} (f). Comparisons between using total measurements $\tau$ in (a), vs.
         primary measurements $p_\theta^*$ estimated by DSE in (b), 
         PhILSCAT in (c), and 
        OV-PhILSCAT in (d). 
        }
        \vspace{-0.6cm}
        \label{fig:line_profile_hor_comp_3D_parallel_CT_sagittal}
    \end{figure}

The FBP $p$-reconstruction of a test phantom
and the absolute reconstruction errors in HU obtained by the algorithms are shown in Fig.~\ref{fig:recon_comp_3D_parallel_CT_sagittal}.
Line profiles extracted from the reconstructions at the same location as in Fig.~\ref{fig:recon_comp_3D_parallel_CT_sagittal} are compared in Fig.~\ref{fig:line_profile_hor_comp_3D_parallel_CT_sagittal}. 

As expected from the high scatter/primary ratios in the projections,
uncorrected $\tau$-reconstructions   displaying large magnitude scatter artifacts such as decreased contrast for the axial and sagittal slices, are the worst.
DSE corrects these magnitude errors to a significant extent, which is reflected as an overall improvement on the metrics compared to $\tau$-reconstructions. However, as seen in Fig.~\ref{fig:recon_comp_3D_parallel_CT_sagittal}, with DSE there are remaining artifacts around the object edges. Also, DSE does not suppress the streaking artifacts, density errors over highly attenuating objects, and errors in the background very well. 
This is also seen in the line profiles  in Fig. \ref{fig:line_profile_hor_comp_3D_parallel_CT_sagittal}. 

Both PhILSCAT and OV-PhILSCAT suppress the streaks better than DSE, with better reduction of errors around the object edges, which is especially visible for the titanium slabs in Fig.~\ref{fig:recon_comp_3D_parallel_CT_sagittal}. 
As a consequence, they both achieve substantially better metrics than DSE. 
Finally, 
compared to the total measurement reconstructions, both proposed algorithms reduce the peak error by  $\approx70\%$, whereas DSE by only  $\approx 20\%$.  

\subsection{3D Cone Beam CT  - Ti Rod Phantoms}
\label{sec:ti_rods_CBCT}
In this subsection, we study using both monochromatic and polychromatic sources,
a setting more typical of a non-destructive evaluation (NDE) application, with phantoms  consisting of randomly placed titanium rods as described in \ref{sec:data_generation}. 
These phantoms include  high object densities resulting in high ray attenuations. Moreover, the scatter signals corresponding to various views have higher frequency content and there is an increased dependence of the scatter signal on the angle of the measurement. These factors all have the potential to make the problem more challenging. Several authors (e.g.,\cite{maier2018deep}) estimated scatter at reduced spatial resolution to take advantage of its  characteristic smoothness compared to the primary. However, having high contrast sharp objects as in Sections \ref{sec:mono_parallel_CT} and \ref{sec:ti_rods_CBCT}, we have chosen not to do so since smoothness of the scatter may not hold uniformly.

DSE and PhILSCAT were each trained on 27 and tested on 3 such phantoms, respectively, using $K=360$ uniformly spaced views with $P=10^8$ photons each for each phantom. 

As expected with this high density object scenario, the peak scatter-to-primary ratios $s(t,\theta)/p(t,\theta)$ are extremely high due to photon scarcity along highly attenuating paths for both source settings. 
For the test phantoms, the overall average ratio $\operatorname{AVG}_{t,\theta}s(t,\theta)/p(t,\theta)$, and the corresponding average over areas in the projections where $p_\theta < \sqrt{I_0}$ were found to be $10\%$ and $54\%$, respectively for the  monochromatic case,  and $17\%$ and $74\%$, respectively, for the polychromatic case.  These high scatter levels are indicative of the difficulty of the reconstruction task in these experiments. 

For the monochromatic setting, the observed peak error between the $\tau$-reconstructions and the true $p$-reconstructions is 3711 HU, or 42\% of the peak density of 8761 HU in the $p$-reconstructions. For the polychromatic setting, the corresponding values
are 4229 HU or 40\% out of the peak  $p$-reconstruction density of 10560 HU. The HU reconstructions for all methods in the polychromatic setting were obtained after performing HU calibration to adjust the water linear attenuation level to 0 HU in the $p$-reconstructions, which were computed as explained in Sec. \ref{sec:data_generation}.

These numbers indicate a stronger effect of scatter in these reconstructions than in the other CBCT experiments.  
 The average reconstruction accuracy metrics are reported in Tables~\ref{table:recon_acc_mono_3D_recon_CBCT_ti_rods} and~\ref{table:recon_acc_poly_3D_recon_CBCT_ti_rods}, for the monochromatic and polychromatic cases, respectively.

\begin{table}[h!]
\centering
\begin{tabular}{@{}lccc@{}}
\toprule
\multicolumn{1}{c}{  }&\multicolumn{1}{c}{Uncorrected}& \multicolumn{1}{c}{DSE}& \multicolumn{1}{c}{PhILSCAT} \\
\cmidrule(r){1-1}\cmidrule(lr){2-2}\cmidrule(lr){3-3}\cmidrule(l){4-4}
     PSNR (dB) & $35.8 \pm 0.5$ & $49.8 \pm 2.7$ & {$\bold{51.6 \pm 2.6}$} \\
\cmidrule(r){0-0}
     SSIM & $0.980 \pm 0.001$ & $0.995 \pm 0.002$ & {$\bold{0.997 \pm 0.001}$} \\
\cmidrule(r){1-1}
     MAE (HU) & $30.1 \pm 1.3$ & $8.8 \pm 1.3$ & {$\bold{8.3 \pm 1.4}$} \\
\cmidrule(r){1-1}
Peak Error (HU) & $3327$ & $756$ & $\bold{700}$ \\
\bottomrule
\end{tabular}
\caption{\small Monochromatic CBCT: average reconstruction accuracy results and standard deviations for different algorithms for 3 Ti rod test phantoms.}
\vspace{-0.4cm}
\label{table:recon_acc_mono_3D_recon_CBCT_ti_rods}
\end{table}

\begin{table}[h!]
\centering
\begin{tabular}{@{}lccc@{}}
\toprule
\multicolumn{1}{c}{  }&\multicolumn{1}{c}{Uncorrected}& \multicolumn{1}{c}{DSE}& \multicolumn{1}{c}{PhILSCAT} \\
\cmidrule(r){1-1}\cmidrule(lr){2-2}\cmidrule(lr){3-3}\cmidrule(l){4-4}
     PSNR (dB) & $36.8\pm0.4$ & $49.9\pm1.8$ & {$\bold{51.7\pm0.9}$} \\
\cmidrule(r){0-0}
     SSIM & $0.988 {\small \pm6.10^{-4}}$ & {$\bold{0.997{\small \pm6.10^{-4}}}$} & {$\bold{0.997 {\small\pm4.10^{-4}}}$} \\
\cmidrule(r){1-1}
     MAE (HU) & $35.0\pm1.5$ & $13.3\pm1.4$ & {$\bold{11.9\pm1.0}$} \\
\cmidrule(r){1-1}
Peak$\,$Error$\,$(HU) & $4229$ & $1538$ & {$\bold{954}$} \\
\bottomrule
\end{tabular}
\caption{\small Polychromatic CBCT: average reconstruction accuracy results and standard deviations for different algorithms for 3 Ti rod test phantoms.}
\vspace{-0.2cm}
\label{table:recon_acc_poly_3D_recon_CBCT_ti_rods}
\end{table}

\begin{figure*}[!htp]
\centering
\setlength{\tabcolsep}{-0.05cm}
\renewcommand{\arraystretch}{0.01}
\vspace{-0.5cm}
\begin{tabular}{cccccc}
\hspace{-8mm} \textbf{Reference Recon} & \multicolumn{3}{c}{\textbf{Error Magnitudes}} & \multicolumn{2}{c}{\textbf{Tighter HU Window Recon}}
\\[-2mm]
\includegraphics[width=.18\linewidth]{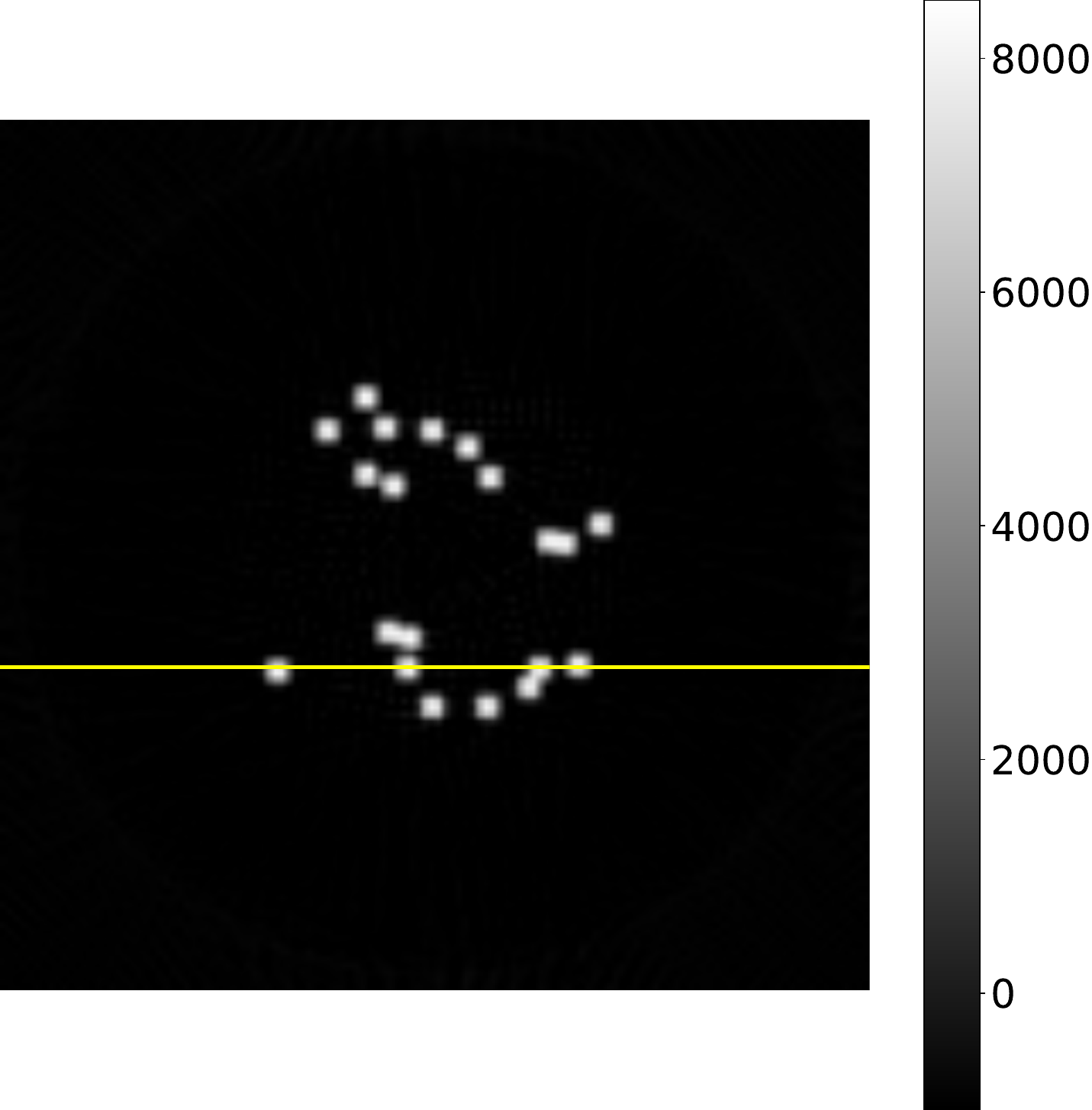} & \hspace{0.0cm}
\includegraphics[width=.18\linewidth]{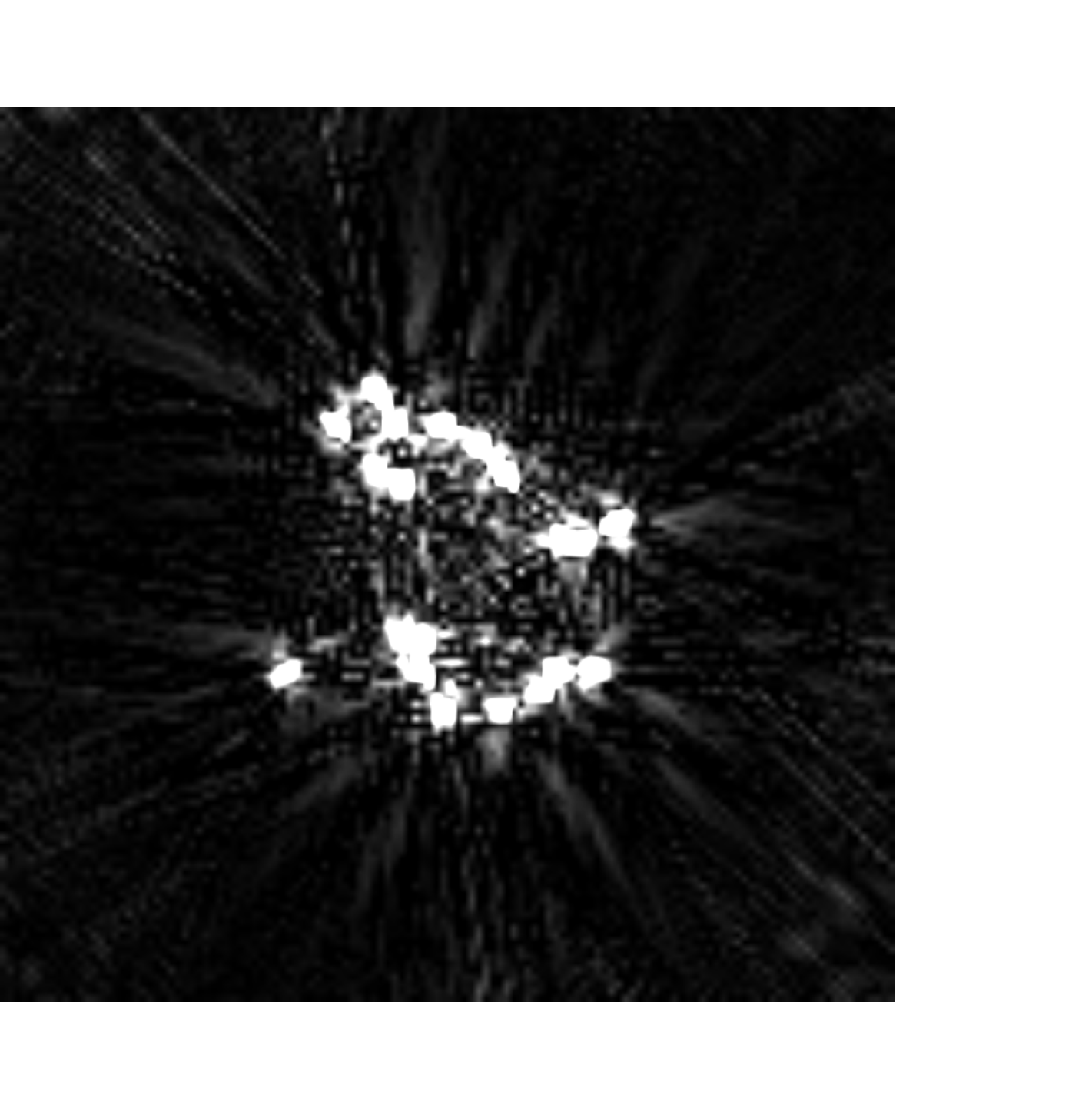} & \hspace{-0.6cm}
\includegraphics[width=.18\linewidth]{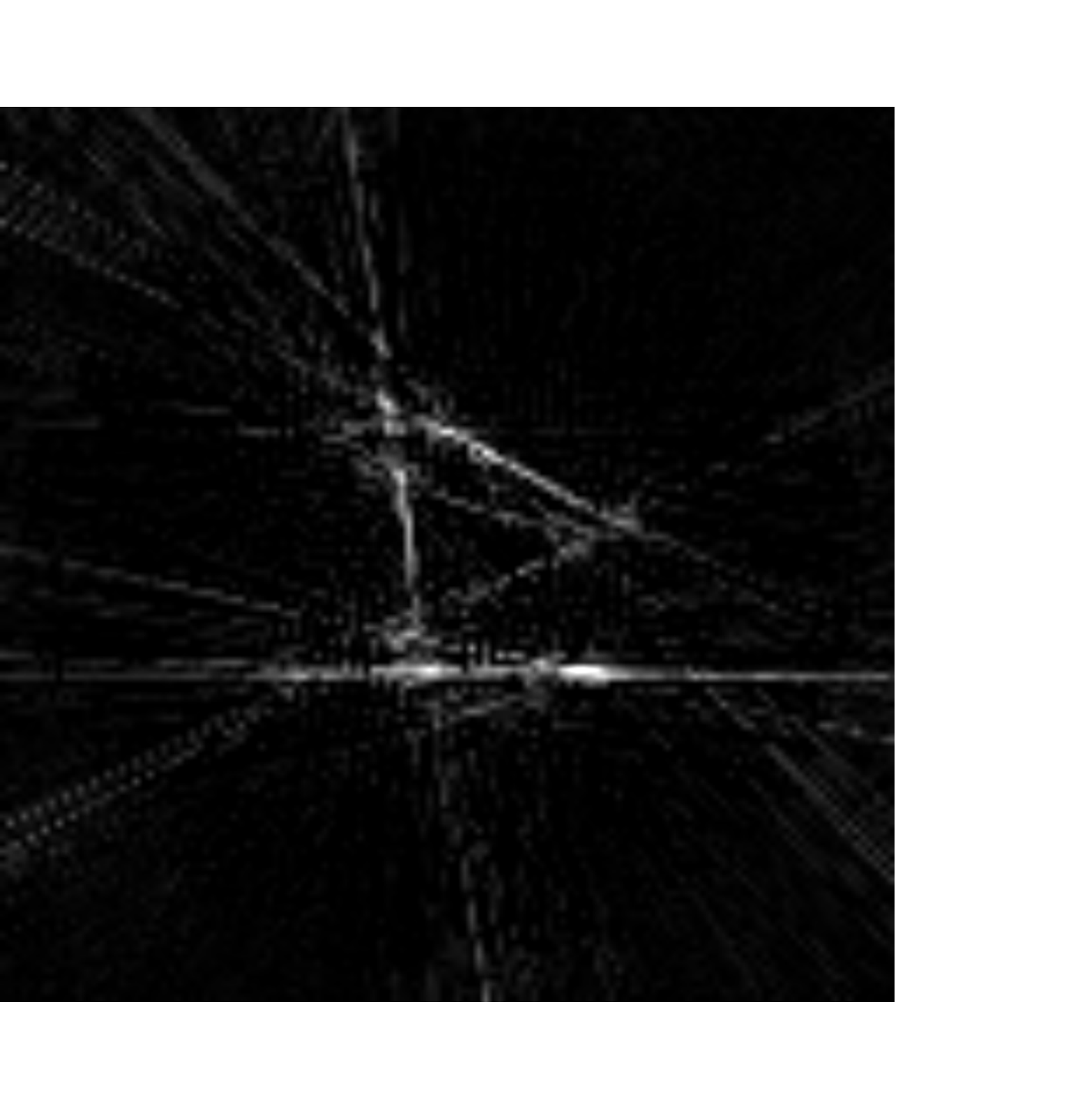} & \hspace{-0.6cm}
\includegraphics[width=.18\linewidth]{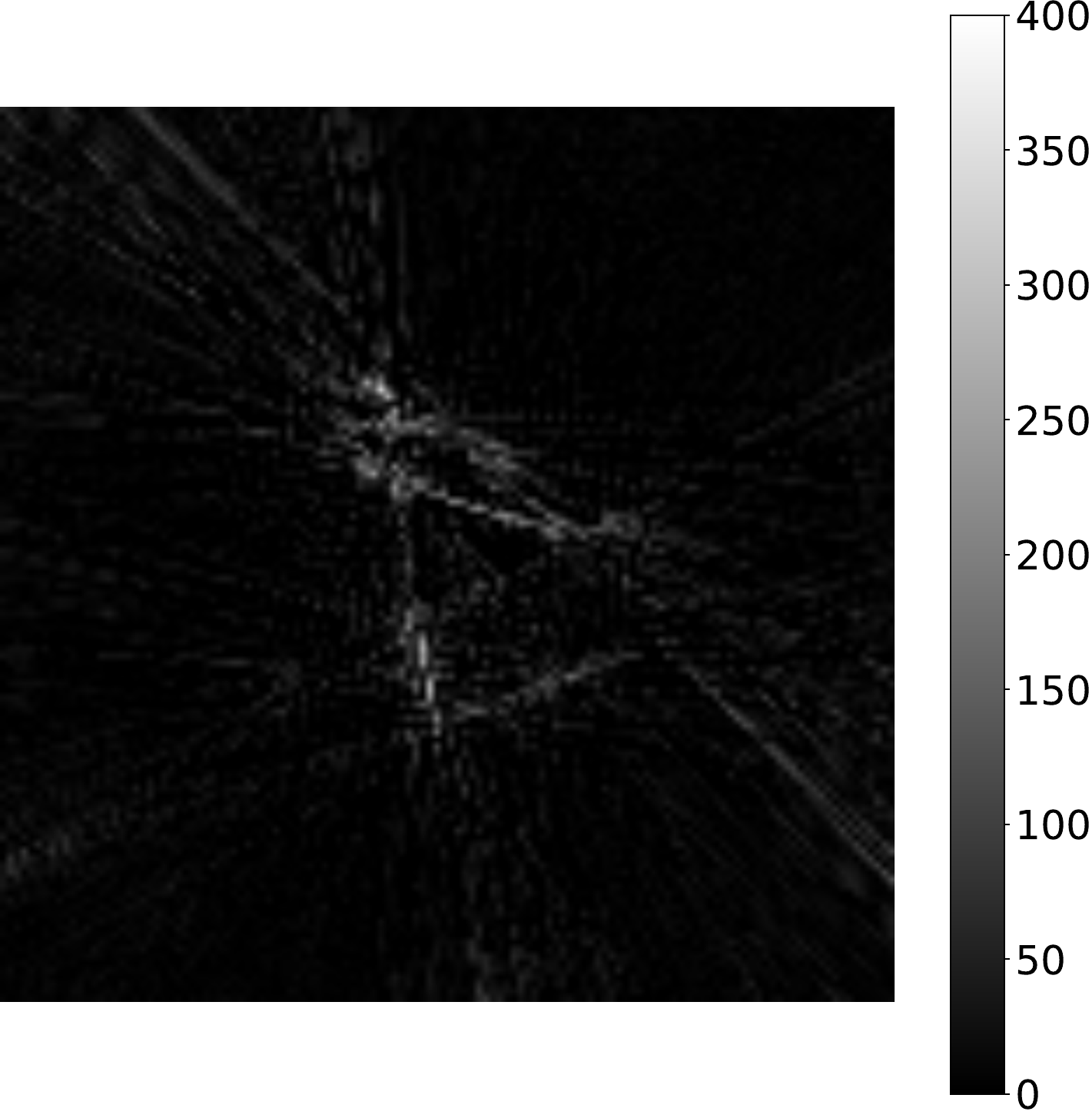} & \hspace{0.3cm}
\includegraphics[width=.18\linewidth]{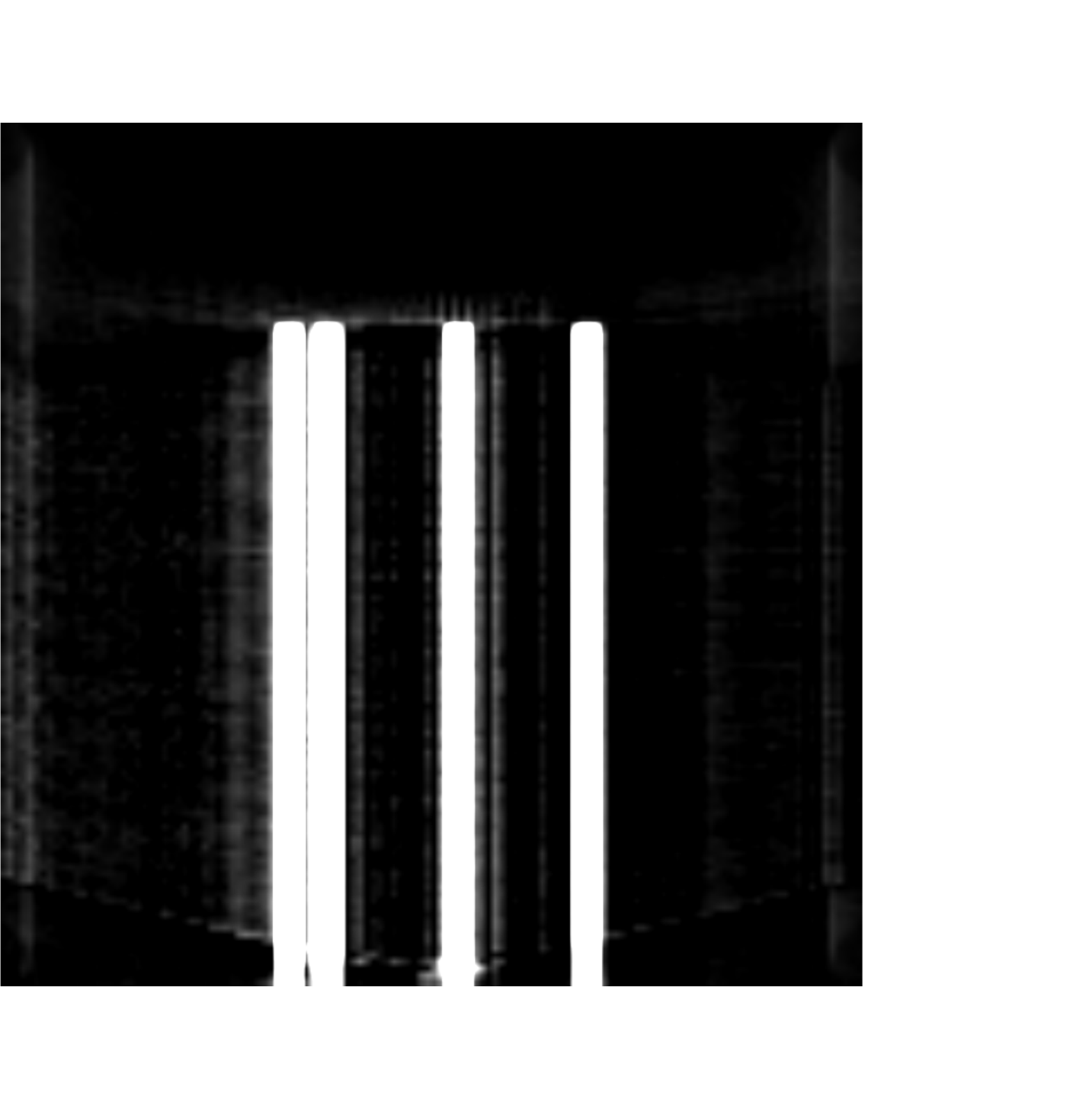} & \hspace{-0.6cm}
\includegraphics[width=.18\linewidth]{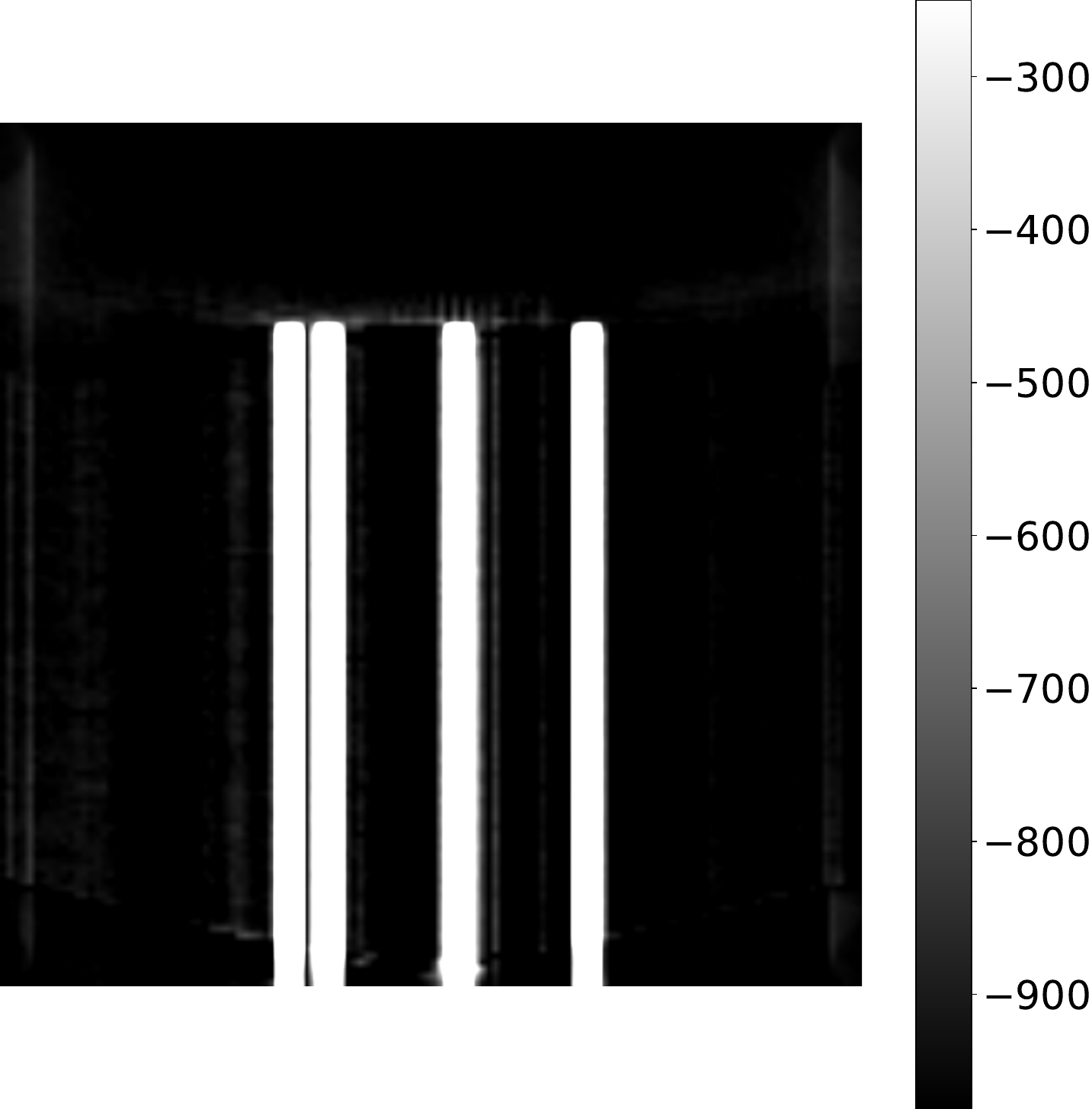}
\vspace{-0.2cm}
\\
\hspace{-0.7cm} (a) Axial & \hspace{-0.5cm} (c) Total & \hspace{-1cm} (e) DSE & \hspace{-1cm} (g) PhILSCAT &
\hspace{-0.5cm} (i) Sagittal & \hspace{-1.5cm} (k) Total 
\\

\includegraphics[width=.18\linewidth]{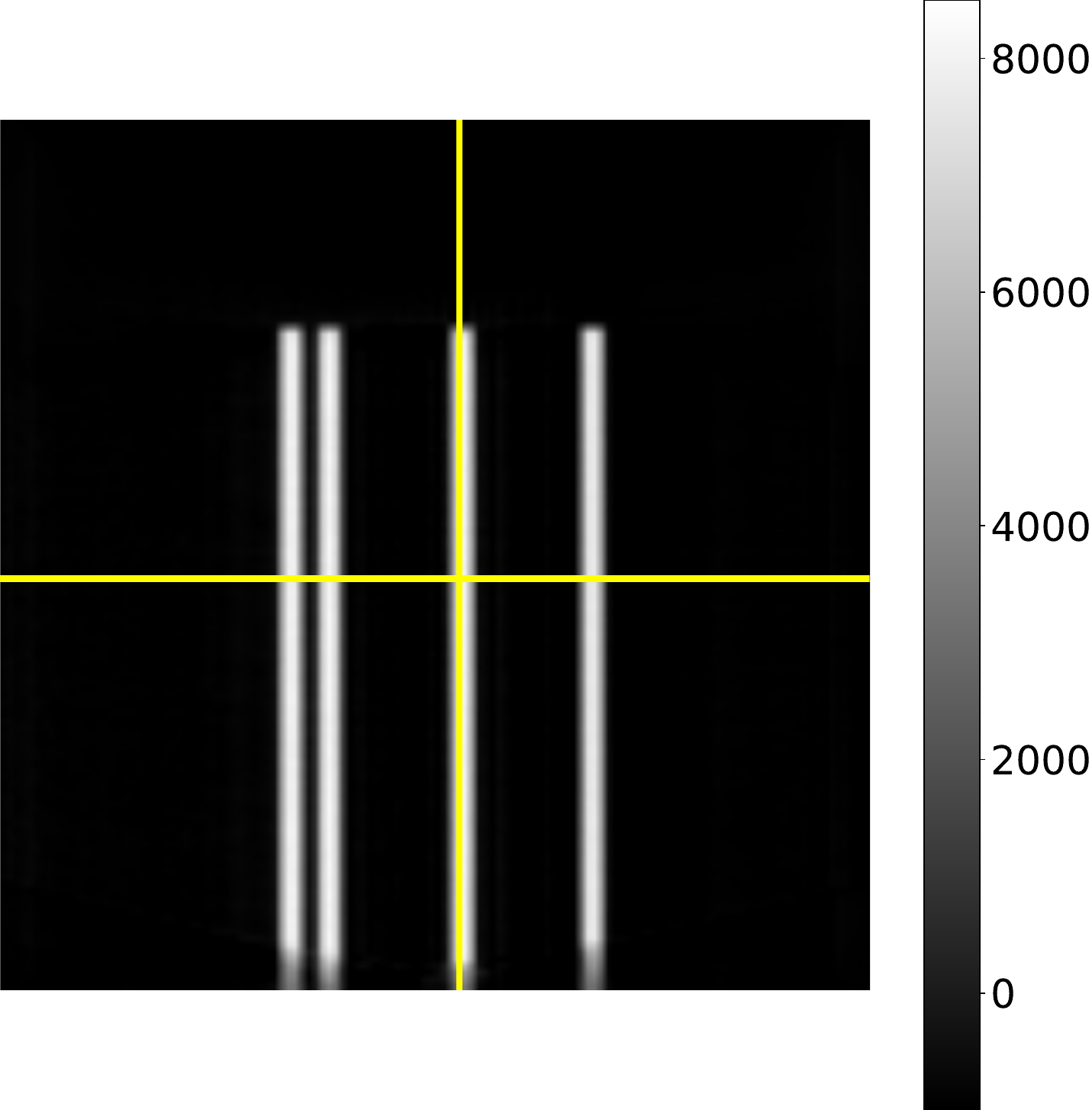} & \hspace{0.0cm}
\includegraphics[width=.18\linewidth]{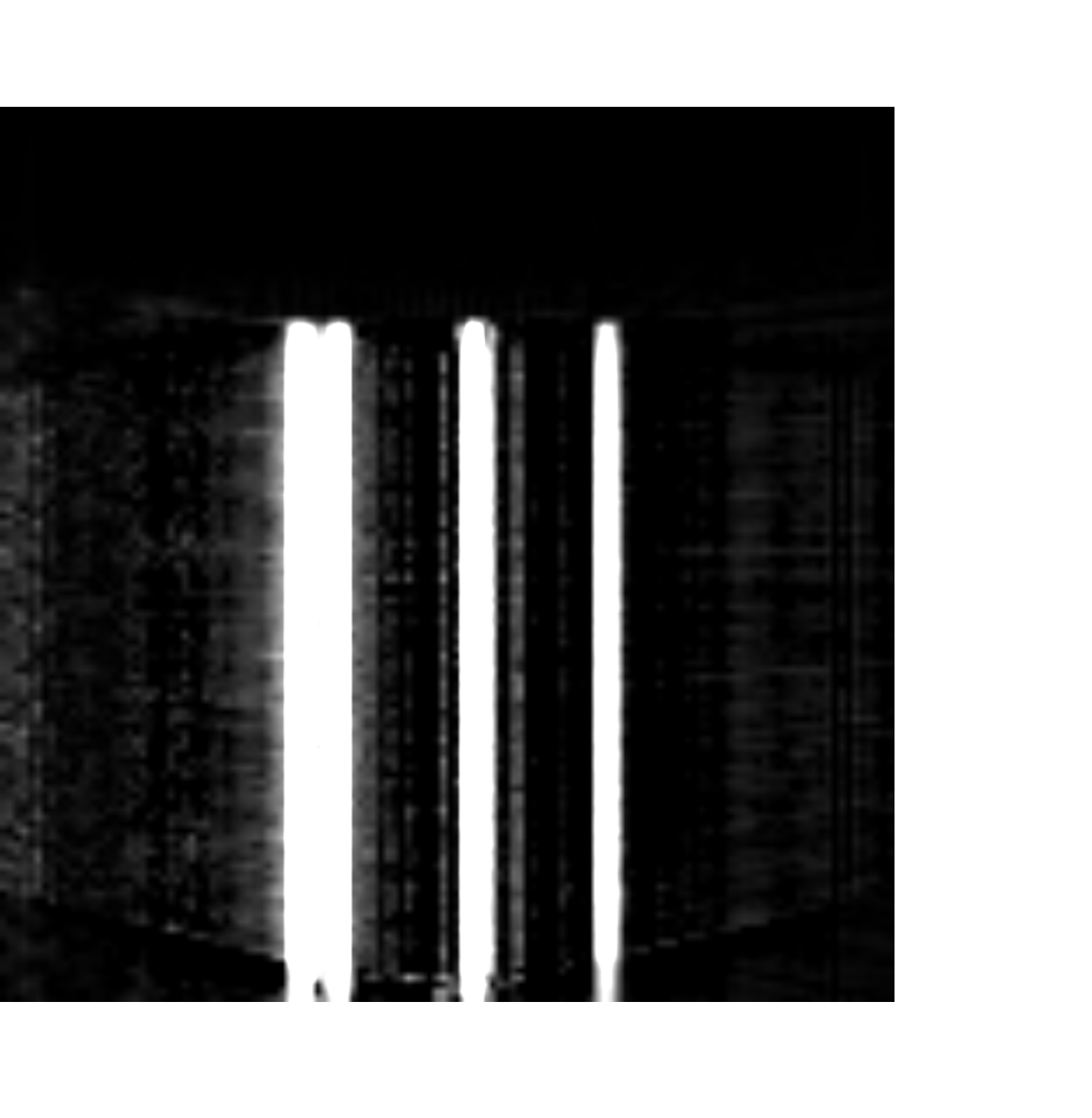} & \hspace{-0.6cm}
\includegraphics[width=.18\linewidth]{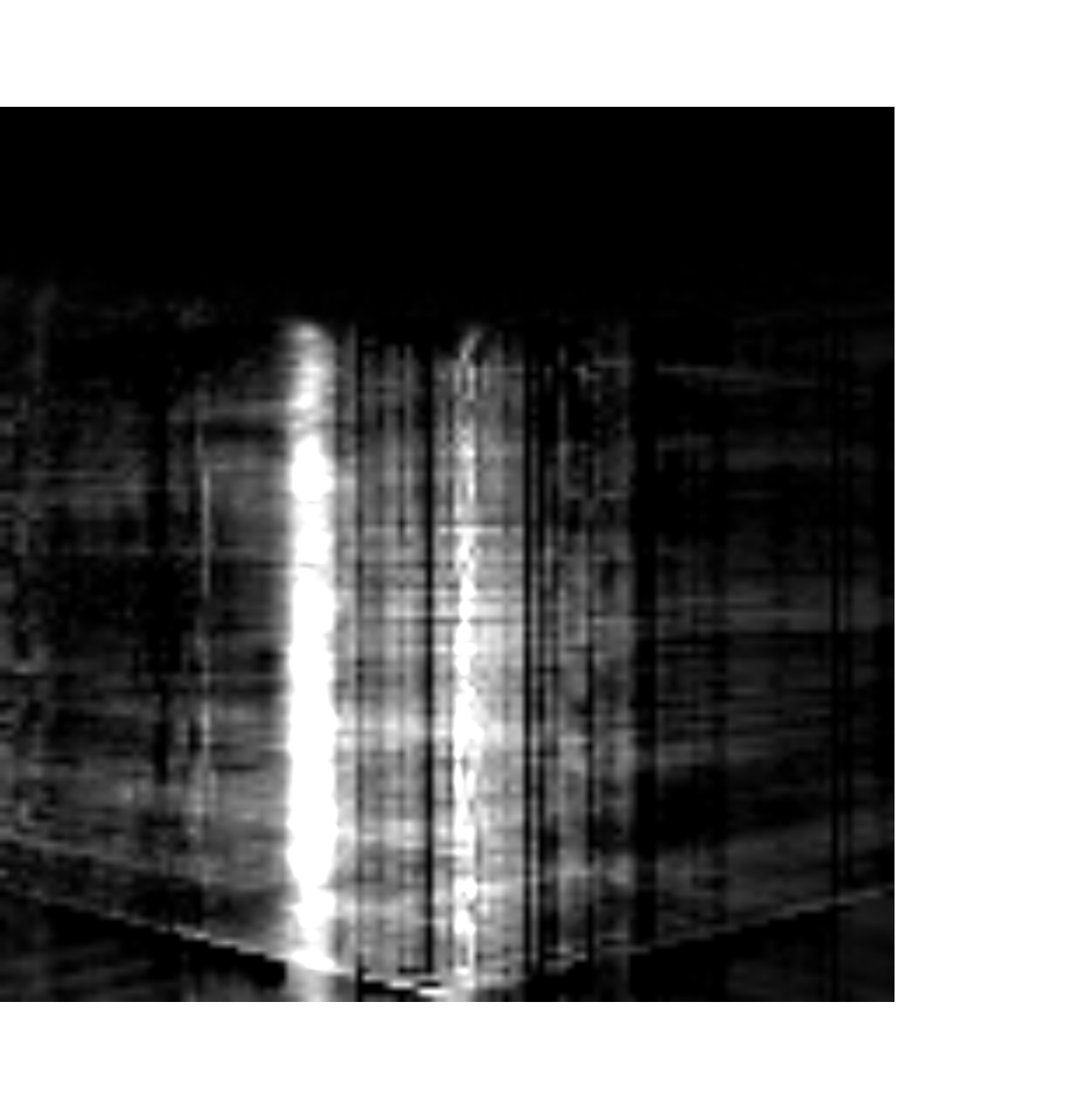} & \hspace{-0.6cm}
\includegraphics[width=.18\linewidth]{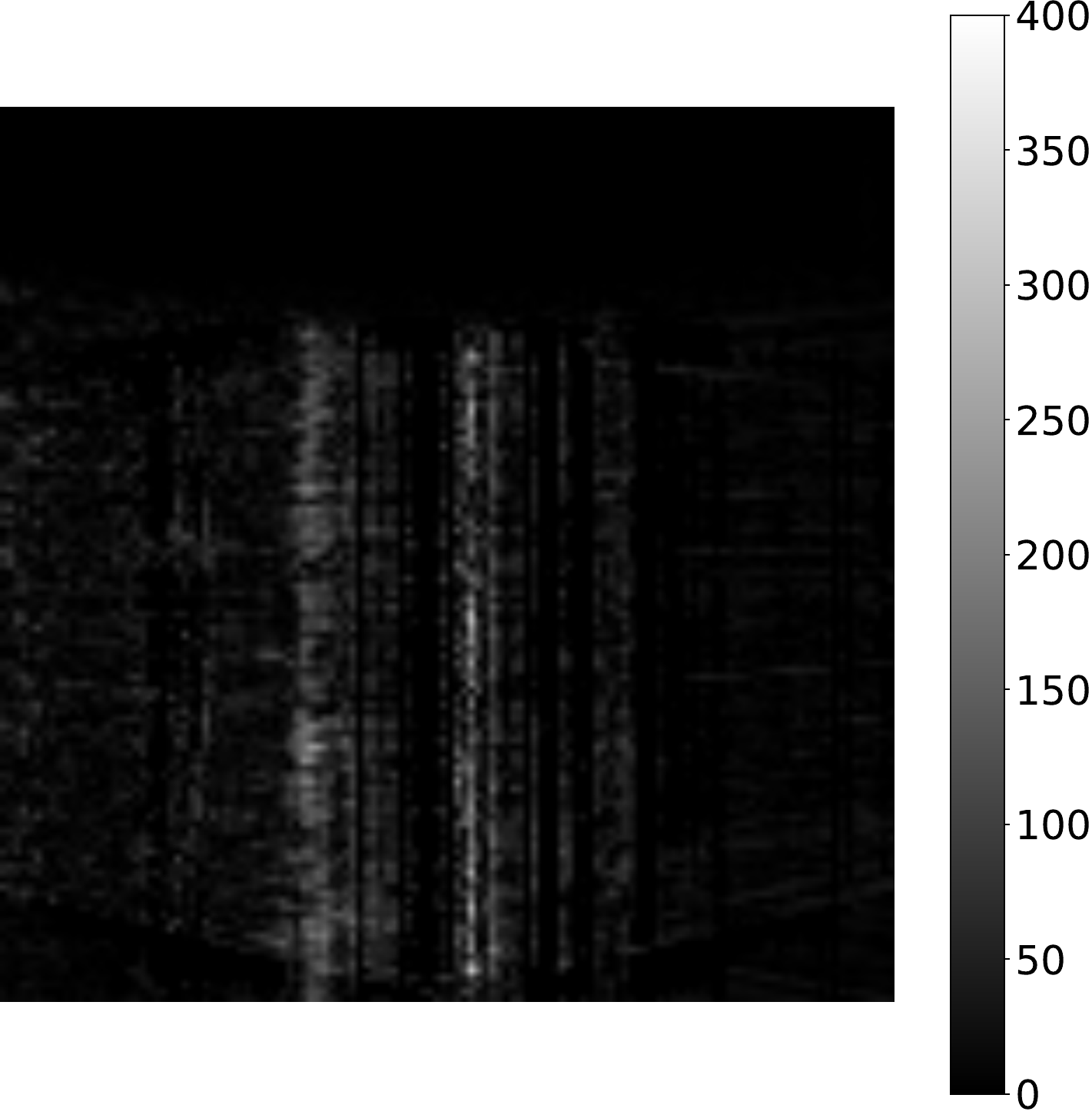} &
\hspace{0.3cm}
\includegraphics[width=.18\linewidth]{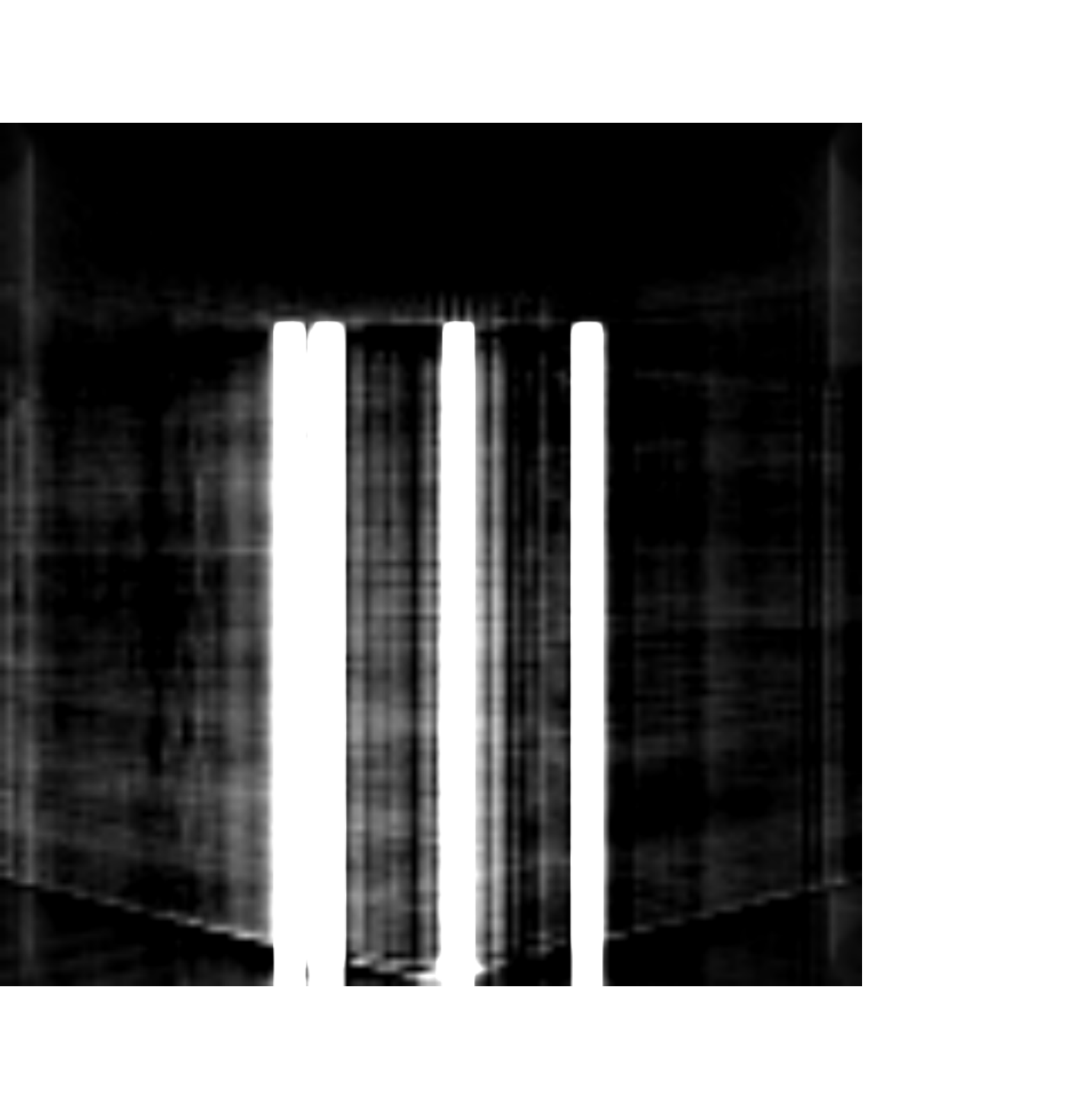} & \hspace{-0.6cm}
\includegraphics[width=.18\linewidth]{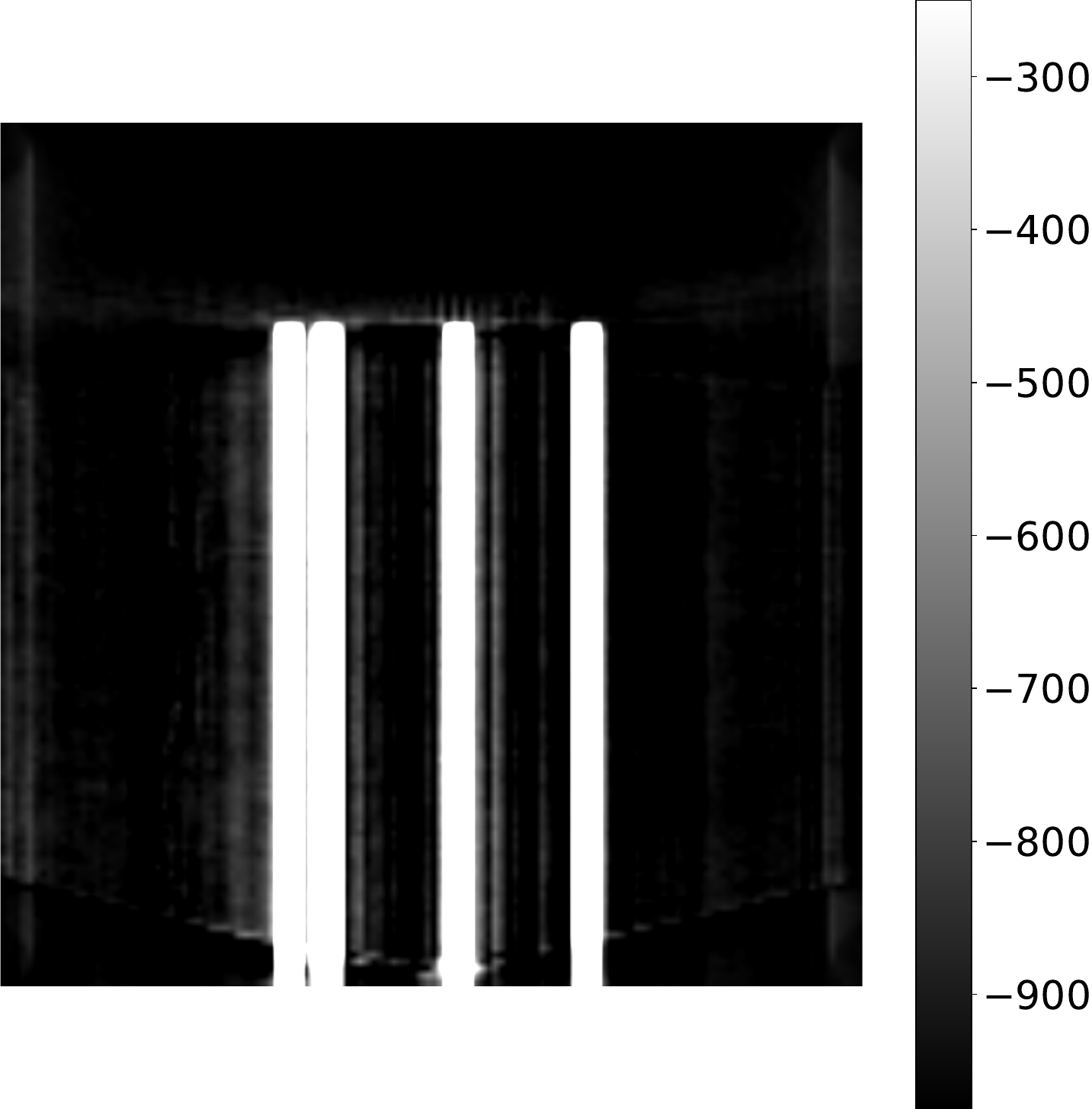}
\vspace{-0.2cm}
\\ 
\hspace{-0.7cm} (b) Sagittal & \hspace{-0.5cm} (d) Total & \hspace{-1cm} (f) DSE & \hspace{-0.9cm} (h) PhILSCAT &
\hspace{-0.5cm}(j) DSE & \hspace{-1.5cm}(l) PhILSCAT \\

\end{tabular}
\caption{\small Monochromatic CBCT  reconstructions of Ti rod test phantom:
(a) (b) Central axial and sagittal slices, with tighter HU window in (i) of
 $p_\theta$-reconstructions
 ;
(c), (d) error magnitude and tighter HU window reconstruction (k) using total measurements $\tau_\theta$; 
vs. (e), (f) and (j) using primary measurements estimated by  DSE; and (g),(h) and (l) estimated by PhILSCAT. Display windows in HU 
are indicated by the colorbars.}
\vspace{-0.2cm}
    \label{fig:mag_err_comp_3D_CBCT_ti_rods} 
\end{figure*}

\begin{figure*}[t!]
\centering
\setlength{\tabcolsep}{-0.05cm}
\renewcommand{\arraystretch}{0.1}
\begin{tabular}{ccccc} \hspace{-8mm} \textbf{Reference Recon}
& \multicolumn{3}{c}{\textbf{Error Magnitudes}} & \textbf{Error MIPs}
\\[-3mm]
\includegraphics[width=.18\linewidth]{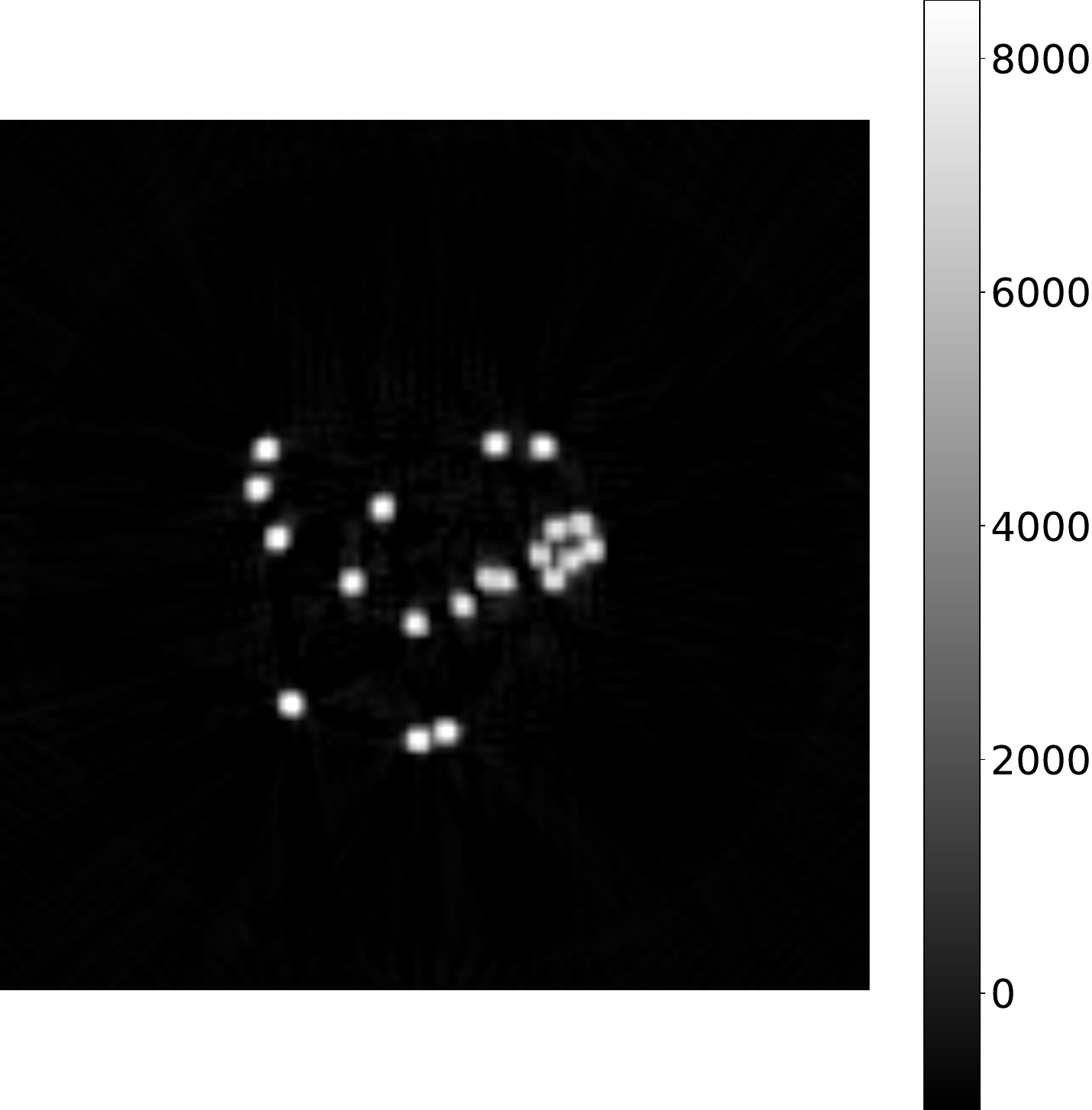} & \hspace{-0.0cm}
\includegraphics[width=.18\linewidth]{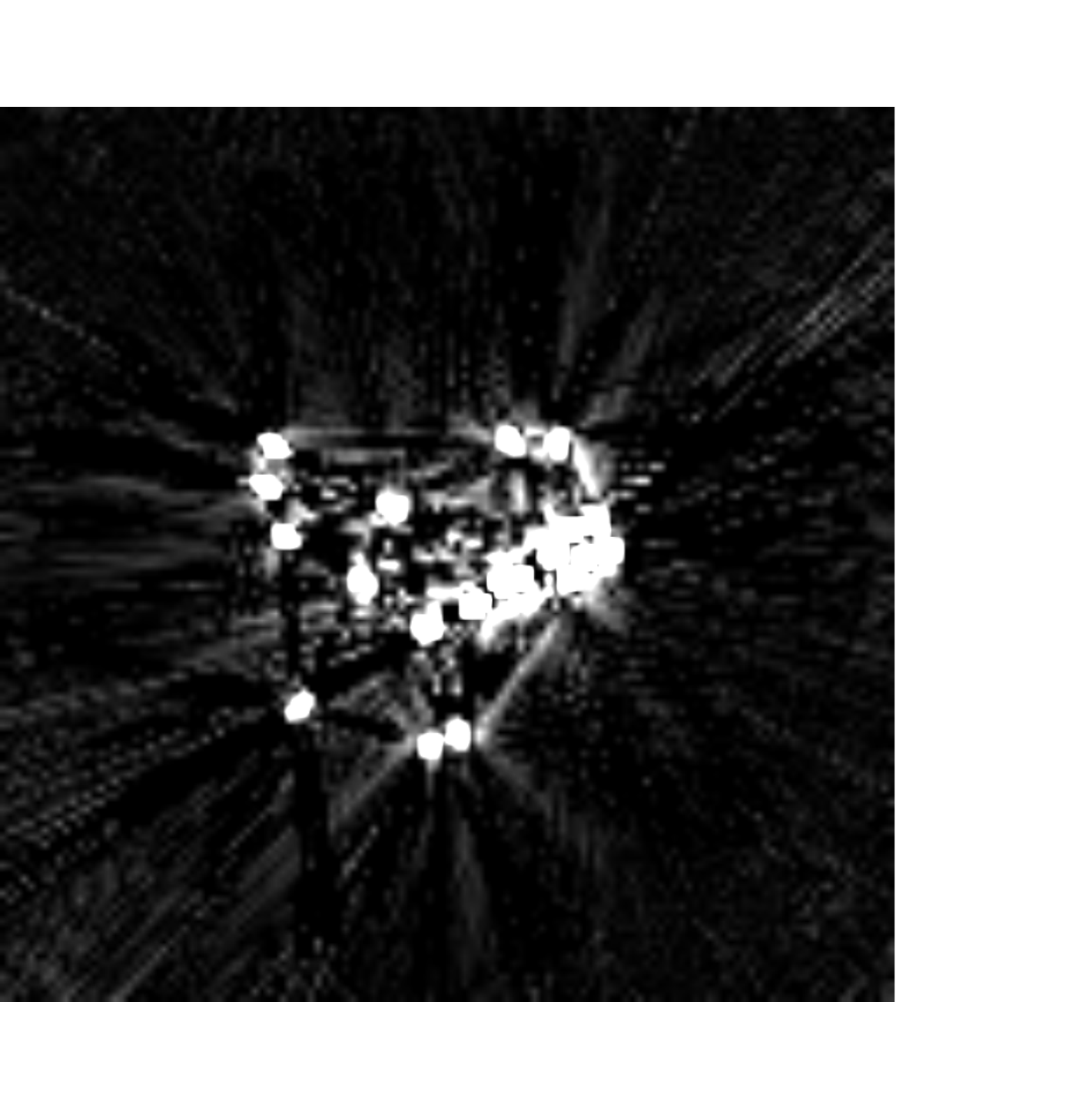} & \hspace{-0.6cm}
\includegraphics[width=.18\linewidth]{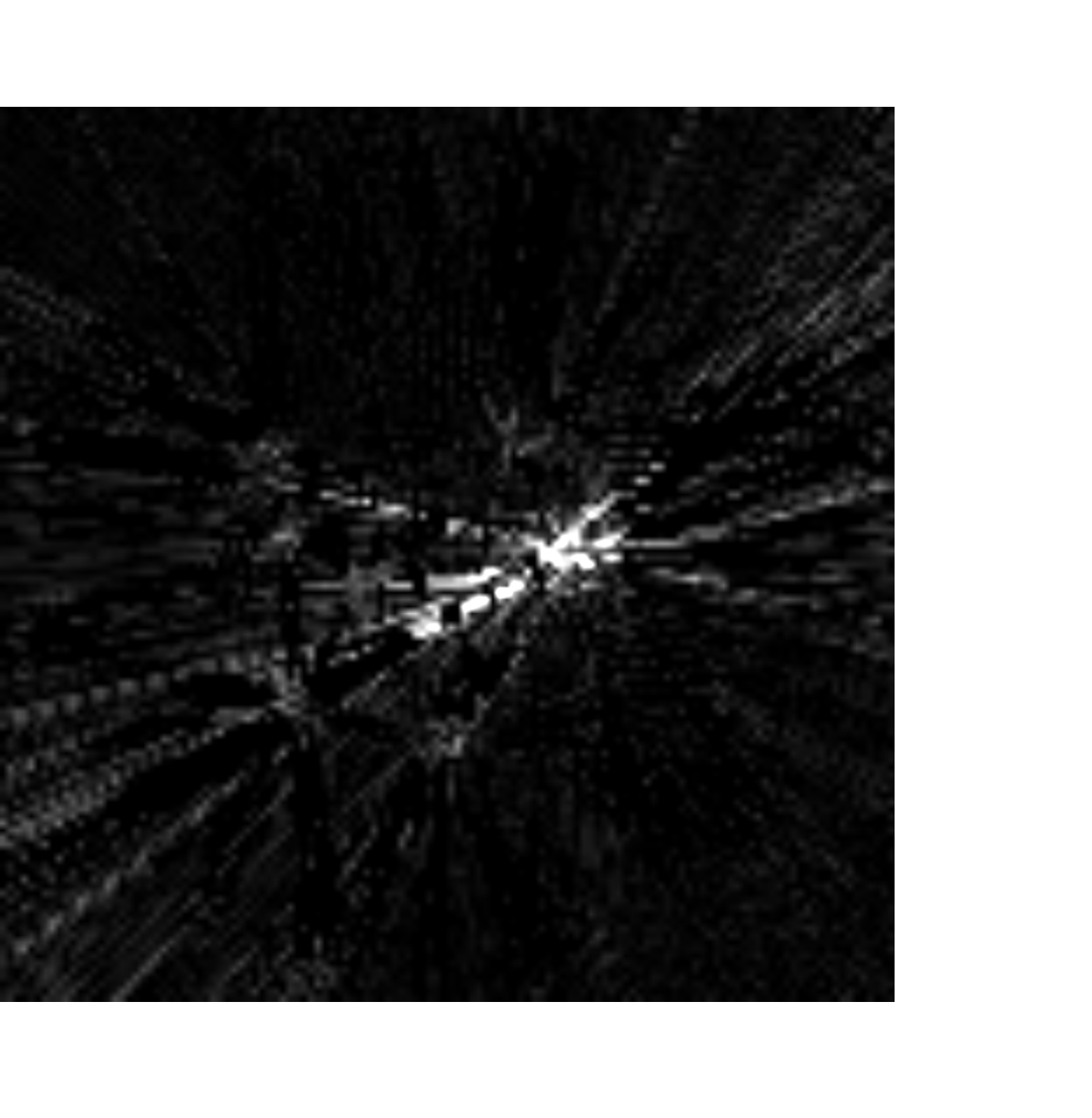} & \hspace{-0.6cm}
\includegraphics[width=.18\linewidth]{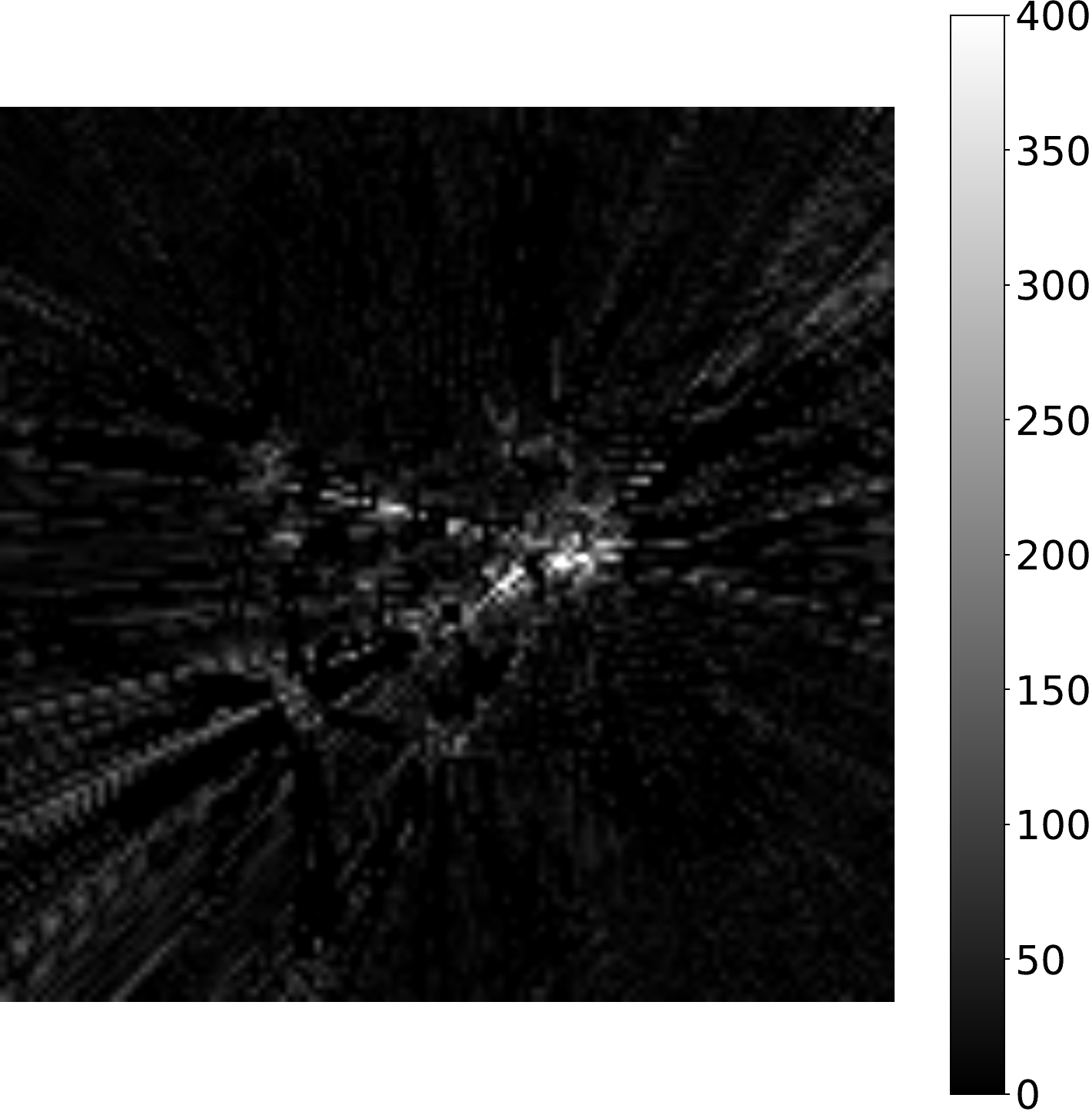} &
\hspace{0.1cm}
\includegraphics[width=.18\linewidth]{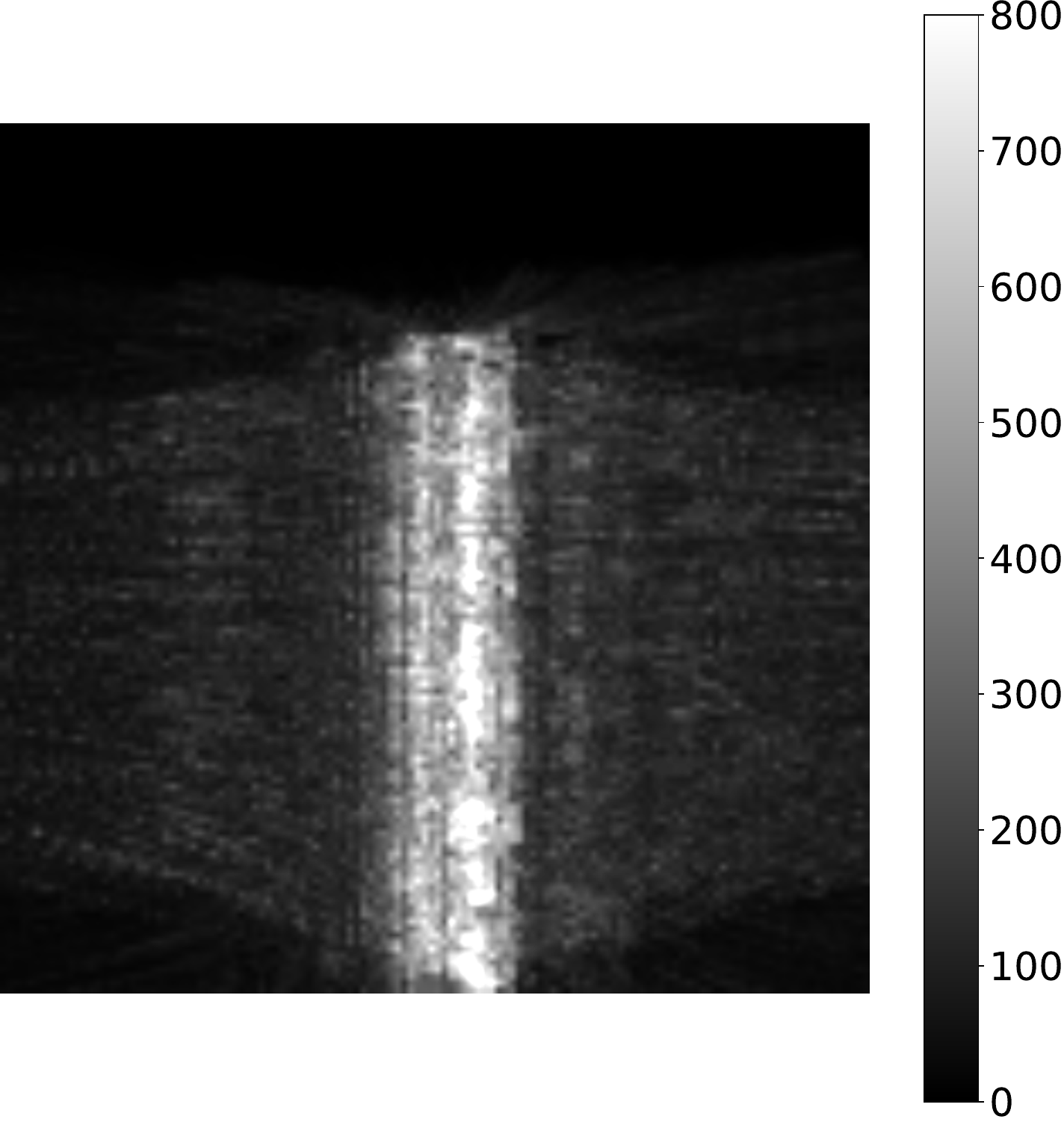} 
\vspace{-0.2cm}
\\
\hspace{-0.7cm} {\small (a) Axial} & \hspace{-0.5cm} {\small (c) Total} & \hspace{-1cm} {\small (e) DSE} & \hspace{-0.9cm} {\small (g) PhILSCAT} & \hspace{-0.5cm} 
{\small (i) DSE error MIP}
\vspace{0.3cm}
\\[-2mm]
& \multicolumn{3}{c}{\textbf{Tighter HU Window Reconstructions}} &
\\[-4mm]
\includegraphics[width=.18\linewidth]{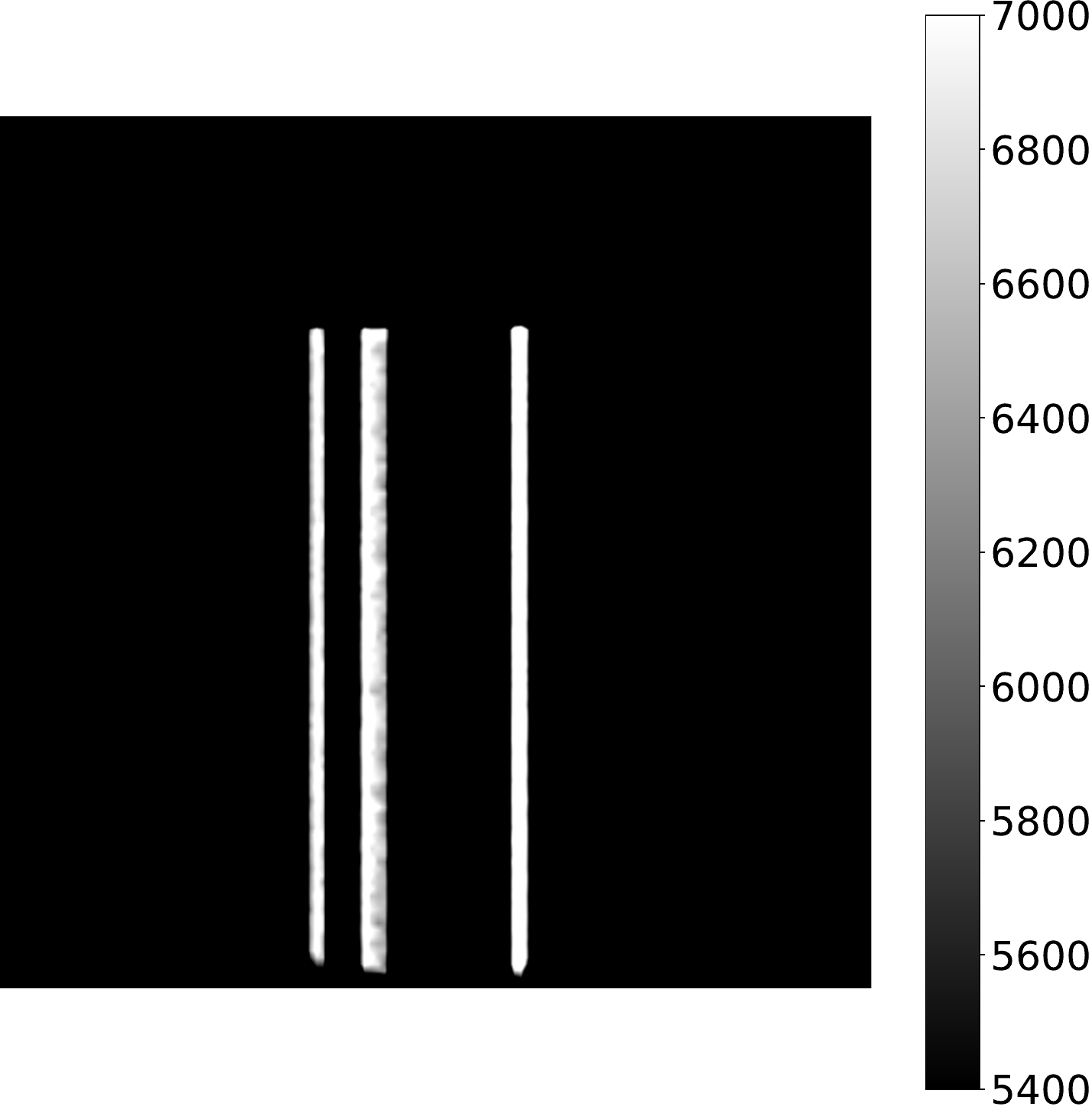} & \hspace{-0.0cm}
\includegraphics[width=.18\linewidth]{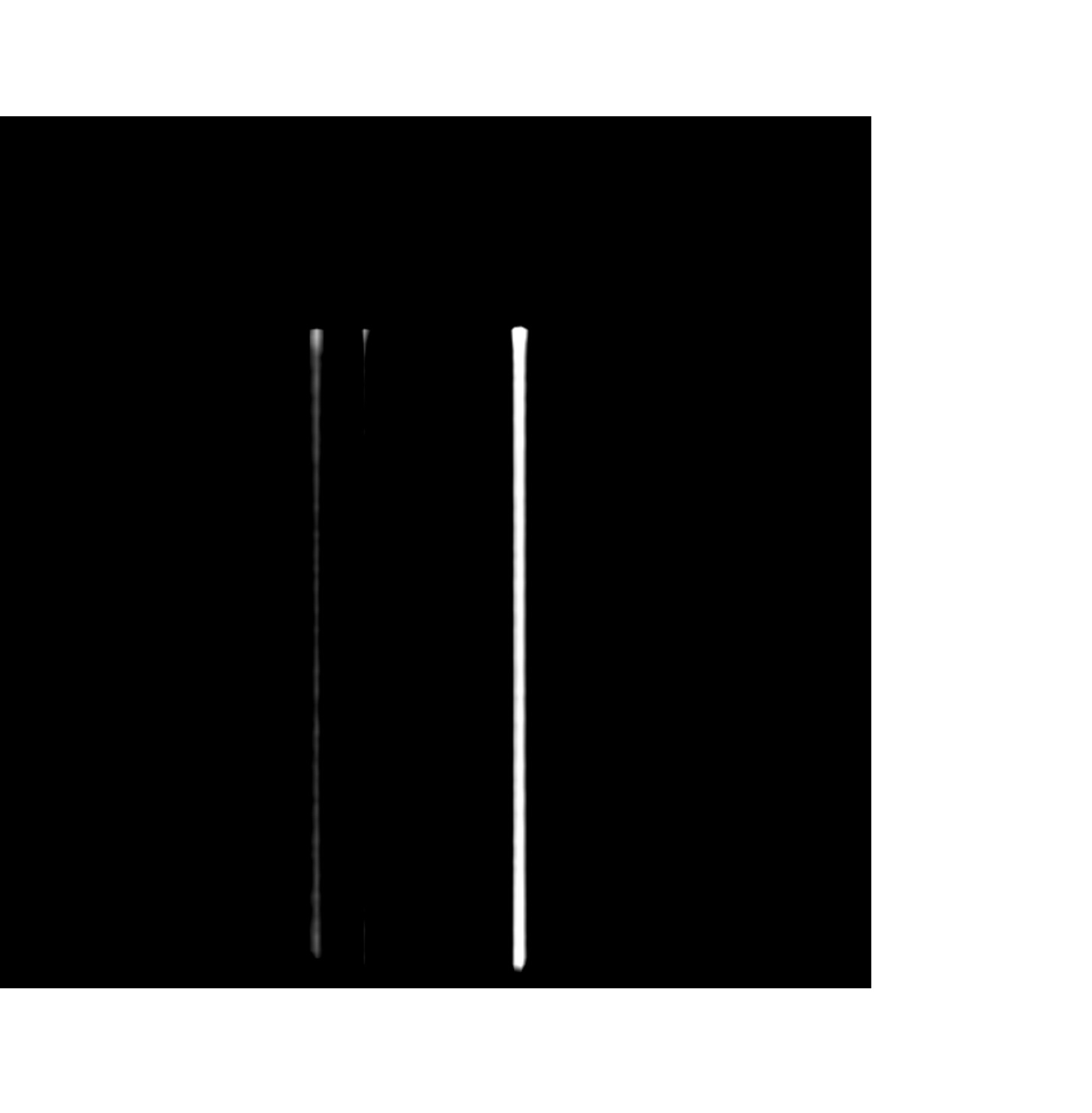} & \hspace{-0.6cm}
\includegraphics[width=.18\linewidth]{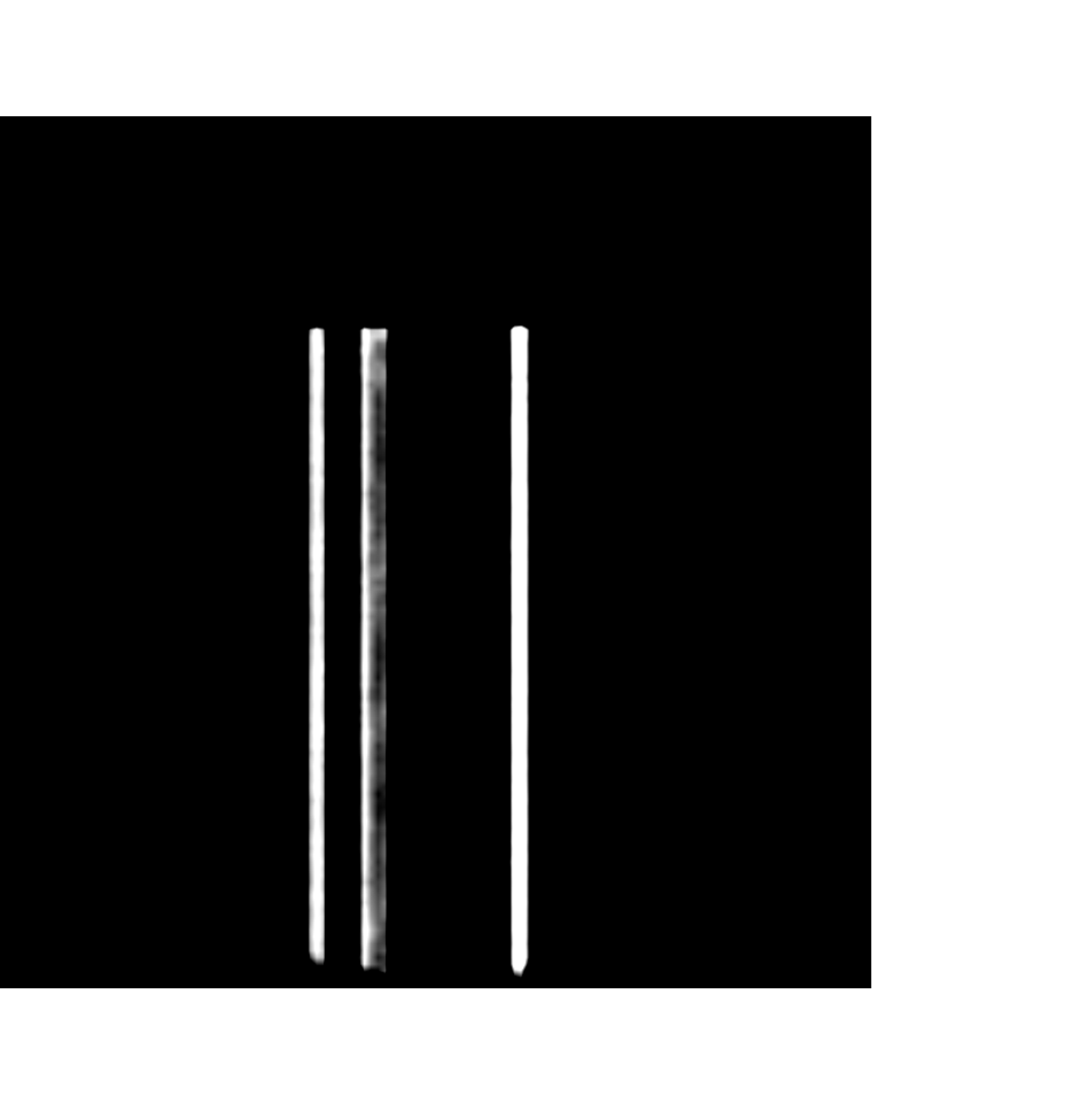} & \hspace{-0.6cm}
\includegraphics[width=.18\linewidth]{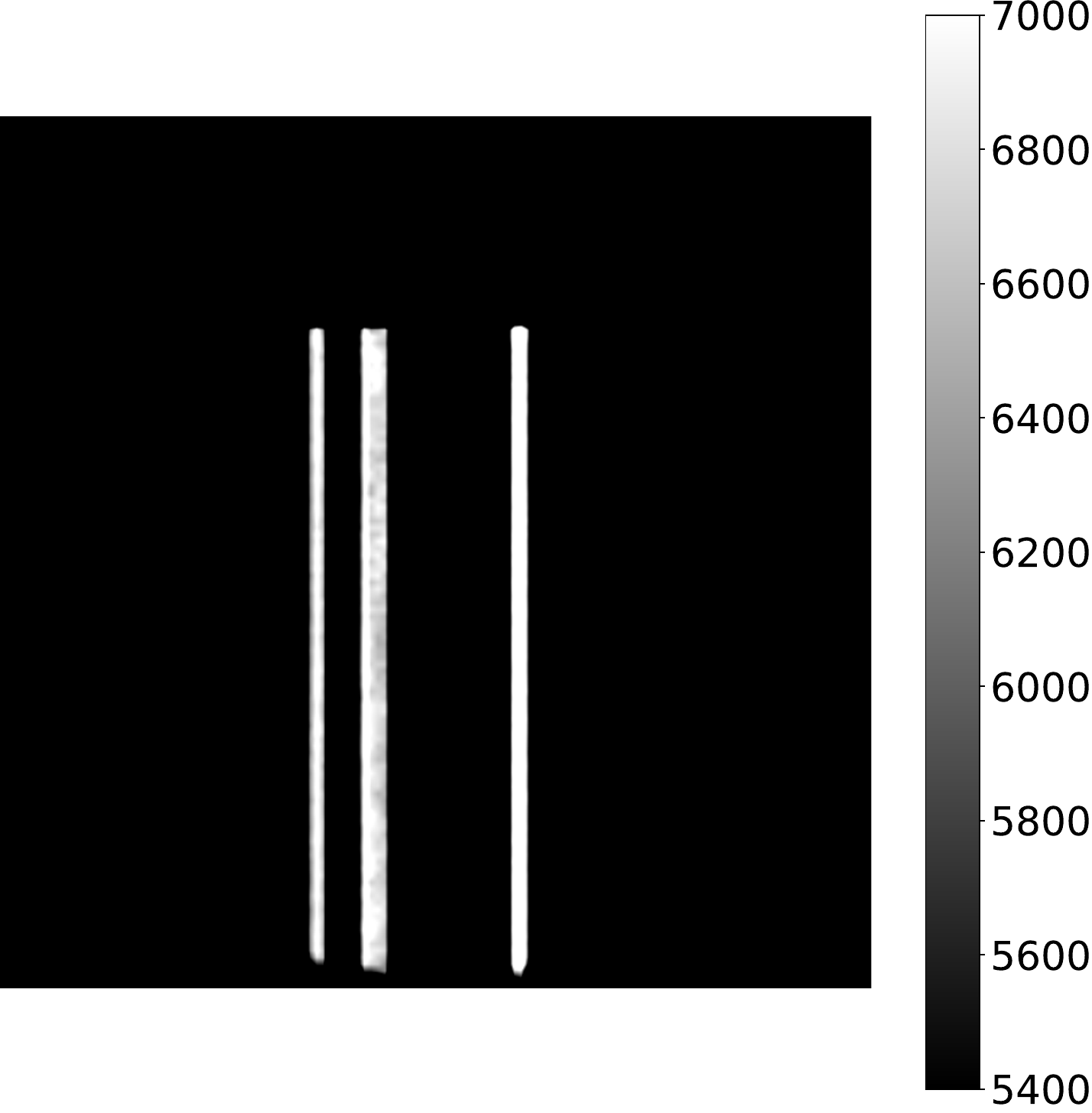} &
\hspace{0.1cm}
\includegraphics[width=.18\linewidth]{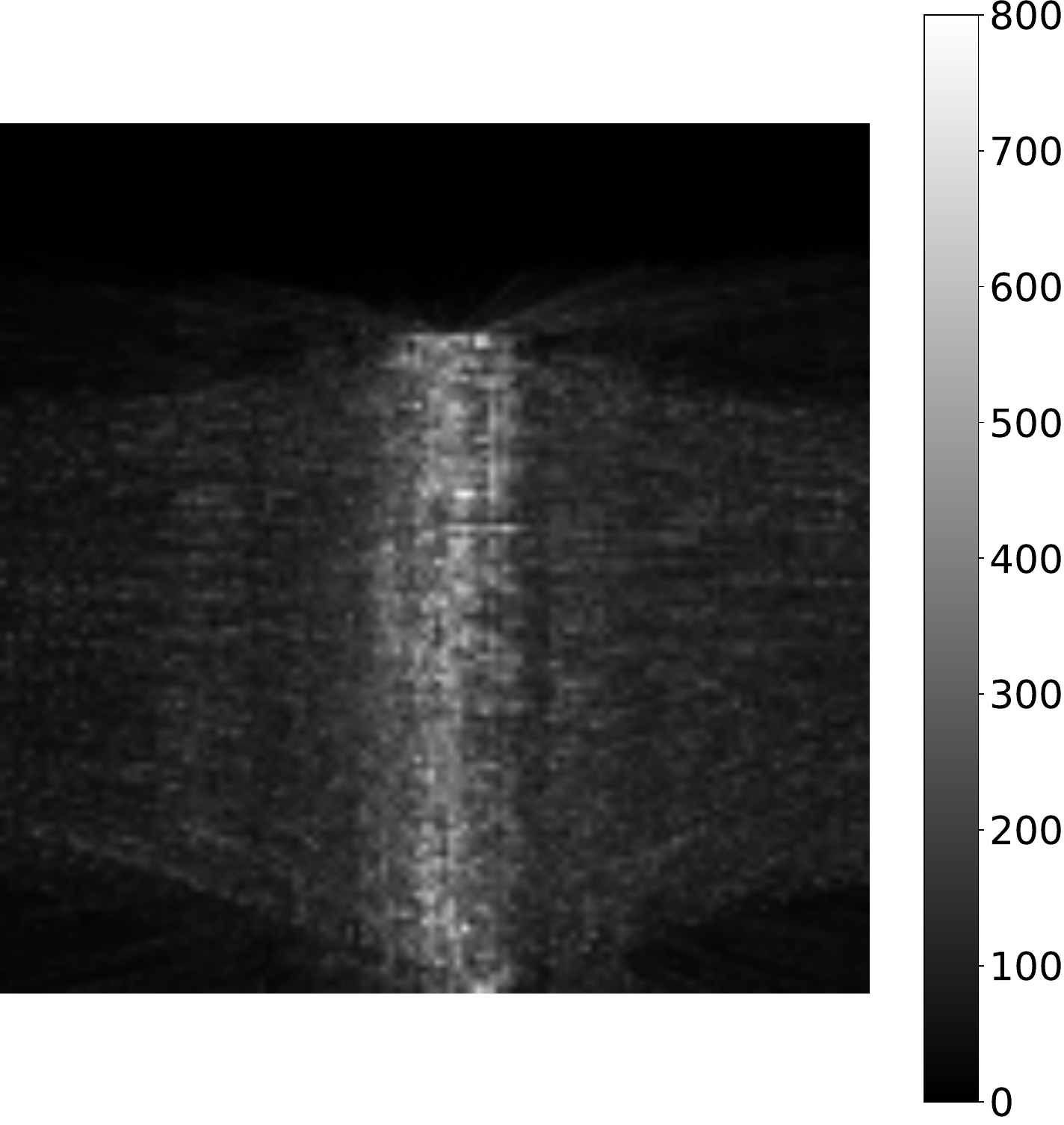}
\vspace{-0.2cm}
\\ 
\hspace{-0.7cm} {\small (b) Sagittal} & \hspace{-0.5cm} {\small (d) Total} & \hspace{-1cm} {\small (f) DSE} & \hspace{-0.9cm} {\small (h) PhILSCAT} & \hspace{-0.5cm} {\small (j) PhILS. error MIP}
\\

\end{tabular}
\caption{\small Polychromatic CBCT reconstruction of Ti rod test phantom  -- central axial and sagittal slices.  (Numerical values in HU). (a), (b)
$p$-Reconstruction
; (c) error magnitudes and (d) reconstruction using total measurements $\tau$, 
vs. (e), (f) using primary measurements estimated by  DSE; or (g), (h) estimated by PhILSCAT. (i) and (j) Sagittal slice maximum intensity projections (MIPs) of the error volumes using $p^*$ estimated by DSE  vs. estimated by PhILSCAT, respectively.}
\vspace{-0.5cm}
    \label{fig:mag_err_comp_3D_CBCT_ti_rods_poly} 
\end{figure*}

\begin{figure}[!htp]
\centering
\setlength{\tabcolsep}{0.02cm}
\begin{tabular}{ccc}
\includegraphics[width=0.33\linewidth]{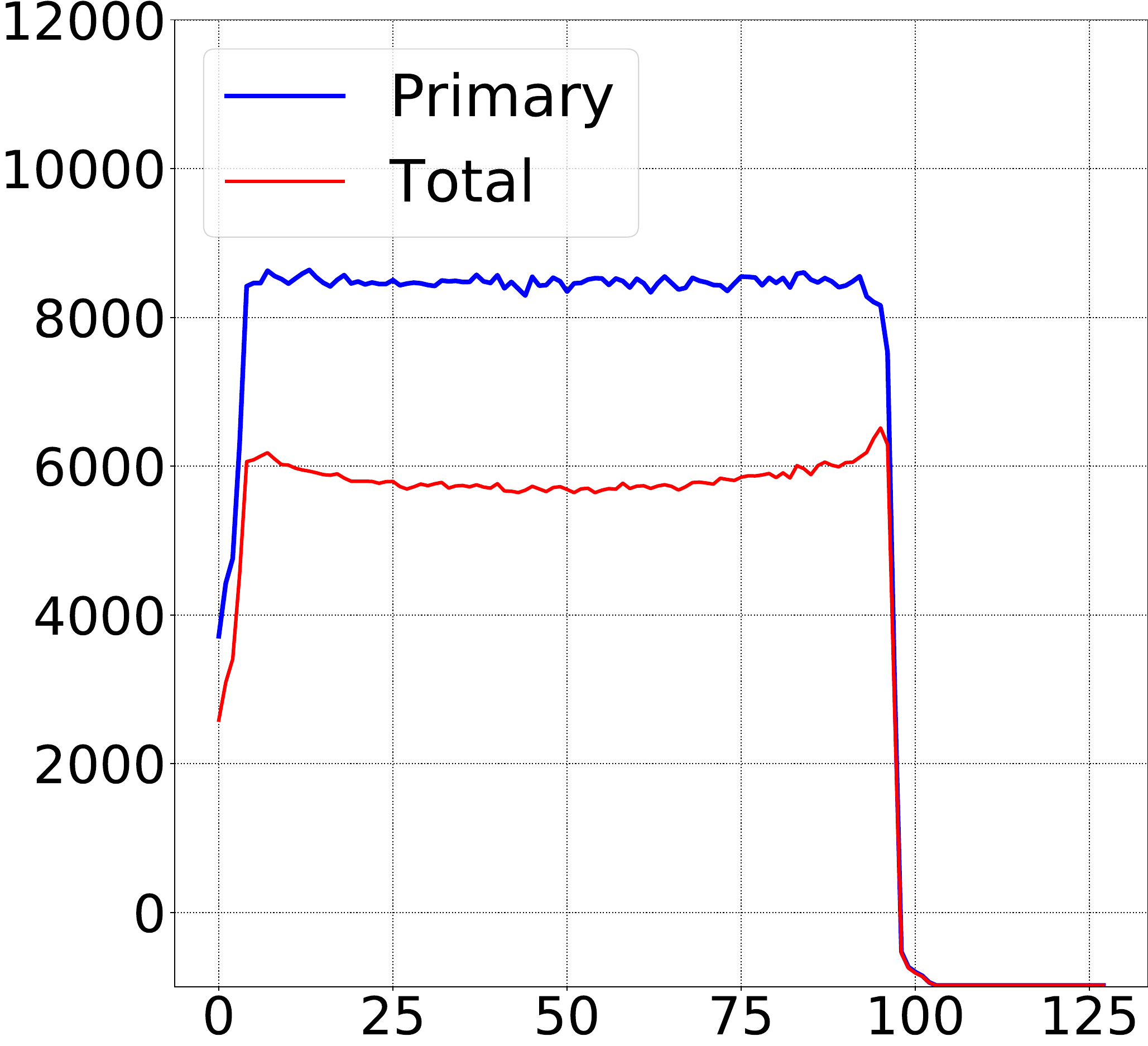}&
\includegraphics[width=0.33\linewidth]{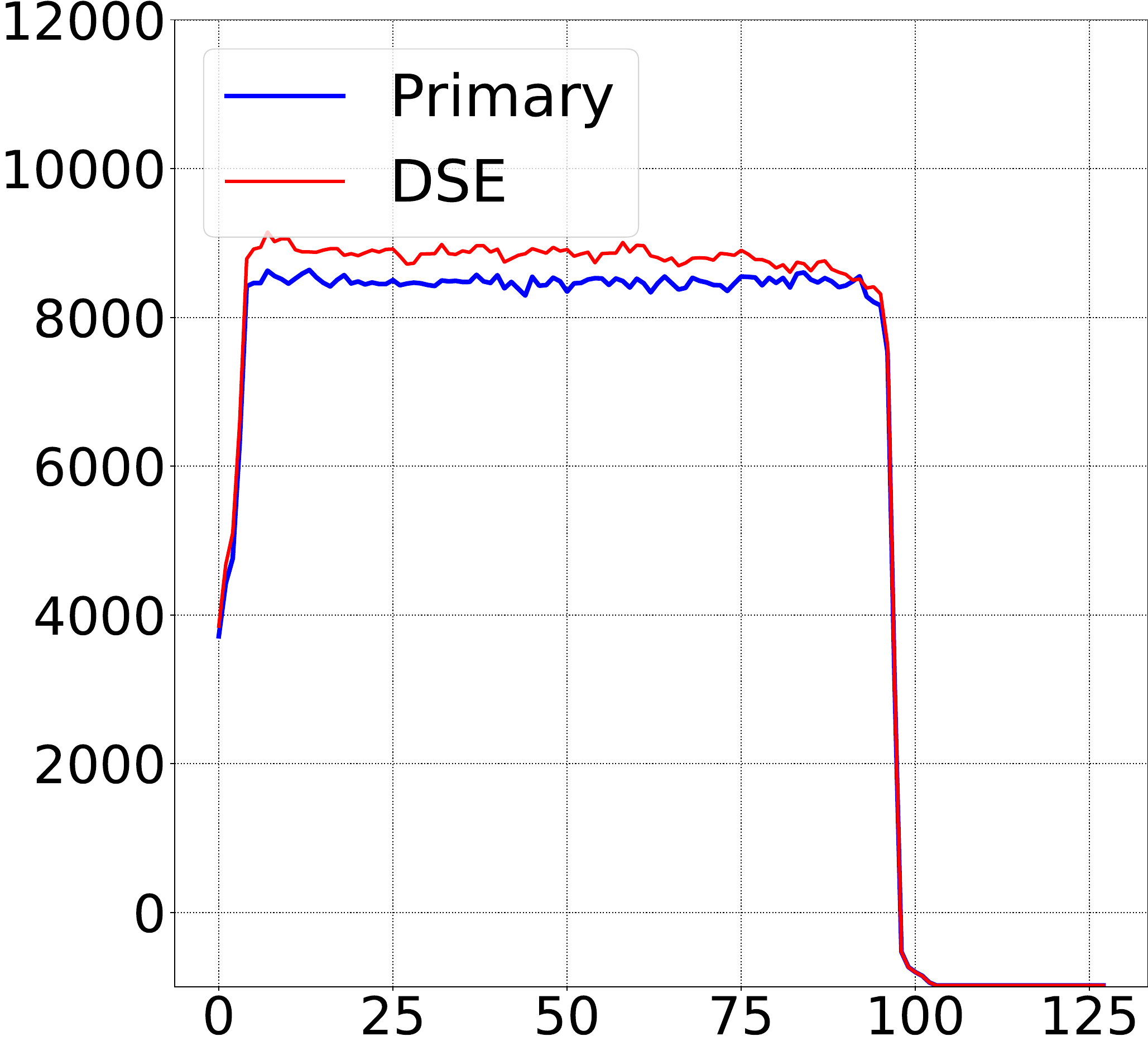}&
\includegraphics[width=0.33\linewidth]{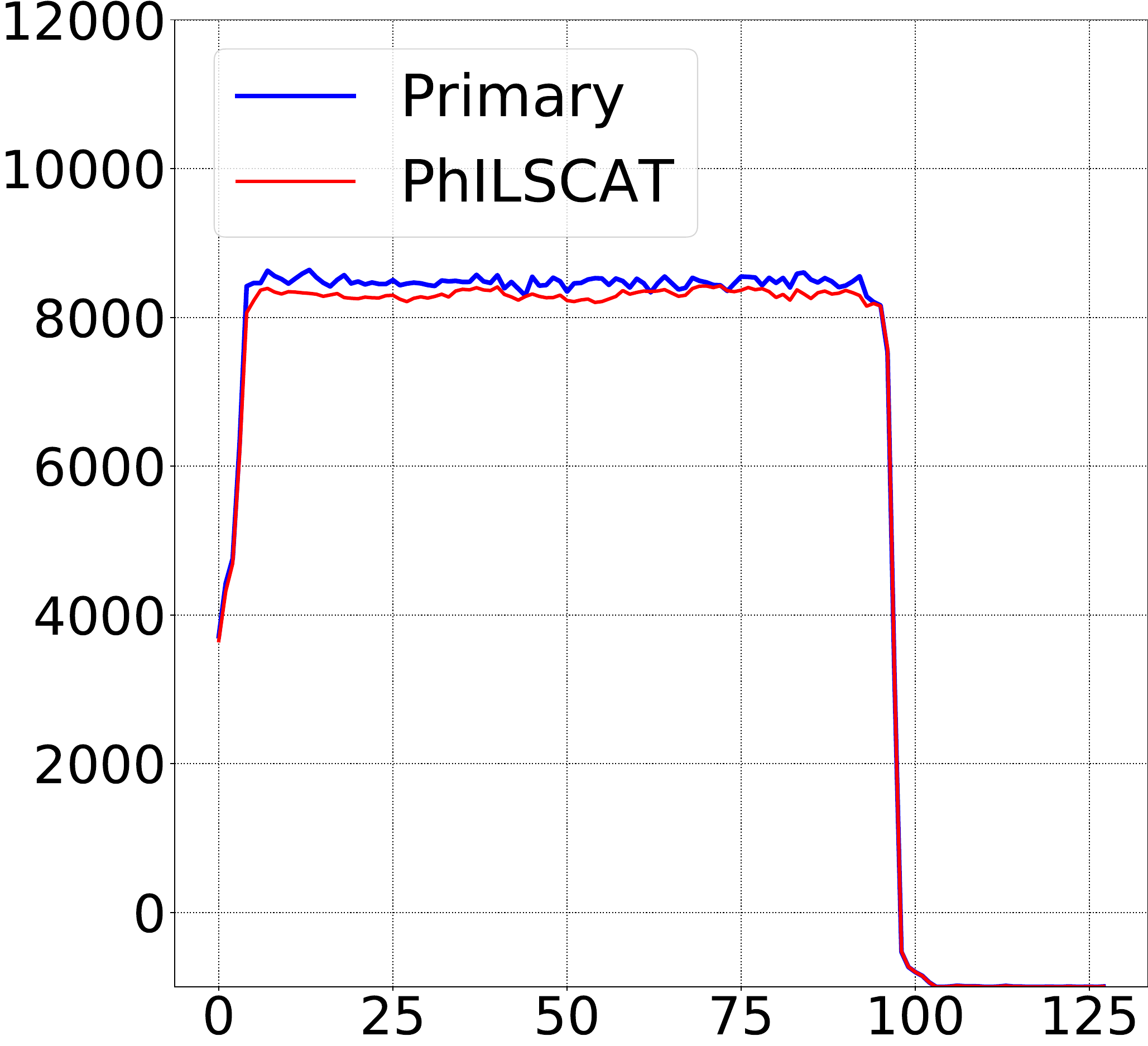}\\
(a) Total & (b) DSE & (c) PhILSCAT
\end{tabular}
\caption{\small Polychromatic  CBCT  reconstructions  of  Ti  rod  test  phantom: horizontal line profiles indicated in Fig. \ref{fig:mag_err_comp_3D_CBCT_ti_rods_poly}:
(a) Using total measurements $\tau_\theta$; vs.
using primary measurements $p_\theta^*$ estimated by (b) DSE; and
(c) by PhILSCAT. }
\label{fig:line_profile_hor_comp_3D_CBCT_ti_rods_sagittal} \vspace{-0.2in}
\end{figure}

Reference reconstructions computed using primary measurements $p$, tighter HU window scatter-corrected reconstructions, and error magnitude plots are given in Figures \ref{fig:mag_err_comp_3D_CBCT_ti_rods} and \ref{fig:mag_err_comp_3D_CBCT_ti_rods_poly}, for the monochromatic and polychromatic cases, respectively.
To facilitate comparison, line profiles for each method and maximum intensity projections (MIP) of the absolute error volumes for DSE and PhILSCAT are provided together in Figures \ref{fig:mag_err_comp_3D_CBCT_ti_rods_poly} and ~\ref{fig:line_profile_hor_comp_3D_CBCT_ti_rods_sagittal} for the polychromatic case. A MIP in a particular direction is obtained by assigning to the projection the maximum absolute values the rays encounter in any voxel they traverse. Since they accumulate peak errors in many slices, it is possible that MIP values have considerable spatial variation, resulting in a noise-like map.

As could be expected from the strong scatter, the $\tau$-reconstructions 
contain large magnitude errors of 1000 -- 2000 HU over the objects. The projection-domain DSE algorithm improves considerably on these magnitude errors. However, there still remain relatively large errors on the highly attenuating paths. PhILSCAT corrects scatter-related artifacts noticeably better over those regions while 
not suffering any setbacks on any other part of the reconstruction
compared to DSE. Additionally, streaking artifacts are suppressed to a greater extent using PhILSCAT. This can also be observed in reconstructions with tighter HU window focusing on lower densities in Fig. \ref{fig:mag_err_comp_3D_CBCT_ti_rods}. Finally, tighter HU window reconstructions for higher HU values in Fig. \ref{fig:mag_err_comp_3D_CBCT_ti_rods_poly} show the better performance of PhILSCAT in reducing shading artifacts and  increasing contrast. Thanks to these differences, PhILSCAT performs better than DSE in terms of PSNR, SSIM and the peak error, as revealed by the metrics reported in Tables~\ref{table:recon_acc_mono_3D_recon_CBCT_ti_rods} and ~\ref{table:recon_acc_poly_3D_recon_CBCT_ti_rods}. 

The peak errors are over the entire reconstructed volume and a single voxel with a large outlier error can determine this quantity. To provide further insight into the larger error values, we note that the total number of voxels for the test phantoms that have error magnitudes larger than 500 HU is reduced by PhILSCAT by
$\approx 8$ fold 
from their number (3,900) in DSE. Also, thanks to the loss function described in Sec. \ref{sec:loss_fct}, which minimizes the error in the reconstruction domain, although  PhILSCAT has $ 12\%$ larger MSE in the estimates of scatter, the \emph{reconstruction} PSNR of PhILSCAT is 1.8 dB better than DSE.

\subsection{Anthropomorphic Phantoms}
\label{sec:poly_med_phantoms}
For the polychromatic 3D CBCT reconstructions, 
DSE and PhILSCAT were each trained on 27 and tested on 3 anthropomorphic phantoms as described in Section \ref{sec:data_generation} with the imaging geometry shown in Fig. \ref{fig:scatter_analytical_level_illustration}-(b). 

The peak error between the $\tau$-reconstructions of the three test phantoms and the $p$-reconstructions was 1699 HU, or 84\% of  the peak density of 2034 HU in the $p$-reconstructions, indicating significant degradation due to uncorrected scatter. 

The average and the peak
scatter-to-primary ratios for the three test phantoms were $\operatorname{AVG}_{t,\theta}s(t,\theta)/p(t,\theta) =29\%$, and $\max_{t,\theta} s(t,\theta)/p(t,\theta) = 212\%$, with $\operatorname{AVG}_{t,\theta}s(t,\theta)/p(t,\theta) =79\%$  over areas in the projections where $p(t,\theta) < \sqrt{I_0}$.
The closer scatter-to-primary ratios in areas with $p(t,\theta) < I_0$ vs. those with $p(t,\theta) < \sqrt{I_0}$ 
indicate that the scatter signals tend to be smoother 
for these phantoms compared to the Ti rod experiments in Section \ref{sec:ti_rods_CBCT}. 
Owing to this characteristic,
the difference in reconstruction quality between PhILSCAT and DSE, as expressed by the average PSNR and SSIM results for the 3 test phantoms in Table \ref{table:recon_acc_poly_3D_recon_CBCT_med}, is reduced compared to Sec.~\ref{sec:ti_rods_CBCT}. Still, PhILSCAT suppresses the peak error $20\%$ better than DSE, and does better in terms of PSNR and MAE on average.

\begin{figure*}[htbp]
\centering
\setlength{\tabcolsep}{-0.05cm}
\renewcommand{\arraystretch}{0.1}
\vspace{-0.5cm}
\begin{tabular}{ccccc}
\hspace{-8mm} \textbf{Reference Recon} &  \multicolumn{3}{c}{\textbf{Error Magnitudes}} & \hspace{-5mm} \textbf{Error MIPs}
\\[-3mm]
\includegraphics[width=.18\linewidth]{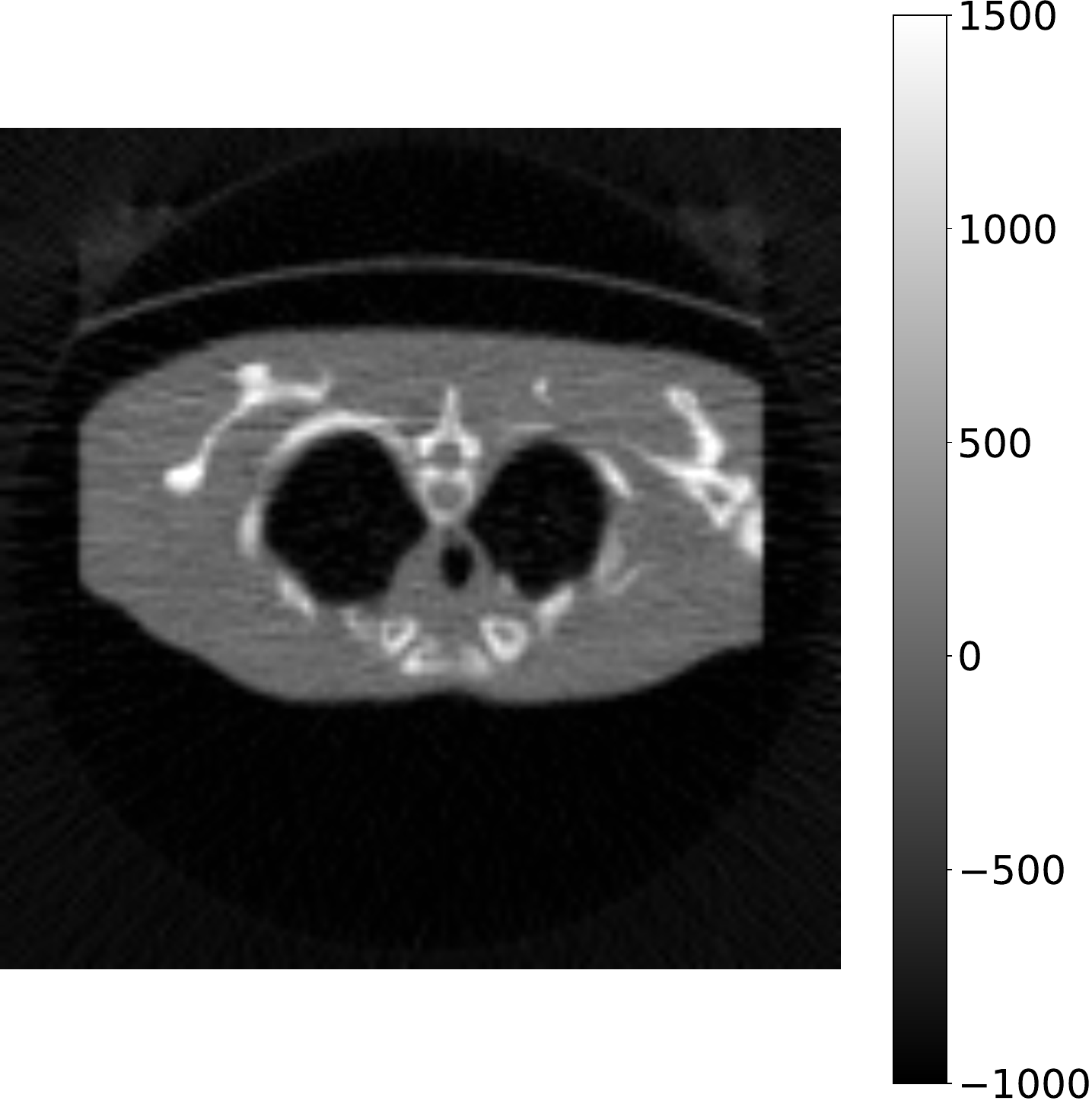} & \hspace{-0.0cm}
\includegraphics[width=.18\linewidth]{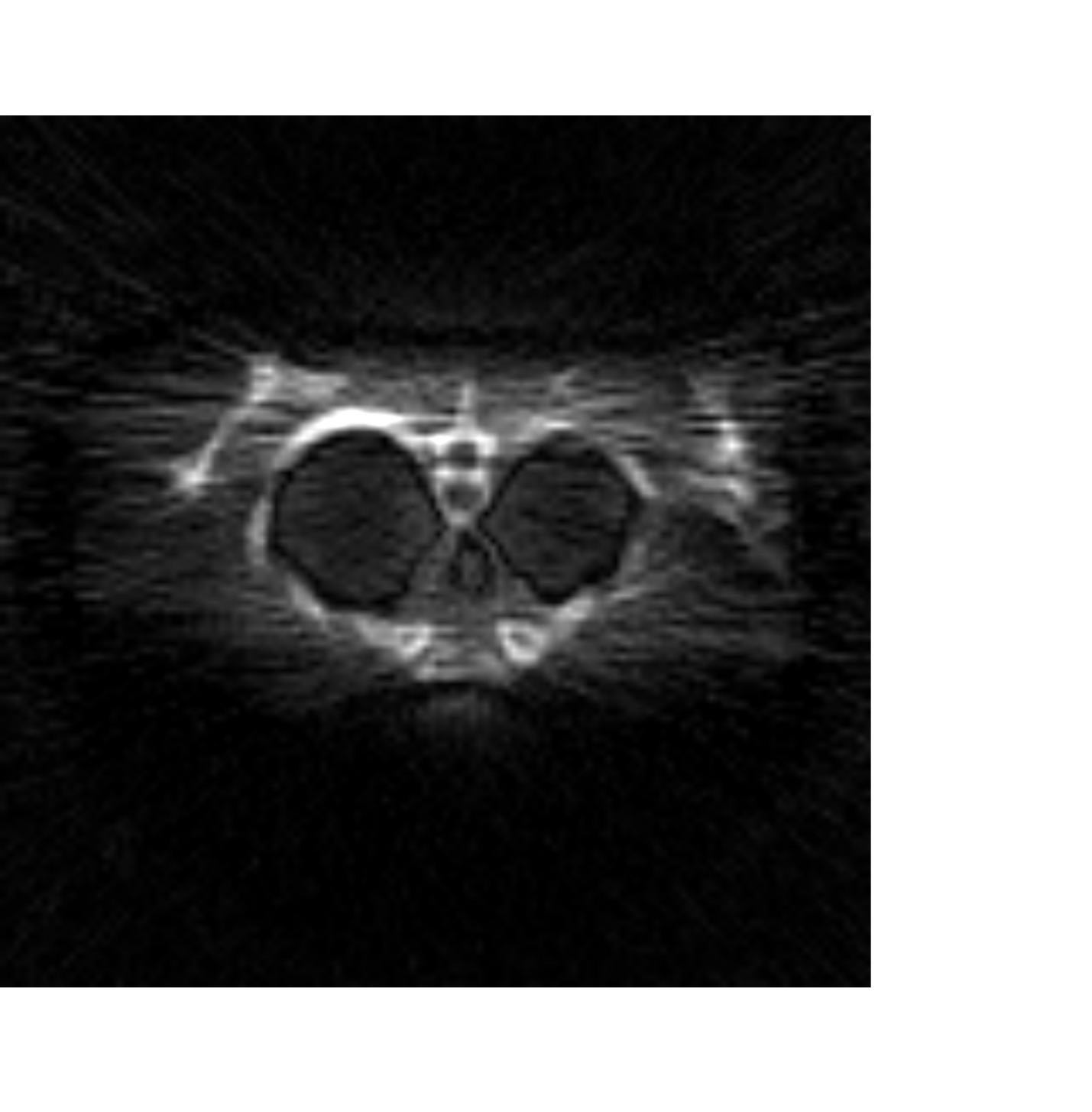} & \hspace{-0.6cm}
\includegraphics[width=.18\linewidth]{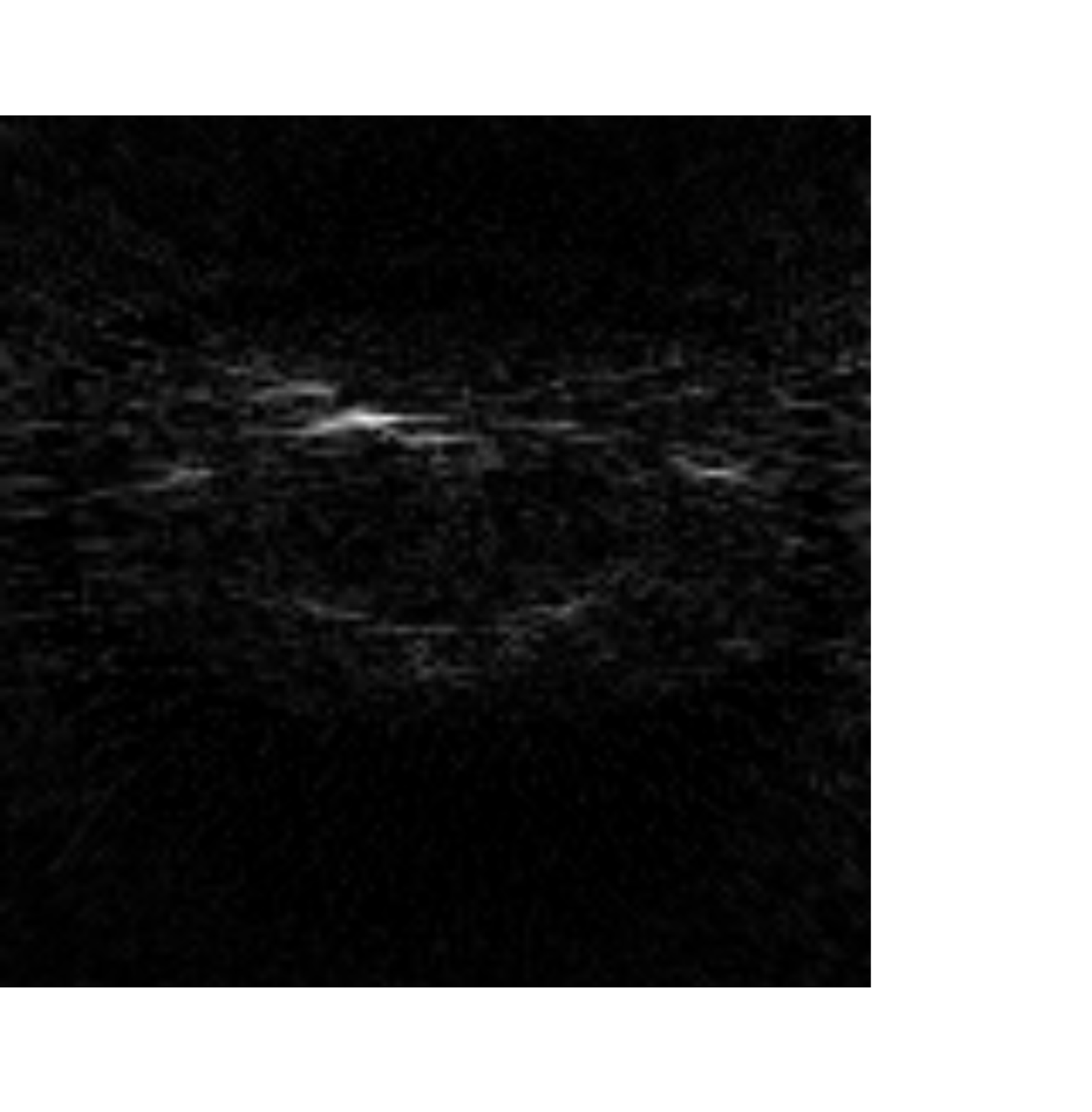} & \hspace{-0.6cm}
\includegraphics[width=.18\linewidth]{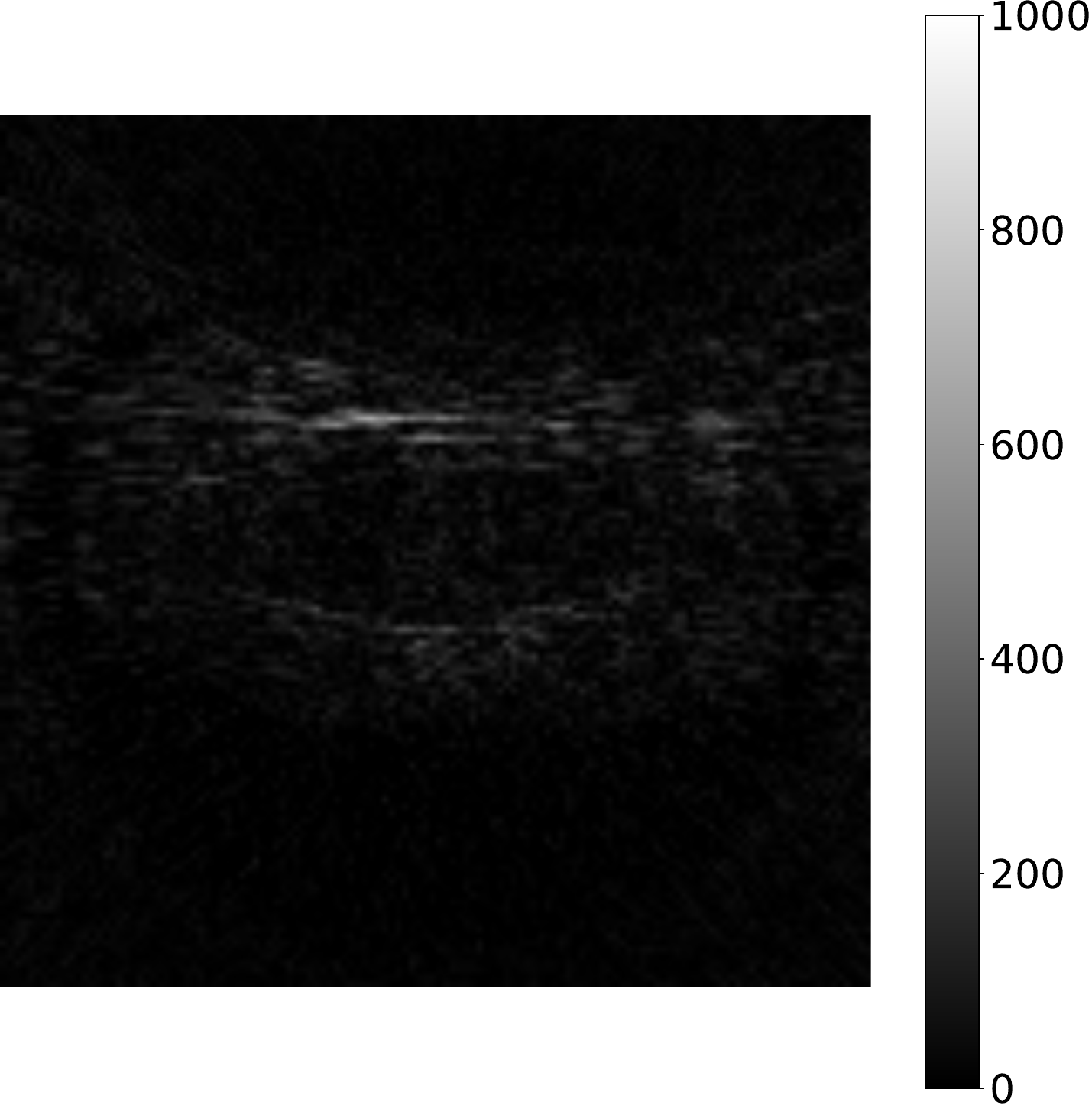} & \hspace{0.1cm}
\includegraphics[width=.18\linewidth]{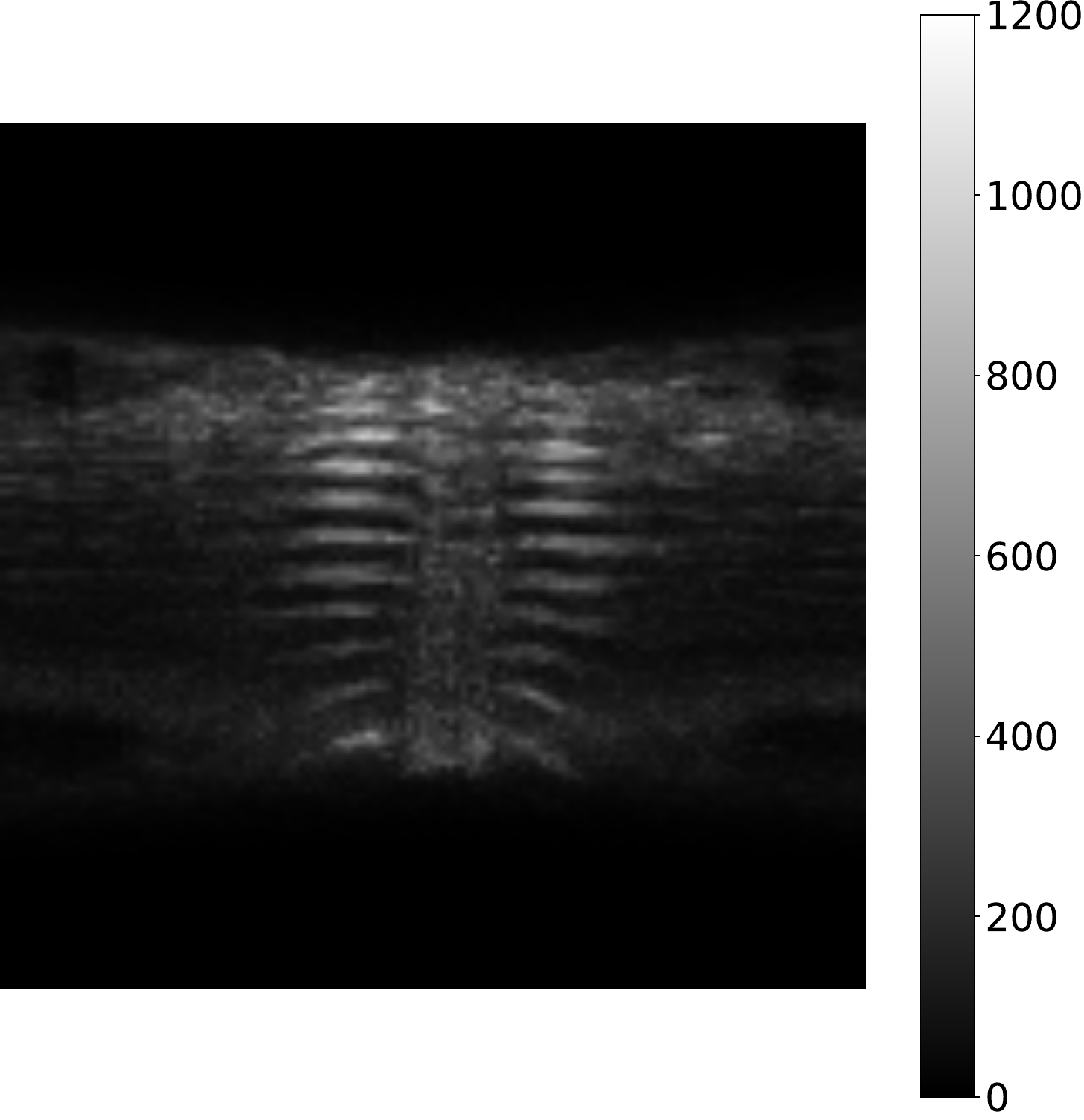}
\vspace{-3mm}
\\
\hspace{-0.7cm} {\small (a) Axial} & \hspace{-0.5cm} {\small (c) Total} & \hspace{-1.2cm} {\small(e) DSE} & \hspace{-1.2cm} {\small(g) PhILSCAT} & \hspace{-0.6cm} (i) {\small DSE MIP}
\\[2mm]
& \multicolumn{3}{c}{\textbf{Tighter HU Window Reconstructions}} &
\\[-3mm]
\includegraphics[width=.18\linewidth]{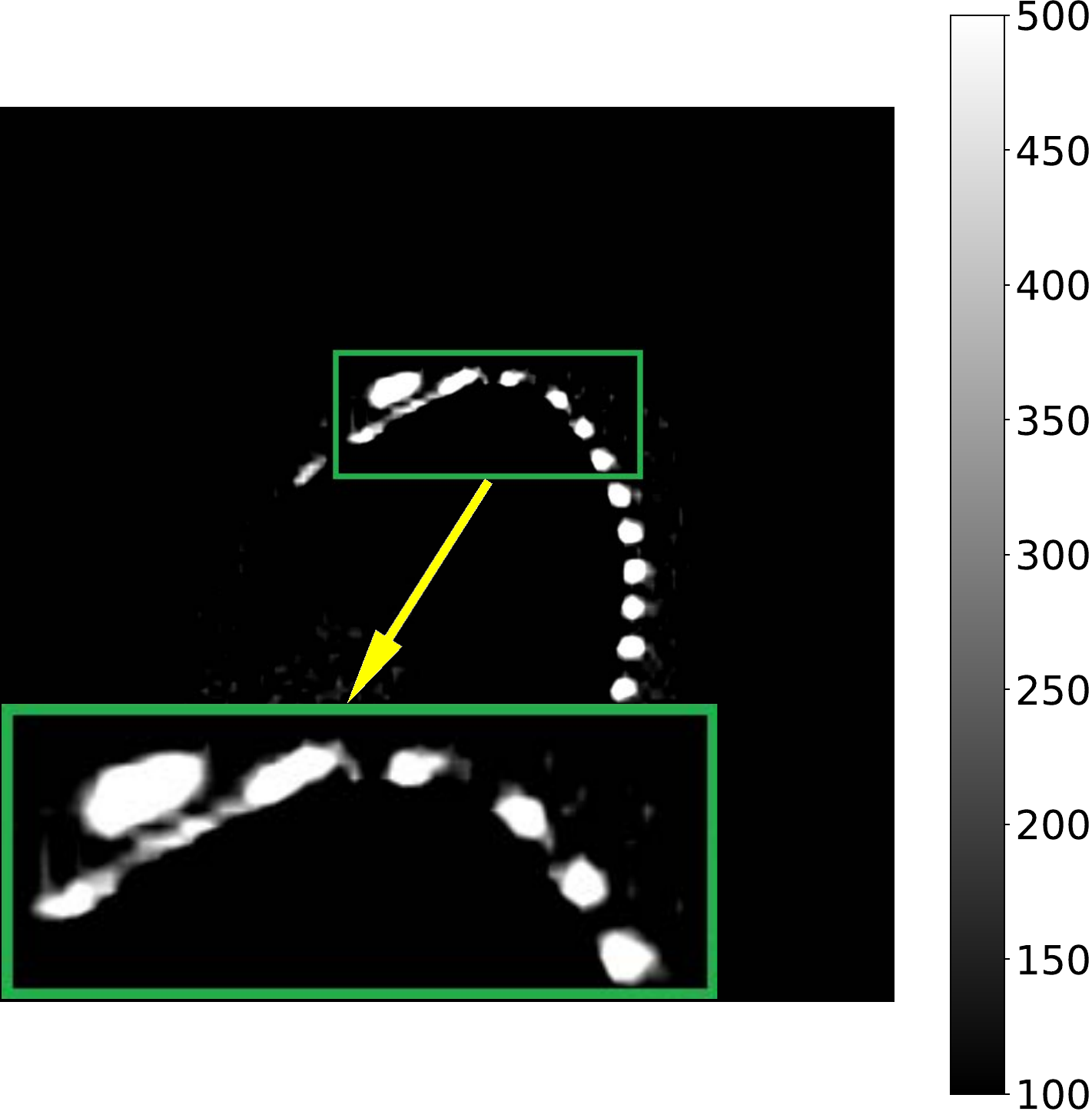} & \hspace{-0.0cm}
\includegraphics[width=.18\linewidth]{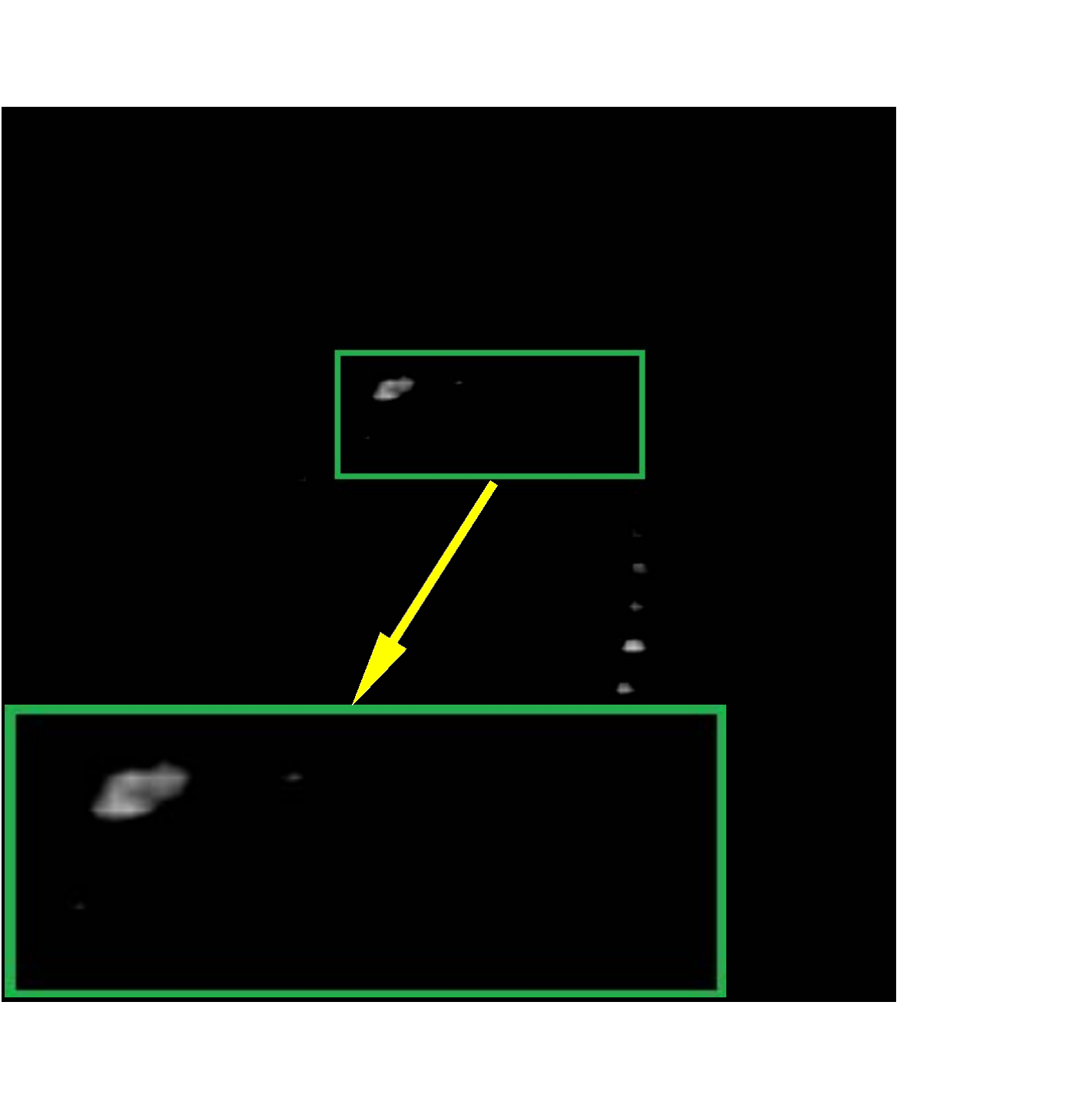} & \hspace{-0.6cm}
\includegraphics[width=.18\linewidth]{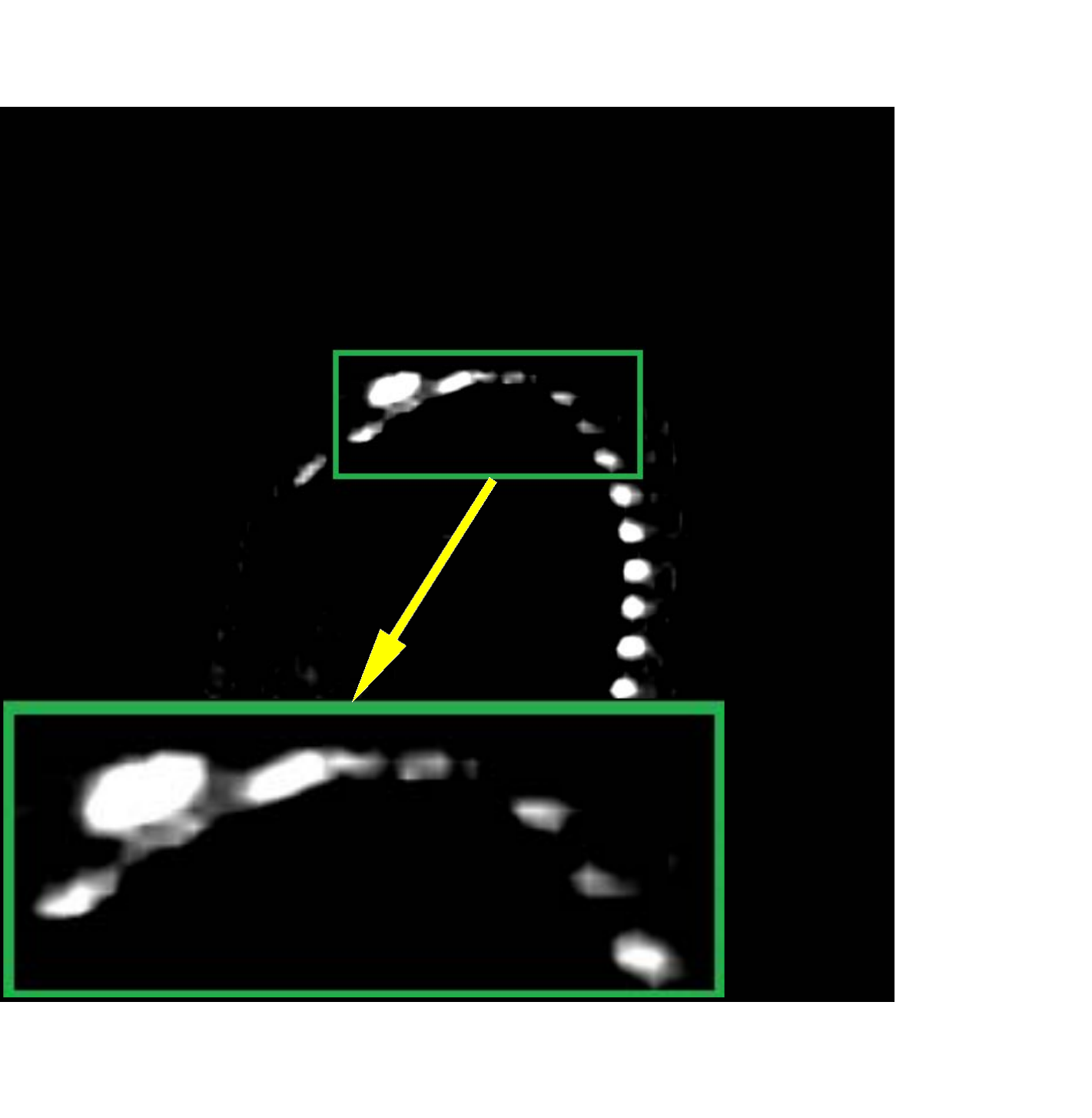} & \hspace{-0.6cm}
\includegraphics[width=.18\linewidth]{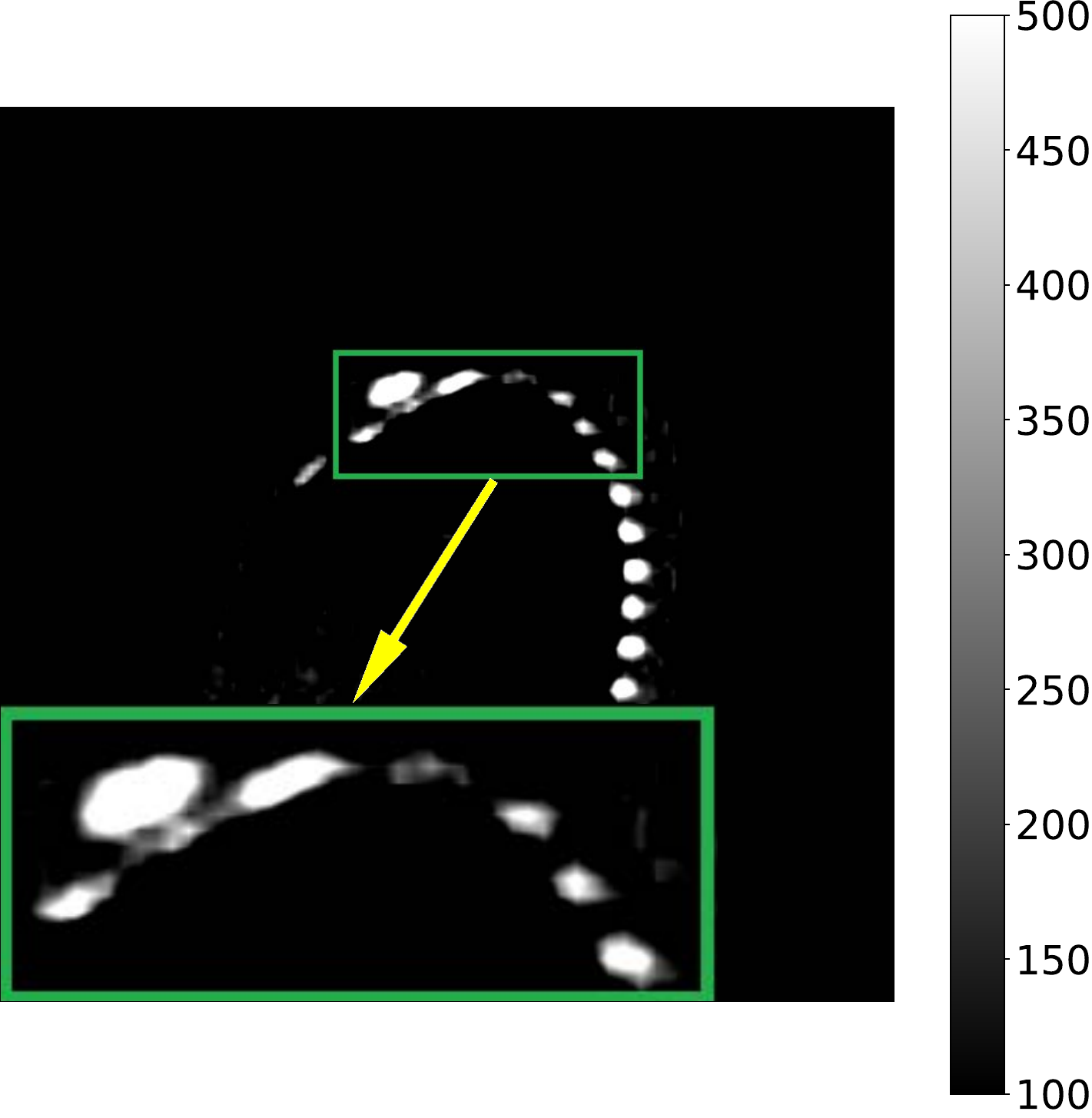} & \hspace{0.1cm}
\includegraphics[width=.18\linewidth]{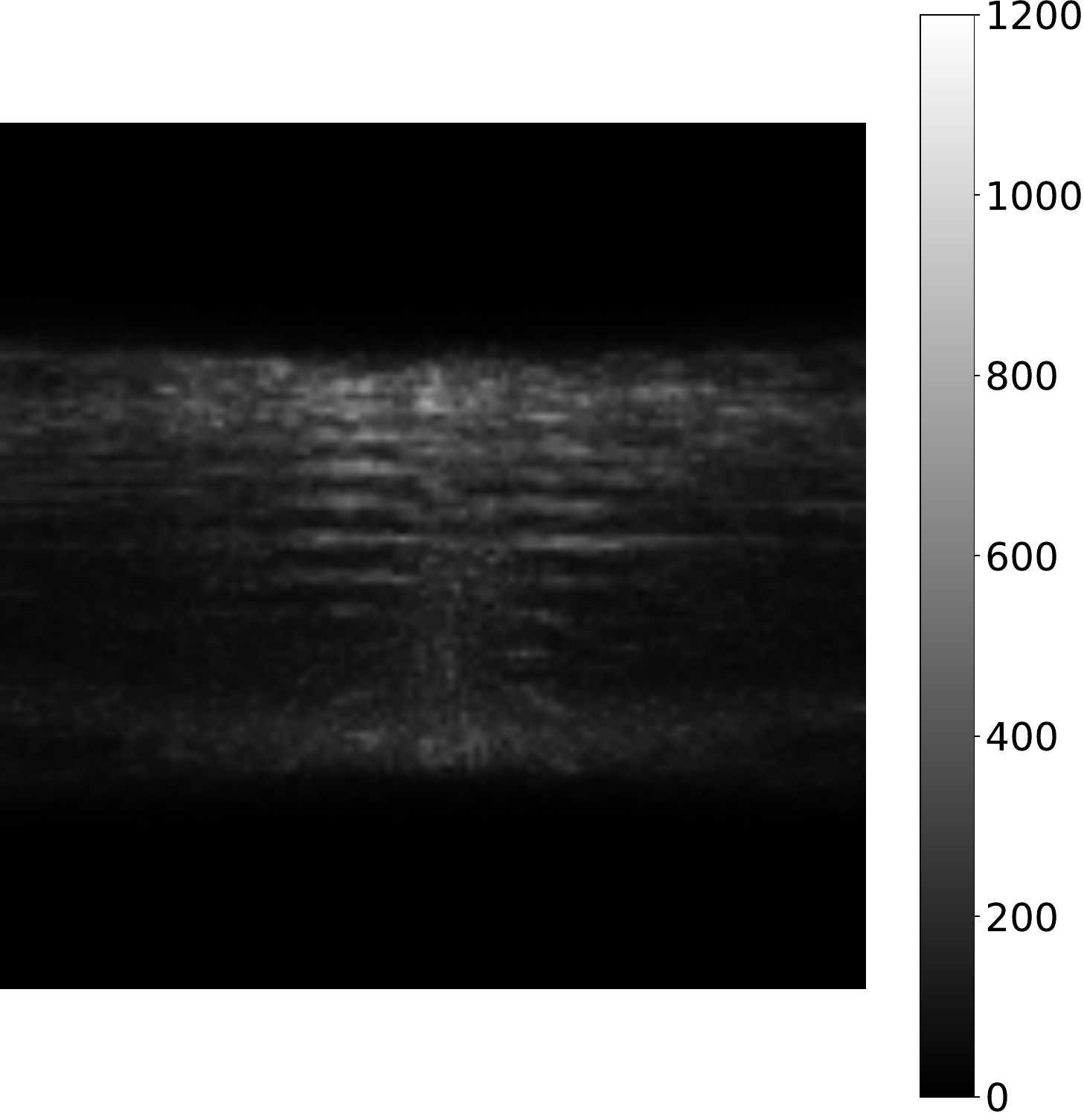}
\vspace{-0.2cm}\\
\hspace{-0.7cm} {\small (b) Sagittal} & \hspace{-0.5cm} {\small (d) Total} & \hspace{-1.2cm} {\small (f) DSE} & \hspace{-1.2cm} {\small (h) PhILSCAT} & \hspace{-0.7cm} {\small (j) PhILSCAT MIP}
\\

\end{tabular}
\caption{\small Polychromatic CBCT reconstructions of  anthropomorphic test phantom  (Numerical values in HU):
(a), (b) $p$-Reconstruction; (c) Error magnitude and (d) reconstruction using total measurements $\tau$; vs. (e), (f) using primary measurements estimated by DSE; or (g), (h) estimated by PhILSCAT. (i) and (j) Sagittal maximum intensity projections (MIPs) of the error volumes using $p^*$ estimated by DSE vs. estimated by PhILSCAT, respectively.}
\vspace{-0.5cm}
\label{fig:mag_err_comp_3D_CBCT_med_poly} 
\end{figure*}

Consistent with the results of the other experiments, as seen in the comparison of the magnitude of the reconstruction errors in Fig. \ref{fig:mag_err_comp_3D_CBCT_med_poly}, PhILSCAT performs visibly better
than DSE, especially in highly attenuating regions of the phantoms. To highlight differences in the sagittal slice reconstructions, they are shown with a tighter HU window, and the green-boxed regions are zoomed-in at the bottom left of  Figs.~\ref{fig:mag_err_comp_3D_CBCT_med_poly}(d) - \ref{fig:mag_err_comp_3D_CBCT_med_poly}(h). It is seen that in a highly attenuating region PhILSCAT recovers bone densities better than DSE.

\begin{table}[htbp]
\centering
\begin{tabular}{@{}lccc@{}}
\toprule
\multicolumn{1}{c}{  } & \multicolumn{1}{c}{Uncorrected} & \multicolumn{1}{c}{DSE} & \multicolumn{1}{c}{PhILSCAT} \\
\cmidrule(r){1-1}\cmidrule(lr){2-2}\cmidrule(lr){3-3}\cmidrule(l){4-4}
     PSNR (dB) & $26.9\pm0.3$ & $36.8\pm0.7$ & $\bold{37.2\pm0.8}$ \\
\cmidrule(r){0-0}
     SSIM & $0.864\pm0.01$ & $\bold{0.975\pm0.004}$ & $\bold{0.975\pm0.005}$ \\
\cmidrule(r){1-1}
     MAE (HU) & $35.8 \pm 1.6$ & $12.9 \pm 1.1$ & $\bold{12.3 \pm 1.2}$ \\
\cmidrule(r){1-1}
Peak$\,$Error$\,$(HU) & $1699$ & $970$ & $\bold{800}$ \\
\bottomrule
\end{tabular}
\caption{\small Polychromatic CBCT reconstructions of anthropomorphic phantoms: average reconstruction accuracy results and standard deviations for 3 test phantoms.}
\vspace{-0.5cm}
\label{table:recon_acc_poly_3D_recon_CBCT_med}
\end{table}

\subsection{Testing on Different $I_0$}
\label{sec:test_on_different_I}
To further test the generalizability, the algorithms trained on vacuum fluence $I_0$ for the polychromatic Ti rods experiment in Sec. \ref{sec:ti_rods_CBCT} were tested on the same test phantoms imaged using $I_0/4$ and average performances are shown in Table \ref{table:ablation_studies}.

Although results deteriorate significantly for both cases, the performance gap between PhILSCAT and DSE enhances, indicating that PhILSCAT is better able to generalize for smaller photon count (lower SNR) settings. This observation is also consistent with the parallel beam CT results where the photon count is considerably lower than CBCT experiments.

\subsection{Ablation Studies}
\subsubsection{Network Architecture}
\label{sec:network_structures_ablation}
To check the advantage of using the proposed network architecture of Sec. \ref{sec:network_structures} over a U-Net architecture \cite{ronneberger2015u} as used in DSE \cite{maier2018deep, maier2019real}, two alternatives were compared. The initial 128 channel reconstruction $\tilde{f}_\theta$ was compressed to a single channel ``image" using two consecutive convolutional layers with 16 and 1 output channels, respectively, and then the single channel ``image" was concatenated with the total measurement $\tau_\theta$, resulting in a 2-channel input to the U-Net. Besides having a 2-channel input, the rest of U-Net was identical to DSE. The proposed loss function in Sec. \ref{sec:loss_fct} was used to train the network. The results, for Polychromatic CBCT Ti rod case shown in Table \ref{table:ablation_studies}
demonstrate the clear advantage of the proposed architecture.

\subsubsection{Input}
\label{sec:poly_CBCT_input_ablation}
To evaluate contribution of  the  initial reconstruction to  PhILSCAT without modifying the architecture of the network, the initial scatter-corrupted reconstruction was replaced with a fixed random input (which, just as the initial reconstruction, is rotated by the view angle for each training projection), and training and testing was performed on the polychromatic Ti rods CBCT setting. Table \ref{table:ablation_studies}, showing an improvement in all metrics, indicates the advantage of using the initial reconstruction as an additional input. Also, we observed $\approx$ two-fold reduction (over the 3 test phantoms) in the number of voxels with errors larger than 500 HU.

\subsubsection{Loss Function} 
\label{sec:poly_CBCT_loss_function}
To evaluate the contribution of the tailored loss function of Sec. \ref{sec:loss_fct} to PhILSCAT, we compared the performance (in the polychromatic Ti rods CBCT setting)  to the same algorithm but without the application of filter $h$, that is, using a standard MSE loss on the scatter estimate.
As seen in Table \ref{table:ablation_studies}, the tailored loss provided an improvement on three of the four metrics. A more dramatic improvement was a $\approx$ two-fold reduction (over the 3 test phantoms) in the number of voxels with errors larger than 500 HU, similar to Sec. \ref{sec:poly_CBCT_input_ablation}.

\begin{table}[htbp]
\centering
\vspace{-0.25cm}
\begin{tabular}{@{}lcccc@{}}
\toprule
\multicolumn{1}{c}{  } & \multicolumn{1}{c}{PSNR$\,$(dB)} & \multicolumn{1}{c}{SSIM} & \multicolumn{1}{c}{MAE$\,$(HU)} &
\multicolumn{1}{c}{P.$\,$Err 
} 
\\
\cmidrule(r){1-1}\cmidrule(lr){2-2}\cmidrule(lr){3-3}\cmidrule(lr){4-4}\cmidrule(l){5-5}
    (a) \makecell[l]{DSE$\,I_0/4$\\PhILSCAT$\,I_0/4$} & \makecell{$43.4 {\small \pm 0.6}$\\$\bold{45.7{\small\pm 0.4}}$} & \makecell{$0.98
    $
    \\
    $\bold{0.99
    }$} 
    & \makecell{$36.3 {\small\pm 1.7}$\\$\bold{25.9 {\small\pm 1.2}}$} & \makecell{$1867$\\$\bold{1578}$} \\
     
    \cmidrule(r){0-0}

 (b) \makecell[l]{PhILSCAT U-Net} & $49.3 {\small\pm 1.4}$ & $0.996
 $ & $13.9 {\small\pm 1.2}$ & $1630$ \\

 (c) \makecell[l]{PhILSCAT random} & $51.1 {\small\pm 0.8}$ & $0.996
 $ & $12.4 {\small\pm 1.0}$ & $1098$ \\
    
 (d) \makecell[l]{PhILSCAT Scat. MSE} & $51.4 {\small \pm 1.2}$ & $0.997
 $ & $12.2 {\small \pm 1.1}$ & $1047$ \\
 (e)   \makecell[l]{PhILSCAT} & $\bold{51.7 {\small\pm 0.9}}$ & $\bold{0.997
 }$ & $\bold{11.9 {\small \pm 1.0}}$ & $\bold{954}$ \\
    
\bottomrule
\end{tabular}
\caption{\small Generalization and ablation studies (on polychromatic CBCT reconstruction of Ti rod phantoms, averages and standard deviations for 3 test phantoms): (a) DSE and PhILSCAT trained on $I_0$ but tested on $I_0/4$; (b) PhILSCAT with U-Net; (c) Initial reconstruction replaced by a random image; (d) Scatter MSE loss (as in DSE) used for training; (e) Original PhILSCAT. Standard deviation on all SSIM entries smaller than next significant digit.
}
\vspace{-0.5cm}
\label{table:ablation_studies}
\end{table}

%% file: conclusions.tex
\section{Conclusions}
\label{sec:conclusion}
We proposed two novel physics-inspired deep learning-based algorithms for scatter correction in X-ray CT images. The empirical results for the proposed methods demonstrate their advantage in various settings to another recent projection-by-projection-based data-driven de-scattering method. The proposed algorithms use scatter-corrupted measurements and an initial reconstruction of the object obtained from these measurements to estimate and correct the scatter in the projection domain. Unlike previous data-driven methods, the proposed algorithms incorporate constraints that are motivated by the physics of the CT imaging. The  cost function for training the algorithms is tailored to  express the norm of an image-domain error in the projection domain, but without the need for using the filtered backprojection. The results of numerical experiments are promising, indicating the potential of the proposed algorithms.

Possible directions for future work include 
theoretical analysis of the factors limiting the performance of  learning-based scatter reduction algorithms, and the use of a neural network architecture search to improve on the network architecture. It will also be of interest to compare the proposed approach to learning-based image-domain methods rather than the projection-domain. Evaluating the proposed approach on real experimental CT data is another important future study. Training and testing on real CT data would also involve scatter from other objects (e.g. detector and backplane), which would be learned and mitigated by the proposed approach (scatter from the detector is modeled in our monochromatic parallel beam experiments).
It would also be of interest to evaluate (in addition to the experiments described) how well a trained network generalizes to testing (use)  scenarios different from the training scenario - such as different objects, anatomies, or different scanner settings. Our experiments reported here, and experience from other deep learning tasks, suggests that limited generalizability is to be expected of any learning-based approach. However, as a practical matter, we note that it is standard practice to use different CT scanning protocols for different classes of subjects - e.g, male, female, or pediatric, even fine tuning the protocol to subject weight, let alone different animals or objects, or CT geometries and settings. We envision that similarly, different network models would be trained for such different classes (or even CT protocols) 
so that the scatter correction network does not have to generalize over an extremely wide range of scenarios.  We do not regard this as a significant limitation of the proposed approach. 

%% file: appendix.tex
\newpage
\section*{Appendix}
\subsection*{Denoising of Parallel-Beam Simulated Data}
We provide here some details of the pre-processing step used for denoising the training data in the parallel beam CT experiments reported in the manuscript in Sec.~\ref{sec:data_generation_parallel_beam}.
The idea of the pre-processing step is to identify the areas in the obtained 2D total measurements $\tau_\theta$ 
that contain only stochastic noise due to limited photon counts and little scatter, and to smooth them to reduce the noise, without affecting the areas that contain object or significant scatter information. 
To this end, we used the analytically calculated primary measurement $p_\theta$ corresponding to each projection to determine the areas in the ``shadow" of the object using the criterion that these areas
only include attenuated fluence $p_\theta < I_0$. 
These ``shadow" regions were used to define an initial binary ``smoothing-exclusion mask" $M_\theta \in \{0,1\}^{d \times d}$, 
and the non-masked part of the measurements were smoothed to reduce the noise. 

To minimize the modification of non-shadow regions containing significant (and informative) scatter by the pre-processing step, we extend the smoothing-exclusion mask to include neighboring
out-of-shadow regions. This is motivated by two observations: (i) scatter outside the object shadow is most significant near the shadow boundaries; and (ii)  further away from the shadow boundaries the scatter is spatially smooth, and will suffer little change when filtered by the noise-smoothing filter. 
To this end, each binary smoothing-exclusion mask was modified by morphological dilation with a $5 \times 5$ pixel structuring element $k$, producing the extended smoothing-exclusion mask $\bar{M}_\theta = k \oplus M_\theta$. 

Finally, the regions in the complement of the mask were smoothed by a 2D filter $G$ with Gaussian impulse response with width parameter $\sigma=2$. The smoothed regions were truncated back to the complement of the support of the smoothing-exclusion mask, and combined with the regions in the support of the smoothing\allowbreak-exclusion mask. 